\documentclass[longauth]{aa}  
\usepackage{graphicx}
\graphicspath{{figures/}}

\usepackage{siunitx}

\usepackage{amsmath}

\usepackage[normalem]{ulem} 

\usepackage{txfonts}
\usepackage{natbib}
\usepackage[draft]{hyperref}
\usepackage{url}
\usepackage{footmisc}
\usepackage{xcolor}

\usepackage[switch]{lineno}

\usepackage{natbib}
\bibpunct{(}{)}{;}{a}{}{,} 

\usepackage{tablefootnote}

\DeclareSIUnit\erg{erg}
\DeclareSIUnit\pe{p.e.}
\DeclareSIUnit\parsec{pc}
\DeclareSIUnit\gauss{G}

\usepackage{multirow}

\begin{document} 

\sisetup{range-phrase=..}
\sisetup{range-units=single}
\sisetup{list-final-separator = \translate{, and }}


   \title{A search for new supernova remnant shells in the Galactic plane with H.E.S.S.}

   \titlerunning{A search for new SNRs in the Galactic plane with H.E.S.S.}

\makeatletter
\renewcommand*{\@fnsymbol}[1]{\ifcase#1\or*\or$\dagger$\or$\ddagger$\or**\or$\dagger\dagger$\or$\ddagger\ddagger$\fi}
\makeatother

\author{\normalsize H.E.S.S. Collaboration
\and H.~Abdalla \inst{1}
\and A.~Abramowski \inst{2}
\and F.~Aharonian \inst{3,4,5}
\and F.~Ait~Benkhali \inst{3}
\and A.G.~Akhperjanian\protect\footnotemark[2] \inst{6,5} 
\and T.~Andersson \inst{10}
\and E.O.~Ang\"uner \inst{21}
\and M.~Arakawa \inst{43}
\and M.~Arrieta \inst{15}
% ...
}

\institute{
Centre for Space Research, North-West University, Potchefstroom 2520, South Africa \and 
Universit\"at Hamburg, Institut f\"ur Experimentalphysik, Luruper Chaussee 149, D 22761 Hamburg, Germany \and 
Max-Planck-Institut f\"ur Kernphysik, P.O. Box 103980, D 69029 Heidelberg, Germany \and 
Dublin Institute for Advanced Studies, 31 Fitzwilliam Place, Dublin 2, Ireland \and 
National Academy of Sciences of the Republic of Armenia,  Marshall Baghramian Avenue, 24, 0019 Yerevan, Republic of Armenia  \and
Yerevan Physics Institute, 2 Alikhanian Brothers St., 375036 Yerevan, Armenia \and
Institut f\"ur Physik, Humboldt-Universit\"at zu Berlin, Newtonstr. 15, D 12489 Berlin, Germany \and
University of Namibia, Department of Physics, Private Bag 13301, Windhoek, Namibia \and
GRAPPA, Anton Pannekoek Institute for Astronomy, University of Amsterdam,  Science Park 904, 1098 XH Amsterdam, The Netherlands \and
Department of Physics and Electrical Engineering, Linnaeus University,  351 95 V\"axj\"o, Sweden \and
Institut f\"ur Theoretische Physik, Lehrstuhl IV: Weltraum und Astrophysik, Ruhr-Universit\"at Bochum, D 44780 Bochum, Germany \and
GRAPPA, Anton Pannekoek Institute for Astronomy and Institute of High-Energy Physics, University of Amsterdam,  Science Park 904, 1098 XH Amsterdam, The Netherlands \and
Institut f\"ur Astro- und Teilchenphysik, Leopold-Franzens-Universit\"at Innsbruck, A-6020 Innsbruck, Austria \and
School of Physical Sciences, University of Adelaide, Adelaide 5005, Australia \and
LUTH, Observatoire de Paris, PSL Research University, CNRS, Universit\'e Paris Diderot, 5 Place Jules Janssen, 92190 Meudon, France \and
Sorbonne Universit\'es, UPMC Universit\'e Paris 06, Universit\'e Paris Diderot, Sorbonne Paris Cit\'e, CNRS, Laboratoire de Physique Nucl\'eaire et de Hautes Energies (LPNHE), 4 place Jussieu, F-75252, Paris Cedex 5, France \and
Laboratoire Univers et Particules de Montpellier, Universit\'e Montpellier, CNRS/IN2P3,  CC 72, Place Eug\`ene Bataillon, F-34095 Montpellier Cedex 5, France \and
DSM/Irfu, CEA Saclay, F-91191 Gif-Sur-Yvette Cedex, France \and
Astronomical Observatory, The University of Warsaw, Al. Ujazdowskie 4, 00-478 Warsaw, Poland \and
Aix Marseille Universit\'e, CNRS/IN2P3, CPPM UMR 7346,  13288 Marseille, France \and
Instytut Fizyki J\c{a}drowej PAN, ul. Radzikowskiego 152, 31-342 Krak{\'o}w, Poland \and
Funded by EU FP7 Marie Curie, grant agreement No. PIEF-GA-2012-332350,  \and
School of Physics, University of the Witwatersrand, 1 Jan Smuts Avenue, Braamfontein, Johannesburg, 2050 South Africa \and
Laboratoire d'Annecy-le-Vieux de Physique des Particules, Universit\'{e} Savoie Mont-Blanc, CNRS/IN2P3, F-74941 Annecy-le-Vieux, France \and
Landessternwarte, Universit\"at Heidelberg, K\"onigstuhl, D 69117 Heidelberg, Germany \and
Universit\'e Bordeaux, CNRS/IN2P3, Centre d'\'Etudes Nucl\'eaires de Bordeaux Gradignan, 33175 Gradignan, France \and
Oskar Klein Centre, Department of Physics, Stockholm University, Albanova University Center, SE-10691 Stockholm, Sweden \and
Wallenberg Academy Fellow,  \and
Institut f\"ur Astronomie und Astrophysik, Universit\"at T\"ubingen, Sand 1, D 72076 T\"ubingen, Germany \and
Laboratoire Leprince-Ringuet, Ecole Polytechnique, CNRS/IN2P3, F-91128 Palaiseau, France \and
APC, AstroParticule et Cosmologie, Universit\'{e} Paris Diderot, CNRS/IN2P3, CEA/Irfu, Observatoire de Paris, Sorbonne Paris Cit\'{e}, 10, rue Alice Domon et L\'{e}onie Duquet, 75205 Paris Cedex 13, France \and
Univ. Grenoble Alpes, IPAG,  F-38000 Grenoble, France \protect\\ CNRS, IPAG, F-38000 Grenoble, France \and
Department of Physics and Astronomy, The University of Leicester, University Road, Leicester, LE1 7RH, United Kingdom \and
Nicolaus Copernicus Astronomical Center, Polish Academy of Sciences, ul. Bartycka 18, 00-716 Warsaw, Poland \and
Institut f\"ur Physik und Astronomie, Universit\"at Potsdam,  Karl-Liebknecht-Strasse 24/25, D 14476 Potsdam, Germany \and
Friedrich-Alexander-Universit\"at Erlangen-N\"urnberg, Erlangen Centre for Astroparticle Physics, Erwin-Rommel-Str. 1, D 91058 Erlangen, Germany \and
DESY, D-15738 Zeuthen, Germany \and
Obserwatorium Astronomiczne, Uniwersytet Jagiello{\'n}ski, ul. Orla 171, 30-244 Krak{\'o}w, Poland \and
Centre for Astronomy, Faculty of Physics, Astronomy and Informatics, Nicolaus Copernicus University,  Grudziadzka 5, 87-100 Torun, Poland \and
Department of Physics, University of the Free State,  PO Box 339, Bloemfontein 9300, South Africa \and
Heisenberg Fellow (DFG), ITA Universit\"at Heidelberg, Germany  \and
GRAPPA, Institute of High-Energy Physics, University of Amsterdam,  Science Park 904, 1098 XH Amsterdam, The Netherlands \and
Department of Physics, Rikkyo University, 3-34-1 Nishi-Ikebukuro, Toshima-ku, Tokyo 171-8501, Japan
% ...
}

\offprints{H.E.S.S.~collaboration,
\protect\\\email{\href{mailto:contact.hess@hess-experiment.eu}{contact.hess@hess-experiment.eu}};
\protect\\\protect\footnotemark[1] Corresponding authors
\protect\\\protect\footnotemark[2] Deceased
}

\authorrunning{H.E.S.S. Collaboration}

   \date{submitted on }

 \abstract
{
A search for new supernova remnants (SNRs) has been conducted using TeV $\gamma$-ray data from the H.E.S.S.\ Galactic plane survey. As an identification criterion, shell morphologies that are characteristic for known resolved TeV SNRs have been used. Three new SNR candidates were identified in the H.E.S.S.\ data set with this method. Extensive multiwavelength searches for counterparts were conducted. A radio SNR candidate has been identified to be a counterpart to HESS\,J1534$-$571. The TeV source is therefore classified as a SNR. For the other two sources, HESS\,J1614$-$518 and HESS\,J1912$+$101, no identifying counterparts have been found, thus they remain SNR candidates for the time being. TeV-emitting SNRs are key objects in the context of identifying the accelerators of Galactic cosmic rays. The TeV emission of the relativistic particles in the new sources is examined in view of possible leptonic and hadronic emission scenarios, taking the current multiwavelength knowledge into account. 
}


   \maketitle

\section{Introduction}
\label{Sect:Introduction}

Most Galactic supernova remnants (SNRs) have been discovered via radio continuum surveys, that is,\ through synchrotron emission of nonthermal electrons with $\sim$ \SI{}{\giga\eV} energies. Only a handful of SNRs have been discovered in the optical band \citep[e.g.,][]{2010AJ....140.1163F}. Several Galactic SNRs have also been detected first in X-ray surveys with the {\it ROSAT} satellite and also with {\it ASCA} \citep{1995ASPC...80..432A,2003ApJ...589..253B,1991A&A...246L..28P,2001ApJS..134...77S,2001PASJ...53L..21B,2004PASJ...56.1059Y}. 
These SNRs (e.g., RX\,J1713.7$-$3946) typically have low radio-surface brightness and/or are in confused regions. Sources are usually classified as SNR candidates from their respective survey data (or serendipitous observations of the field), for example,\ in the radio band, based on their shell-like morphology. An identification of the source as SNR requires an additional independent measurement such as a spectral signature or an identifying detection from another waveband.

This paper presents a systematic search for new Galactic SNR candidates using the TeV $\gamma$-ray band for the discovery of these objects. Previous \SI{}{\tera\eV} SNR studies, for example,\ with the HEGRA Cherenkov telescope system \citep{2001A&A...370..112A,2002A&A...395..803A} or with H.E.S.S.\ \citep{2004Natur.432...75A,special_pop}, focused on the detection of TeV emission from SNRs (or SNR candidates) that have been known from other wavebands. The data that are used for the search presented here stem from the H.E.S.S. Galactic Plane Survey (HGPS) that is presented in detail elsewhere \citep{special_HGPS}. In order to identify SNR candidates in the \SI{}{\tera\eV} band that do not have known counterparts in other wavebands, only morphological signatures, namely a shell-type appearance, can be exploited. There are no TeV spectral (or temporal) characteristics of known SNRs that set them apart from other Galactic TeV source types. Nevertheless, TeV-emitting SNRs are a significant class of identified Galactic \SI{}{\tera\eV} sources. These SNRs are the second most numerous after identified pulsar wind nebulae (PWNe).\footnote{http://tevcat.uchicago.edu/} Therefore, a pure morphological criterion, namely a significant TeV shell-like appearance, can be considered sufficient for classification as a (\SI{}{\tera\eV}) SNR candidate. 

Besides SNRs, \SI{}{\tera\eV} shell-like emission could also stem from superbubbles such as 30\,Dor\,C in the Large Magellanic Cloud. 30\,Dor\,C is a known TeV emitter \citep{2015Sci...347..406H}, but because of limited statistics and the small extension of the source, current TeV data cannot be used to tell if the TeV emission is shell-like, whereas a shell-like morphology is well established in radio and $H_{\alpha}$ emission \citep{1985ApJS...58..197M} and X-rays \citep{2004ApJ...602..257B}. Also other wind-blown gas cavities with surrounding shells into which hadronic particles from central accelerators are moving could appear like \SI{}{\tera\eV} shells. Moreover, given the typical $\sim$ \SI{0.05}{\degree} to $\sim$ \SI{0.1}{\degree} angular resolution of the H.E.S.S.\ telescopes, a chance constellation of several sources could mimic a shell appearance. For the presented work, it has not been attempted to numerically simulate how often these various possible misinterpretations due to non-SNR \SI{}{\tera\eV} shells may occur amongst the selected TeV SNR candidates. Given the possible alternative interpretations, an identification of an individual candidate as a real (or highly likely) SNR needs to come from other wavebands.

In particular, follow-up observations with current X-ray satellites are promising, given that all identified Galactic \SI{}{\tera\eV} SNRs have bright (in terms of current pointed X-ray sensitivity) X-ray counterparts. Current X-ray surveys may not have led to a discovery of the objects because of interstellar absorption (in particular for {\it ROSAT}) or stray-light contamination from nearby, strong, X-ray sources (e.g.,\ for {\it ASCA}). One prominent such example is the identified SNR HESS\,J1731$-$347, which was discovered in the course of the ongoing HGPS \citep{2008A&A...477..353A}. The source has a significant \SI{}{\tera\eV} shell-like appearance \citep{2011A&A...531A..81H}, as well as radio and X-ray counterparts, but was not detected as a SNR candidate in previous radio or X-ray surveys \citep{2008ApJ...679L..85T}.

While the lack of X-ray emission is a hindrance for the identification of the SNR candidate, it might have implications for the interpretation of the TeV source. Supernova remnants  are prime candidates for the acceleration of the bulk of Galactic cosmic rays (CRs). But the TeV emission from many known, TeV-bright SNR shells may be dominated by leptonic emission from inverse Compton scattering of relativistic electrons. In contrast, the nondetection of X-ray emission from one of the new SNR candidates, even at the current pointed satellite sensitivity level, might indicate that the TeV $\gamma$-ray emission stems predominantly from hadronically induced $\pi^0$-decay. This could be inferred by adopting the arguments that have been made to interpret dark \SI{}{\tera\eV} accelerators, that is,\ TeV sources without X-ray and radio counterparts, as evolved SNRs \citep{2006MNRAS.371.1975Y}.

The search presented in this paper has revealed three SNR candidates, \object{HESS\,J1534$-$571} \citep{2015arXiv150903872P}, \object{HESS\,J1614$-$518} \citep{2006ApJ...636..777A}, and \object{HESS\,J1912$+$101} \citep{2008A&A...484..435A}. HESS\,J1534$-$571 was additionally classified as a confirmed SNR in the course of the presented study based on its identification with a radio SNR candidate. For HESS\,J1614$-$518 and HESS\,J1912$+$101, no counterparts have been found yet that would permit firm identification. 

The paper is organized as follows: In Sect.\,\ref{Sect:HessAnalysis}, the \SI{}{\tera\eV} SNR candidate search and identification strategies applied to the HGPS data are presented. A dedicated H.E.S.S.\ data analysis of the three sources that fulfill the candidate criteria is performed, including a \SI{}{\tera\eV} morphological and spectral characterization. For the first time, SNR candidates are established from the TeV band alone. A thorough evaluation of the available multiwavelength (MWL) data toward the sources including nondetections is necessary to permit valid physics conclusions for these TeV-selected sources. In Sect.\,\ref{Sect:RadioGeV}, the search for MWL counterparts in the radio continuum and the GeV bands is presented, which led to positive identifications for HESS\,J1534$-$571 (radio, \SI{}{\giga\eV}) and HESS\,J1614$-$518 (\SI{}{\giga\eV}). Section \ref{Sect:XrayIRSubmm} deals, to the extent necessary, with MWL data (X-ray, infrared, and radio/sub-mm line data) and catalog searches that have not revealed clear counterparts. In the discussion (Sect.\,\ref{Sect:Discussion}), significant results from the MWL searches are summarized. The nondetections of X-ray counterparts are evaluated in view of implications for the relativistic particles that give rise to the TeV emission, specifically for HESS\,J1534$-$571. The possible energy content of relativistic protons is derived for each object, employing the interstellar matter (ISM) line data to gain information on the potential surroundings of the objects.

\subsection*{On the SNR nomenclature}
\label{Subsect:Nomenclature}

Identified SNRs are often classified according to their radio and X-ray morphologies \citep[e.g.,][]{2015A&ARv..23....3D}. Throughout this paper, the term SNR is used synonymously with shell-type SNR, which is applicable to objects with emission morphologies that are identified with SNR shocks. The term TeV SNR (meaning TeV-emitting SNR) is also used for an unresolved \SI{}{\tera\eV} source or a source with unclear morphology, for which the identification of the \SI{}{\tera\eV} emission with a SNR is highly probable, for example, based on spectral identification with known SNRs and the apparent lack of alternative explanations. The SNR Cas\,A belongs to this category \citep{2001A&A...370..112A}. The term \SI{}{\tera\eV} shell does not imply identification as a SNR; it is only used as a morphological description.

A (radio) nebula that is formed by a pulsar wind is often referred to as a  plerionic SNR. In this paper, this approach is not followed. Such a (\SI{}{\tera\eV}) nebula is rather called a PWN, following the nomenclature widely used in the high-energy astrophysics community. In fact, PWNe constitute the most abundant identified Galactic \SI{}{\tera\eV} source class. These objects are relevant for the work presented here because they are used as a blueprint for the morphological null hypothesis, against which the target shell morphologies are tested. 

Composite SNRs consist of a SNR and a PWN. This nomenclature is maintained for the \SI{}{\tera\eV} band. Mixed-morphology SNRs \citep{1998ApJ...503L.167R} consist of SNRs showing a (nonthermal) radio shell and center-filled thermal X-ray emission. Since such thermal emission processes do not have a correspondence in the \SI{}{\tera\eV} band, this morphological classification is not very important for the characterization of \SI{}{\tera\eV} SNR morphologies.


\section{H.E.S.S.\ data analysis, shell search, and results}
\label{Sect:HessAnalysis}

\subsection{High Energy Stereoscopic System (H.E.S.S.)}
\label{SubSect:HESSdescription}

The \SI{}{\tera\eV} data presented in this paper were taken with the H.E.S.S.\ array in its first phase, in which the array consisted of four identical \SI{12}{\metre} imaging atmospheric Cherenkov telescopes, located in the Khomas Highland of Namibia at an altitude of \SI{1800}{\metre} above sea level. The telescopes are used to detect very high-energy $\gamma$-rays through registering the Cherenkov light that is emitted by a shower of charged particles, initiated by the primary photon entering the atmosphere. The telescopes are equipped with mirrors with a total area of \SI{107}{\square\metre} and cameras of 960 photomultiplier tubes each. The energy threshold of the array is roughly \SI{100}{\giga\eV} at zenith and increases with zenith angle. The direction and energy of the primary photon can be reconstructed with an accuracy of $\leq$ \SI{0.1}{\degree} and $\sim$ \SI{15}{\percent}, respectively. The field of view (FoV) of the cameras of \SI{5}{\degree} in diameter for imaging air showers translates into a celestial FoV for reconstructed $\gamma$-rays with roughly constant $\gamma$-ray acceptance of \SI{2}{\degree} ($\gtrsim$ \SI{90}{\percent} of peak acceptance) or \SI{3}{\degree} ($\gtrsim$ \SI{70}{\percent}) in diameter. This makes the H.E.S.S.\ array well suited for studies of extended sources such as the sources presented in this work \citep[for further details see][]{Aharonian:2006}.

For the second phase of H.E.S.S., a fifth telescope with a much larger mirror area of \SI{614}{\square\metre} was added to the array in July 2012 \citep{2015arXiv150902902H}. The additional telescope reduces the energy threshold and improves the point-source sensitivity of the array specifically for soft-spectrum sources, but with the restriction to a smaller FoV in the low-energy range. No relevant data for this paper have been taken in this configuration.

\subsection{H.E.S.S.\ data analysis}
\label{SubSect:HessAnalysis}
The work presented here is based on two not fully identical data sets that were treated with two different analysis chains. For the \SI{}{\tera\eV} shell search on a grid as described in Sect.\ \ref{SubSubSect:TeVShellSearch_MotivationTwoStep}, products from the HGPS data set (Sect. \ref{SubSubSect:HGPSdataset}) published by the \citet{special_HGPS} were used. For the individual source analysis described in Sects.\ \ref{SubSubSect:TeVShellSearch_FittingProcedure} (morphological fits), \ref{SubSect:Surfacebrightness} (surface brightness maps), and \ref{SubSect:Energyspectra} (energy spectra), a dedicated data selection and refined analysis (Sect.\ \ref{SubSubSect:HessDedicatedAnalysis}) was used.

\subsubsection{The HGPS data set}
\label{SubSubSect:HGPSdataset}
Between 2004 and 2013, a survey of the Galactic plane was conducted with the H.E.S.S. telescopes. Initial survey observations \citep{2006ApJ...636..777A} have been followed by deeper observations of regions of interest, and many known objects in the Galactic plane have also been observed individually \citep[e.g.,][]{2011A&A...531A..81H,2016Natur.531..476H,special_1713,special_HGPS}. The H.E.S.S. Galactic Plane Survey (HGPS) data set consists of all data taken in the longitude range between $l = \SI{250}{\degree}$ and $l = \SI{65}{\degree}$ (including the Galactic center) and $|b| \lesssim \SI{3.5}{\degree}$ in latitude. The resulting data set does not provide homogeneous sensitivity. The point source sensitivity in the core survey region is $\lesssim$ \SI{1.5}{\percent} of the Crab nebula integral flux above \SI{1}{\tera\eV}, but better sensitivities are achieved for many individual regions. Details of the data set, analysis methods, and the resulting source catalog are reported in \citet{special_HGPS}. For the work presented in this paper, only a morphological analysis of the HGPS sky map has been performed. Therefore, only sky map products of the HGPS have been used, namely a sky map of $\gamma$-ray event candidates after image-shape based background rejection, a sky map of estimated remaining background level, and an exposure sky map. The search for SNR candidates has been performed on an equidistant sky grid and not, for example,\ on preselected HGPS source positions.

\subsubsection{Dedicated H.E.S.S.\ data analysis of the new SNR candidates}
\label{SubSubSect:HessDedicatedAnalysis}

\noindent\textit{Data sets}\\
After the identification of SNR candidates in the HGPS data, for which the procedure is described in more detail below, the available H.E.S.S. data in the sky area around the candidates were processed following standard H.E.S.S. procedures for individual source analysis. The same primary analysis chain (calibration of the raw data, $\gamma$-ray reconstruction, and background rejection algorithms) as that used for the HGPS primary analysis was chosen, but there are some differences due to optimization toward the individual sources rather than for a survey. In particular, the sky map data selection leads to a more homogeneous exposure at and around the source positions than in the HGPS sky maps. For spectral analysis, the choice of background control regions is particularly important and was optimized for the individual sources. Also, slightly more data could be used for the analysis than what was available for the HGPS analysis. In the following, the individual source data analysis is described in more detail.

\vspace{1ex}\noindent\textit{Observations and quality selection}\\
The data for this work were taken between April 2004 and May 2013. The data from all runs (sky-tracking observations of typically \SI{28}{\minute} duration) were calibrated with standard H.E.S.S. procedures \citep{Bolz:2004,Aharonian:2006}. Runs used for the analysis of each source were then selected based on the distance of tracking to source position and on run quality selection cuts \citep{Aharonian:2006,Hahn:2014}. For this run selection, source positions of the previous H.E.S.S. publications of HESS\,J1614$-$518 \citep{2006ApJ...636..777A} and HESS\,J1912$+$101 \citep{2008A&A...484..435A} were used, while the HGPS pipeline source position was used for HESS\,J1534$-$571.\footnote{It was verified that redoing the run selection using the shell centroids that are derived later in this work (Tab.\,\ref{Table:morphologyresults}) would have a negligible effect.} Two different cut levels were used for detection and morphological analysis and for spectral analysis, respectively. For the morphological studies, all runs with a maximum offset of \SI{3}{\degree} were used. For the spectral analysis, the maximum offset was reduced to \SI{1.5}{\degree}, stricter (spectral) quality cuts were used, and sufficient background control regions need to be available in each run. Resulting acceptance-corrected\footnote{That is,\ dead-time-corrected, corrected for acceptance change depending on off-optical axis angle, and normalized to a standard offset of \SI{0.5}{\degree}} observation times are shown in Table \ref{Table:obstimes}.

\renewcommand{\arraystretch}{1.5}

\begin{table}[t]
\caption{Acceptance-corrected H.E.S.S observation time used for the dedicated analysis of the new SNR candidates.}
\label{Table:obstimes}
\centering
\begin{tabular}{lcc}
\hline\hline
Source & {Sky maps} & {Energy spectra} \\
\hline
HESS\,J1534$-$571 & \SI{61.8}{\hour}  & \SI{25.4}{\hour} \\
HESS\,J1614$-$518 & \SI{34.2}{\hour}  & \SI{10.0}{\hour} \\
HESS\,J1912$+$101 & \SI{121.1}{\hour} & \SI{43.2}{\hour} \\
\hline\hline
\end{tabular}
\end{table}

\renewcommand{\arraystretch}{1.0}

\vspace{1ex}\noindent\textit{Data analysis}\\
Event direction and energy reconstruction was performed using a moment-based Hillas analysis as described in \citet{Aharonian:2006}. Gamma-ray-like events were selected based on the image shapes with a boosted decision tree method \citep{Ohm:2009}. The residual background (from hadrons, electrons, and potentially from diffuse $\gamma$-ray emission in the Galactic plane) is estimated from source-free regions in the vicinity of the studied sources. For sky maps and the morphological studies using these maps, the background at each sky pixel is estimated from a ring around the pixel position \citep{2007A&A...466.1219B}. An adaptive algorithm is applied to optimize the size of the ring which blanks out known sources or excesses above a certain significance level from the rings, requiring thus an iterative or bootstrapping process \citep{special_HGPS}. To ensure a homogeneous acceptance for background events, particularly in the Galactic plane with high optical noise, an image amplitude cut of 160 photoelectrons (p.e.) is applied for sky maps. Together with the energy reconstruction bias cut that is derived from Monte Carlo simulations and discards events with a bias above \SI{10}{\percent}, the presented data sets have an energy threshold of about \SI{600}{\giga\eV}.

To derive energy spectra, the image amplitude cut is usually reduced to achieve a wider spectral energy range. For HESS\,J1534$-$571 and HESS\,J1614$-$518, an amplitude cut of \SI{60}{\pe} was set, leading to a spectral energy threshold of around \SI{300}{\giga\eV}. The spectral background is derived from background control regions that are defined run-wise and are chosen to have the same offset to the camera center as the source region \citep{2007A&A...466.1219B}, to ensure a nearly identical spectral response to background events in the source and background control regions. Similar to the process above, sky areas with known sources or with excess events above a certain significance threshold have to be excluded. After lowering the threshold to \SI{60}{\pe} for a spectral analysis of HESS\,J1912$+$101, this process failed to deliver background estimates with acceptable systematic errors as quoted below. Therefore the \SI{160}{\pe} amplitude cut was maintained for the spectral analysis of this source. Possible reasons are that the sky area around HESS\,J1912$+$101 contains significant soft sub-threshold sources or diffuse emission, or that it suffers from high optical stellar noise.

All three data sets contain large portions of runs that were performed under nonoptimal conditions for spectral analysis, i.e.,\,they were not taken in wobble run mode with a fixed offset between source and tracking position. Given the extensions of the sources of up to \SI{\sim 1}{\degree} diameter, many runs could even not be used at all for spectral analysis because no suitable background control regions were available. The large sizes of the sources also lead to substantial susceptibility of the spectral results to potential errors of the background estimate. All results were cross-checked using an independent calibration and simulation framework, combined with an alternative reconstruction technique based on a semi-analytical description of the air shower development by \citet{Naurois:2009}. From the comparison between the two different analyses and from the results obtained when varying cuts and background estimates, systematic errors for the spectral results of \num{\pm 0.2} for photon indices and \SI{30}{\percent} for integrated flux values were estimated; these values are slightly higher than those typically obtained in H.E.S.S.\ analyses.

\subsection{\SI{}{\tera\eV} shell search: Method and results}
\label{SubSect:TeVShellSearch}

\subsubsection{Motivation of the approach}
\label{SubSubSect:TeVShellSearch_Motivation}

The presented work focuses on searching for new \SI{}{\tera\eV} SNRs in the HGPS data set based on a shell-like morphological appearance of $\gamma$-ray emission regions. This is motivated by the shell-type appearance of most known resolved \SI{}{\tera\eV} SNRs, morphologically also matching their shell-like counterparts in radio and nonthermal X-rays, for example,\ RX\,J1713.7$-$3946, RX\,J0852.0$-$4622 (Vela Jr.), RCW\,86, or HESS\,J1731$-$347 \citep{2011A&A...531A..81H,special_velajr,2016arXiv160104461H,special_1713}. All morphological investigations presented in the following employ a forward-folding technique to compare an expected morphology to the actually measured sky map, by including the $\gamma$-ray and background acceptance changes across the sky map in the respective model.

To examine the shell-type appearance of the \SI{}{\tera\eV} emission region, a shell-type morphological model is fit to the data. The model is a three-dimensional spherical shell, homogeneously emitting between $R_{\mathrm{in}}$ and $R_{\mathrm{out}}$ and projected onto the sky. The emissivity in Cartesian sky coordinates $(x,y)$ is then

\begin{equation}\label{Equation:ShellModel}
    M(r) = A \times \left\{
    \begin{array}{ll}
      \sqrt{R_{\mathrm{out}}^{2}-r^{2}} - \sqrt{R_{\mathrm{in}}^{2}-r^{2}}, & r<R_{\mathrm{in}}, \\
      \sqrt{R_{\mathrm{out}}^{2}-r^{2}}, & R_{\mathrm{in}}<r<R_{\mathrm{out}},\\
      0, & r>R_{\mathrm{out}},\\
    \end{array} \right.
 \end{equation}

\noindent
where $r^{2} = (x-x_{0})^2+(y-y_{0})^2$ is the (squared) distance to the source center at ($x_{0}, y_{0}$). Before fitting, the model is convolved numerically with the point spread function (psf) of the H.E.S.S.\ data set under study, which is derived from Monte Carlo simulations taking the configuration of the array and the distribution of zenith angles into account \citep{Aharonian:2006}. The fitted parameters are $A, x_{0}, y_{0}, R_{\mathrm{in}}$, and $R_{\mathrm{out}}$. During the search and identification procedure, no attempt was made to model deviations from this assumed emission profile, for example, azimuthal variations such as those known from the bipolar morphology of the \SI{}{\tera\eV}-emitting SNR of SN\,1006 \citep{2010A&A...516A..62A}. 

The goodness of the fit is not considered as stand-alone criterion for a shell-like source morphology. Because of limited statistics, acceptable fits to the data might be obtained but without any discriminating power. Rather, the shell fit quality is compared to a fit result of a simpler default model (referred to as null hypothesis) that is chosen to represent a typical Galactic \SI{}{\tera\eV} source different from known SNR shell sources. The most abundant identified Galactic source class of this character comprises PWNe, which are usually well described by a centrally peaked morphology. As a default model, a two-dimensional symmetric Gaussian with variable width was chosen, which represents typical PWNe and the majority of other known Galactic H.E.S.S.\ sources (including point-like sources). The Gaussian model is convolved analytically with the H.E.S.S.\ psf before fitting. By default, the H.E.S.S.\ psf is represented by a sum of three Gaussians \citep{special_HGPS}, well sufficient for our purpose.

The chosen approach ensures a homogeneous treatment of the HGPS sky map and of the sky maps of the four individual candidates, which are introduced in Sect.\,\ref{SubSubSect:TeVShellSearch_MotivationTwoStep}. Limitations from the adopted target morphology, source confusion (thus possibly mimicking a shell appearance), and different sensitivities across the HGPS are not treated, the completeness of the search to a particular sensitivity level is not assessed. 
The threshold above which a shell hypothesis is accepted should be treated with caution, since the limited range of tested null hypothesis morphologies may lead to an increased false-positive rate the effect of which has not been numerically quantified.

\subsubsection{Comparison of the non-nested models}
\label{SubSubSect:TeVShellSearch_ComparisonModels}

Adopting the assumption that the chosen morphological models for shell ($H_1$) and null hypothesis ($H_0$) represent the true \SI{}{\tera\eV} source populations sufficiently well, the improvement in the fit quality (i.e., the likelihood that the shell model describes the given data set better than the Gaussian model) can be interpreted in a numerically meaningful way. However, there is the (purely analytical) issue that the two compared models are non-nested, i.e.,\ one cannot smoothly go from $H_{0}$ to $H_{1}$ with a continuous variation of the parameters \citep[for a rigorous definition see][]{1971smep.book.....E}.
In this case, a likelihood ratio test (LRT) cannot be applied \citep{2002ApJ...571..545P}. One way to overcome this problem is to adopt the \textit{Akaike Information Criterion} (AIC) \citep{Akaike:1974}. For a given model, AIC is computed as

\begin{equation}\label{eq:AIC}
\mathrm{AIC} = 2 k - 2 \ln\left(\mathcal{L}_{\mathrm{ML}}\right)
,\end{equation}

\noindent
where $k$ is the number of parameters and $\mathcal{L}_{\mathrm{ML}}$ is the maximum likelihood value for that model. Testing a set of models on the same data set,

\begin{equation}\label{eq:pAIC}
\mathcal{L}_{\mathrm{AIC},i} = C \cdot \exp\left(-\frac{\mathrm{AIC}_{i}-\mathrm{AIC}_\mathrm{min}}{2}\right)
\end{equation}

\noindent
gives a likelihood or relative strength of model $i$ with respect to the best available model, i.e.,\ the model found to have the minimum AIC ($\mathrm{AIC}_{\mathrm{min}}$) \citep{book_burnham}. In order to quantify if and how $\mathcal{L}_{\mathrm{AIC},H_{0}}$ translates into a probability that the improvement obtained with the shell fit over the Gaussian model is due to statistical fluctuations, a limited number of simulations have been performed using the parameters of the H.E.S.S.\ data set of HESS\,J1534$-$571 (the source with the lowest shell over Gaussian likelihood; see Table\ \ref{Table:morphologyresults}). The number of false-positives (type I errors), i.e.,\ simulated Gaussians misinterpreted as shells, behaves roughly as a null-hypothesis probability in the relevant \SI{90}{\percent} to \SI{99}{\percent} probability range with $C$ (Eq.\,\ref{eq:pAIC}) set to 1, whereas the LRT produces $\sim$3 times too many false positives compared to expectation. In turn, $\sim$\SI{10}{\percent} false-negative (type II) errors, i.e., nondetected shells, are estimated when using $\mathcal{L}_{\mathrm{AIC}}$ for a \SI{99}{\percent} significance threshold, ensuring sufficient sensitivity of the chosen method. Table\ \ref{Table:morphologyresults} also lists $\mathcal{L}_{\mathrm{AIC},H_{0}}$ for HESS\,J1614$-$465 and HESS\,J1912$+$101. While the correspondence to a chance probability was not verified for these two sources with analogous simulations, the resulting values ensure to sufficient degree of certainty a low probability of a chance identification as a shell for these two sources as well. 

\subsubsection{Motivation for two-step approach, grid search setup, and results}
\label{SubSubSect:TeVShellSearch_MotivationTwoStep}

To perform an as unbiased as possible search for new shell morphologies in the HGPS data set, a shell ($H_1$) versus Gaussian ($H_0$) morphology test has been performed on a grid of Galactic sky coordinate test positions covering the HGPS area with equidistant spacings of \SI{0.02}{\degree} $\times$ \SI{0.02}{\degree}. To be computationally efficient, also a grid of tested parameters was defined for both $H_1$ and $H_0$; the parameters for $H_1$ (radius and width of the shell) broadly encompass the parameters of the known \SI{}{\tera\eV} SNR shells. The parameters are listed in Table\ \ref{Table:gridparameters}. At each test position $(x_0,y_0)$, the test statistics difference between the best-fitting shell and the best-fitting Gaussian has been derived and stored into a sky map. In such a map, the signature of a shell candidate is an isolated peak surrounded by a broad ring-like artifact.\footnote{For this first-step map, a test statistics difference was used for simplicity, while AIC was used to derive final likelihood values; cf.\ Sect.\,\ref{SubSubSect:TeVShellSearch_FittingProcedure}.} 

\renewcommand{\arraystretch}{1.5}

\begin{table}[t]
\caption{List of tested $H_{0}$ and $H_{1}$ parameters for the grid search; $w = R_{\mathrm{out}} - R_{\mathrm{in}}$ is the shell width. The first value for $w$ represents a thin shell with zero width; the quoted value was chosen for computational reasons only.}
\label{Table:gridparameters}
\centering
\begin{tabular}{lc}
\hline\hline
\multicolumn{2}{c}{Shell ($H_1$) parameters} \\
\hline
$R_{\mathrm{in}}$ & \SI{0.1}{\degree}, \SI{0.2}{\degree}, \SI{0.3}{\degree}, \SI{0.4}{\degree}, \SI{0.5}{\degree}, \SI{0.6}{\degree}, \SI{0.7}{\degree}, \SI{0.8}{\degree} \\
$w$               & $10^{-5} \times R_{\mathrm{in}}$, $0.1 \times R_{\mathrm{in}}$, $0.2 \times R_{\mathrm{in}}$\\
\hline
\multicolumn{2}{c}{Gaussian ($H_0$) parameters} \\
\hline
$\sigma$          & \SI{0}{\degree}, \SI{0.05}{\degree}, \SI{0.1}{\degree}, \SI{0.2}{\degree}, \SI{0.4}{\degree} \\
\hline\hline
\end{tabular}
\end{table}
\renewcommand{\arraystretch}{1}

The procedure has revealed the known \SI{}{\tera\eV} SNR shells covered by the survey, RX\,J1713.7$-$3946, RX\,J0852.0$-$4622, HESS\,J1731$-$347, and RCW\,86 with high confidence. At further four positions, significant signatures for shell morphologies were revealed. Three positions are clearly identified with known H.E.S.S.\ sources. HESS\,J1614$-$465 was discovered during the initial phase of the H.E.S.S. survey in 2006 \citep{2006ApJ...636..777A} and is so far an unidentified source. HESS\,J1912$+$101 was discovered in 2008 \citep{2008A&A...484..435A} and an association with an energetic radio pulsar was suggested in a PWN scenario. However, there has been no support for this scenario by the detection of an X-ray PWN and energy-dependent \SI{}{\tera\eV} morphology, which is used to identify resolved \SI{}{\tera\eV} sources as PWNe with high confidence \citep{2006A&A...460..365A,2012A&A...548A..46H}. \object{HESS\,J1023$-$577} was discovered in 2007 \citep{2007A&A...467.1075A} and is associated with the young open stellar cluster Westerlund 2; the source of the \SI{}{\tera\eV}-radiating particles is however not firmly identified. 

A fourth TeV source at Galactic coordinates $l \sim$ \SI{323.7}{\degree}, $b \sim$ \SI{-1.02}{\degree} has not been published before. A test using a two-dimensional Gaussian source hypothesis against the background (no source) hypothesis as used for the HGPS source catalog yields a test statistics difference of $TS_{\mathrm{diff}} = 39$, which is well above the HGPS source detection threshold of $TS_{\mathrm{diff}} = 30$. The newly discovered source is named HESS\,J1534$-$571, corresponding to the center coordinates of the fitted shell morphology as derived in the following subsection. 

The grid search for new shells described above has several limitations. The limited number of tested null-hypothesis models and the restriction of keeping the same centroid for the null-hypothesis model as for the shell may lead to a nonoptimum $H_0$ fit and therefore to overestimating the likelihood of the $H_1$ versus\ $H_0$ improvement. Final likelihoods for the individual candidates are therefore estimated with an improved method as described in the following Sect.\,\ref{SubSubSect:TeVShellSearch_FittingProcedure}. A further limitation due to the restricted number of shell models has become obvious when dealing with HESS\,J1023$-$577. The refined morphological analysis as described in the following subsection has revealed that the best-fitting shell morphology of HESS\,J1023$-$577 is center-filled, i.e.,\ $R_{\mathrm{in}} \simeq 0$. This is not the morphology the search is targeting. The source has consequently been removed from the list of \SI{}{\tera\eV} SNR candidates.

The search has been designed to be independent of the HGPS source identification mechanism, which could have also been used to define test positions and regions of interest. In principle, shell morphologies could be present that cover two or more emission regions identified as independent sources by the HGPS procedure. However, no significant such structures have been found and all candidate \SI{}{\tera\eV} SNRs are also identified as individual \SI{}{\tera\eV} sources following the HGPS source catalog prescription.

\subsubsection{Morphology fitting procedure and results from individual source analyses}
\label{SubSubSect:TeVShellSearch_FittingProcedure}

\renewcommand{\arraystretch}{1.5} 

\begin{table}[t]
\caption{Results from the morphological study of the three new \SI{}{\tera\eV} shells. $^a$: \SI{}{\tera\eV} discovery status. $^b$: Source detection significance from excess counts $N_{\mathrm{excess}}$ detected inside $R_{\mathrm{out}}$, following \citet{1983ApJ...272..317L}. $^c$: Likelihood $\mathcal{L}_{\mathrm{AIC},H_{0}}$ as defined in equation \ref{eq:pAIC} used as a measure of whether the fit improvement of the shell ($H_1$) over the Gaussian ($H_0$) is due to fluctuations, 
using the \textit{Akaike Information Criterion}. $^d$: Shell fit results; ($l_0$, $b_0$) are the center coordinates, $R_{\mathrm{in}}$ and $R_{\mathrm{out}}$ are the inner and outer radii of the homogeneously emitting spherical shell, respectively.}
\label{Table:morphologyresults}

\centering
\begin{tabular}{lrrr}
\hline\hline
& \vtop{\hbox{\strut HESS}\hbox{\strut J1534$-$571}}    & \vtop{\hbox{\strut HESS}\hbox{\strut J1614$-$518}} & \vtop{\hbox{\strut HESS}\hbox{\strut J1912$+$101}} \\
\hline
Discovery$^a$                & $TS_{\mathrm{diff}} = 39$      & (1)                      & (2)                     \\
Excess$^b$ & \SI{9.3}{\sigma}                & \SI{25.2}{\sigma}         & \SI{17.3}{\sigma}              \\
$\mathcal{L}_{\mathrm{AIC},H_{0}}$$^c$  & $5.9 \times 10^{-3}$            & $3.1 \times 10^{-6}$            & $1.7 \times 10^{-6}$           \\
$l_0$$^d$                 & \SI{323.70}{\degree}$^{\SI[retain-explicit-plus]{+0.02}{\degree}}_{\SI{-0.02}{\degree}}$
                                                         & \SI{331.47}{\degree}$^{\SI[retain-explicit-plus]{+0.01}{\degree}}_{\SI{-0.01}{\degree}}$
                                                                                           & \SI{44.46}{\degree}$^{\SI[retain-explicit-plus]{+0.02}{\degree}}_{\SI{-0.01}{\degree}}$ \\
$b_0$$^d$                 & \SI{-1.02}{\degree}$^{\SI[retain-explicit-plus]{+0.03}{\degree}}_{\SI{-0.02}{\degree}}$
                                                         & \SI{-0.60}{\degree}$^{\SI[retain-explicit-plus]{+0.01}{\degree}}_{\SI{-0.01}{\degree}}$
                                                                                           & \SI{-0.13}{\degree}$^{\SI[retain-explicit-plus]{+0.02}{\degree}}_{\SI{-0.02}{\degree}}$ \\
$R_{\mathrm{in}}$$^d$     & \SI{0.28}{\degree}$^{\SI[retain-explicit-plus]{+0.06}{\degree}}_{\SI{-0.03}{\degree}}$  
                                                            & \SI{0.18}{\degree}$^{\SI[retain-explicit-plus]{+0.02}{\degree}}_{\SI{-0.02}{\degree}}$  
                                                                                              & \SI{0.32}{\degree}$^{\SI[retain-explicit-plus]{+0.02}{\degree}}_{\SI{-0.03}{\degree}}$ \\
$R_{\mathrm{out}}$$^d$    & \SI{0.40}{\degree}$^{\SI[retain-explicit-plus]{+0.04}{\degree}}_{\SI{-0.12}{\degree}} $ 
                                                            & \SI{0.42}{\degree}$^{\SI[retain-explicit-plus]{+0.01}{\degree}}_{\SI{-0.01}{\degree}}$ 
                                                                                              & \SI{0.49}{\degree}$^{\SI[retain-explicit-plus]{+0.04}{\degree}}_{\SI{-0.03}{\degree}}$ \\
\hline\hline
\end{tabular}
\tablebib{
(1)~\citet{2006ApJ...636..777A}; (2) \citet{2008A&A...484..435A}.
}
\end{table}

\renewcommand{\arraystretch}{1} 

To overcome the limitations of the grid search (see Sect.\,\ref{SubSubSect:TeVShellSearch_MotivationTwoStep}) and from the survey data analysis (cf.\ Sect.\,\ref{SubSect:HessAnalysis}), final morphological parameters and shell identification likelihoods were derived using the H.E.S.S.\ data sets introduced in Sect.\,\ref{SubSubSect:HessDedicatedAnalysis} and the analysis as described in the following.\footnote{The refined analysis is more conservative than the grid search regarding the false positive rates of the shell likelihoods, therefore the initial selection of the candidates from the grid search is not treated as an additional trial.}

Morphological fits were performed on uncorrelated \textit{on-counts} (i.e.,\ $\gamma$-candidates after $\gamma$-hadron separation, not background-subtracted or flatfielded) sky maps with \SI{5}{\degree} $\times$ \SI{5}{\degree} size and with \SI{0.01}{\degree} $\times$ \SI{0.01}{\degree} bin size. Entries in these maps have pure Poissonian statistical errors. The model fit function $On_{i}$ is constructed as

\begin{equation}\label{Equation:OnCtsModel}
  \mathrm{On}_{i} = A \times \mathrm{Bkg}_{i} + \left(\mathrm{psf} \ast M_{i}\right) \times N_{\mathrm{ref},i}.
\end{equation}

\noindent
$\mathrm{Bkg}_{i}$ is the estimated background event map derived from the ring-background method \citep{2007A&A...466.1219B}, $A$ is a normalization factor that is fitted. The term $(\mathrm{psf} \ast M_{i})$ is the morphological model (shell or Gaussian) map, folded with the H.E.S.S. psf, with freely varying parameters in the fit. The values $N_{\mathrm{ref},i}$ are the expected $\gamma$-ray counts that are derived under the assumption of a source energy spectrum following a power-law distribution (see \citet{special_HGPS} for a detailed explanation), while $i$ runs over the bins. For HESS\,J1614$-$518, another source is visible in the field of view with high significance \citep[HESS\,J1616$-$508; cf.][]{2006ApJ...636..777A}. For the fitting of the map containing HESS\,J1614$-$518, HESS\,J1616$-$508 is modeled as a Gaussian component, which is added as additional source model component to increase the stability of the fit. 

The employed fitting routines are based on the \textit{cstat} implementation of the Cash statistics\footnote{\url{http://cxc.harvard.edu/sherpa/statistics/}} \citep{1979ApJ...228..939C} available in the \textit{Sherpa}\footnote{\url{http://cxc.harvard.edu/sherpa/index.html}} package. To quantify the improvement of the fit quality between two models, the \textit{Akaike Information Criterion} as discussed in Sect.\ \ref{SubSubSect:TeVShellSearch_ComparisonModels} was used. Table \ref{Table:morphologyresults} lists the results for the three new \SI{}{\tera\eV} shells.

During the evaluation of the fit improvement when applying the shell model, the assumption of spherical symmetry of the respective shell model was preserved. Nevertheless, after identification of the shell sources, the symmetry of the \SI{}{\tera\eV} sources (Fig.\,\ref{Fig:tevsurfacebrightnesses}) was investigated using azimuthal profiles of the shells (see Appendix \ref{Sect:AppendixTeV} for figures). Applying a $\chi^2$-test, HESS\,J1534$-$571 and HESS\,J1912$+$101 are statistically consistent with a flat azimuthal profile. However, HESS\,J1614$-$518 significantly deviates from azimuthal symmetry. Adding another Gaussian source component (or additional source) to the shell plus HESS\,J1616$-$508 Gaussian model in order to model the excess on top of the northern part of the shell (see Fig.\,\ref{Fig:tevsurfacebrightnesses}), indeed improves the quality of the morphological fit. The AIC was used again to quantify the improvement of the goodness of fit. The parameters of the additional Gaussian component are however not consistent within statistical errors when modifying analysis configurations or using the cross-check analysis. Also, no consistent significant result could be established using the main and cross-check analysis when attempting to model the apparent excess in the south of the HESS\,J1614$-$518 shell as additional Gaussian component. Therefore, fit results for models including these additional components are not given here.

\subsection{\SI{}{\tera\eV} surface brightness maps}
\label{SubSect:Surfacebrightness}

\begin{figure}[h!]
\centering
{\includegraphics[width=0.23\paperheight]{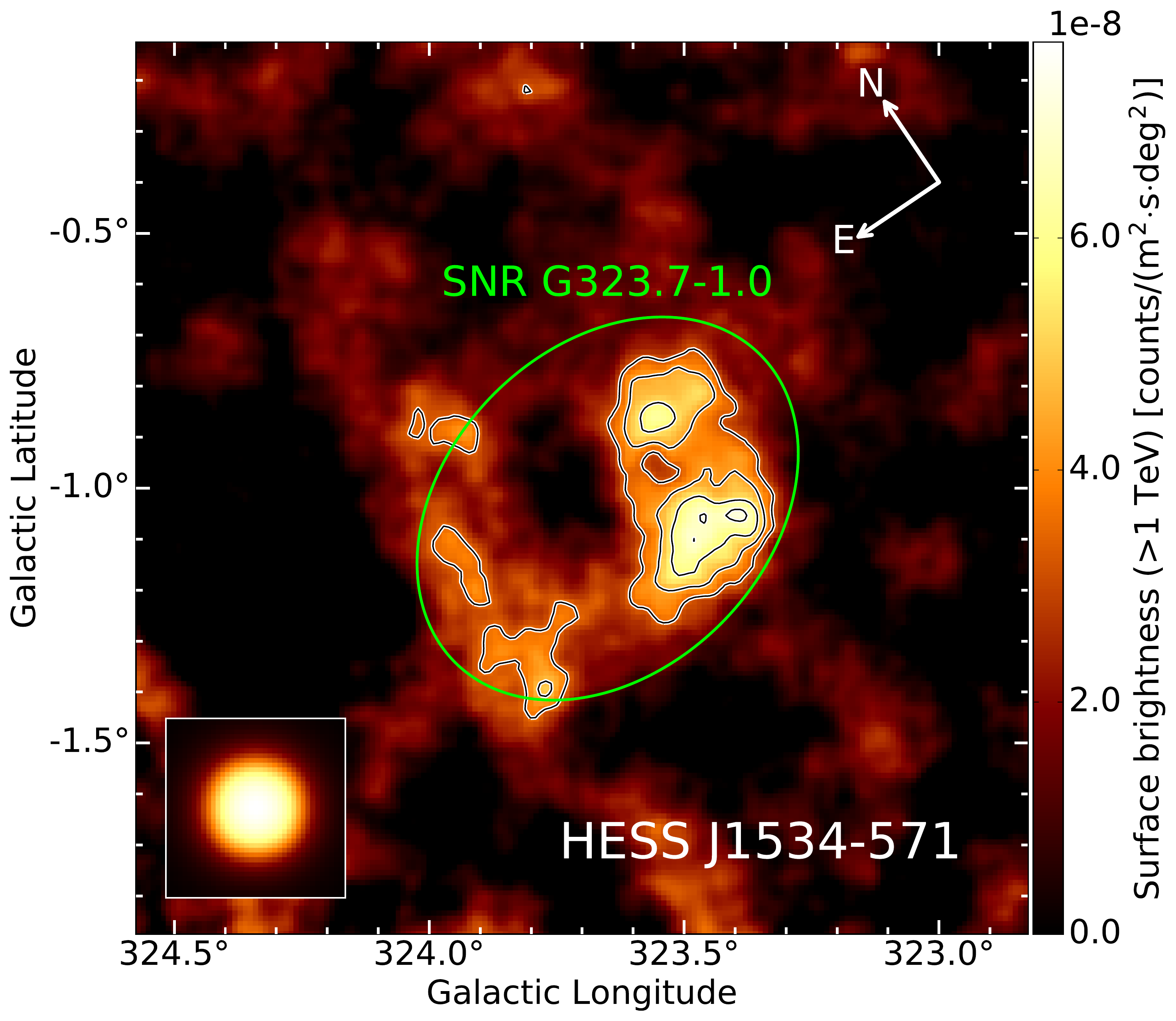}}  \\
{\includegraphics[width=0.23\paperheight]{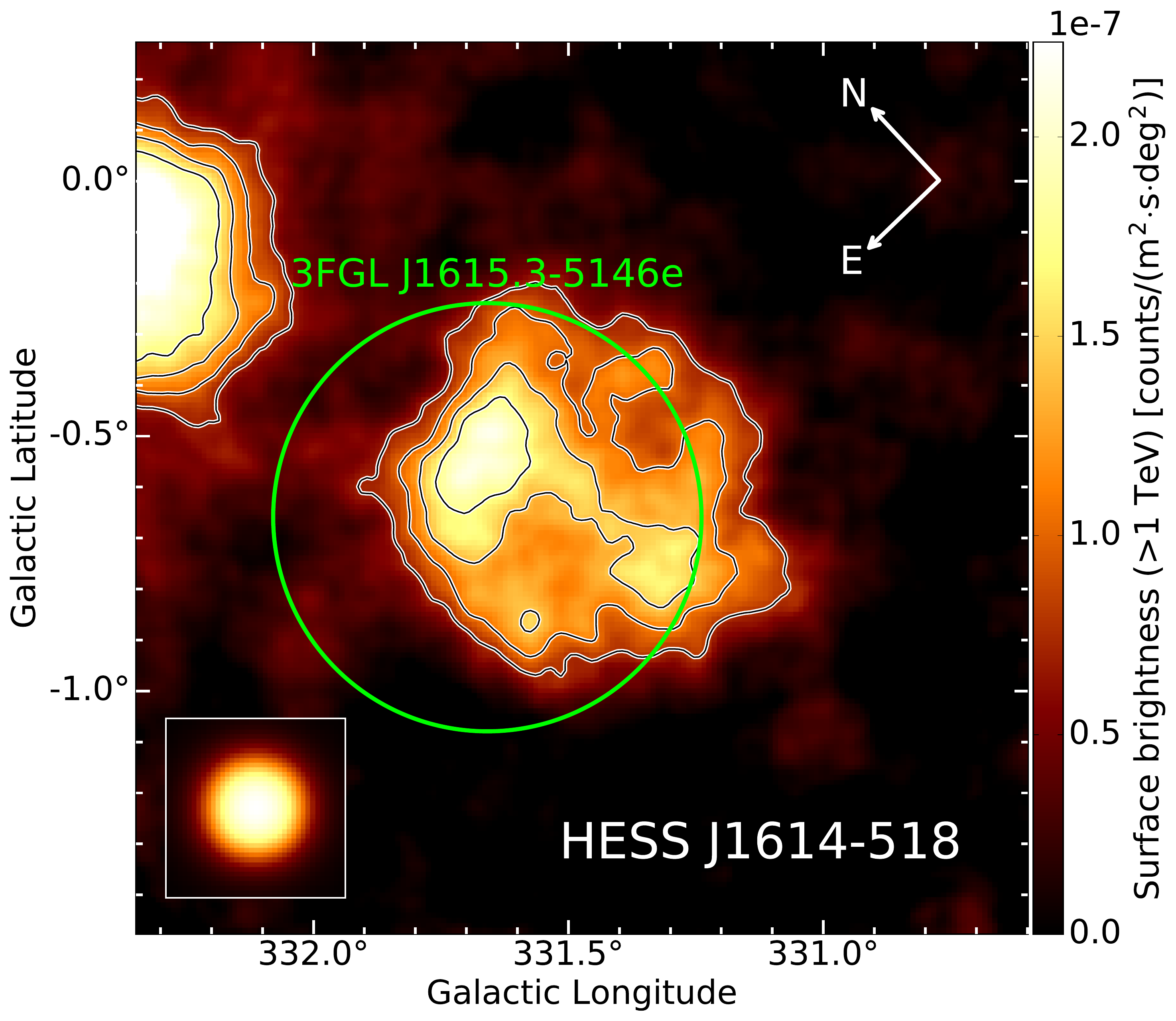}}  \\
{\includegraphics[width=0.23\paperheight]{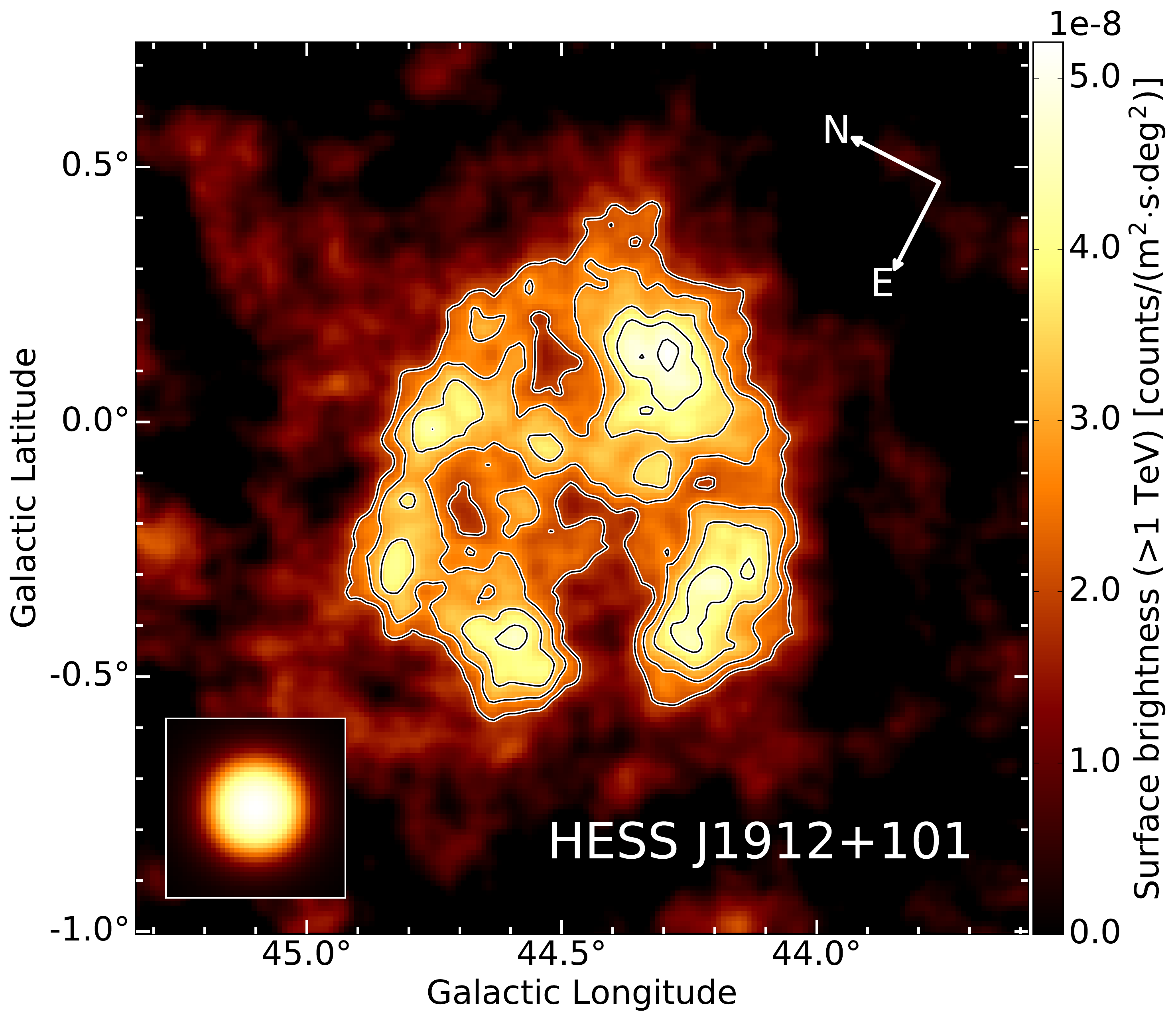}} 
\caption{\SI{}{\tera\eV} surface brightness maps of HESS\,J1534$-$571, HESS\,J1614$-$518, and HESS\,J1912$+$101 in Galactic coordinates. The maps were calculated with a correlation radius of \SI{0.1}{\degree}. An additional Gaussian smoothing with $\sigma=$ \SI{0.01}{\degree} was applied to remove artifacts. The surface brightness is expressed in units of counts above \SI{1}{\tera\eV}, assuming a power law with index $\Gamma_{\mathrm{ref}}$. The inlets show the point spread function of the specific observations after applying the same correlation radius and smoothing, respectively. {\it Top panel:} HESS\,J1534$-$571, assumed $\Gamma_{\mathrm{ref}} = 2.3$. The green ellipse denotes the outer boundary of the radio \object{SNR\,G323.7$-$1.0} from \citet{2014PASA...31...42G}. Contours are 3, 4, 5, 6 $\sigma$ significance contours (correlation radius \SI{0.1}{\degree}). {\it Middle panel:} HESS\,J1614$-$518, assumed $\Gamma_{\mathrm{ref}} =  2.4$. The green circle denotes the position and extent of 3FGL\,J1615.3$-$5146e from \citet{2015ApJS..218...23A}. Contours are 5, 7, 9, 11 $\sigma$ significance contours (correlation radius \SI{0.1}{\degree}). {\it Bottom panel:} HESS\,J1912$-$101, assumed $\Gamma_{\mathrm{ref}} = 2.7$. Contours are 3, 4, 5, 6, 7 $\sigma$ significance contours (correlation radius \SI{0.1}{\degree}).
}
\label{Fig:tevsurfacebrightnesses}
\end{figure}

The morphological fits presented in Sect.\,\ref{SubSect:TeVShellSearch} are based on a forward-folding technique. For the visualization of sky maps, depending on the properties of the data set, it may be necessary to correct acceptance changes in the sky maps themselves, specifically in the case of large extended sources. If H.E.S.S.\ sky maps are dominated by observations expressly targeted at the source of interest, the observations themselves ensure a nearly flat exposure and background level at and around the source, and $\gamma$-ray excess or significance maps are often an appropriate representation of the surface brightness of the sources. For the sources presented in this paper, however, all available observations of the respective source were used, whether they were part of an observation of the particular source or a nearby source. This led to an uneven exposure with differences up to \SI{30}{\percent} on the regions of interest. Therefore, surface brightness maps were constructed for the three sources. 

For this, flux maps were derived following the procedure described in \cite{special_HGPS}. A correlation radius of $r_{\mathrm{int}}=\SI{0.1}{\degree}$ is used, $\Gamma_{\mathrm{ref}}$ is chosen per source. To derive the surface brightness, the flux is divided by the area of the correlation circle $\pi r_{\mathrm{int}}^2$ for every bin of the sky map. The grid size is \SI{0.01}{\degree} $\times$ \SI{0.01}{\degree}. From these maps, the integral flux of a source can be obtained by integrating over the radius of a region of interest, provided that the integration region is large compared to the psf. The surface brightness maps were checked for each of the three sources by comparing the integral flux from the maps with the respective result of the spectral analysis.

In Fig.\,\ref{Fig:tevsurfacebrightnesses}, the TeV surface brightness maps of HESS\,J1534$-$571, HESS\,J1614$-$518, and HESS\,J1912$+$101 are shown using an energy threshold set to 1\,TeV.\footnote{Per construction, the surface brightness maps contain the information from the entire energy range after event selection (cf.\ Sect.\,\ref{SubSubSect:HessDedicatedAnalysis}). The displayed energy range can be freely chosen.} The assumed spectral indices for HESS\,J1614$-$518 and HESS\,J1912$+$101 were fixed independently of the spectral results discussed below and were taken from previous publications \citep[$\Gamma_{\mathrm{ref}} = 2.4$ and $2.7$;][]{2006ApJ...636..777A,2008A&A...484..435A}. For HESS\,J1534$-$571, a typical value for Galactic sources, $\Gamma_{\mathrm{ref}} = 2.3$, was assumed. In fact, the assumed spectral index does not influence the appearance of the map. The reference flux, which is calculated based on the spectral index, only affects the overall scale of the displayed surface brightness (less than 5\% for $\Delta \Gamma = 0.2$; \citealt{special_HGPS}). Radial and azimuthal profile representations of these surface brightness maps including the morphological fit results can be found in Appendix \ref{Sect:AppendixTeV}.

\subsection{Energy spectra}
\label{SubSect:Energyspectra}

\renewcommand{\arraystretch}{1.5}
\begin{table*}[ht!]
\caption{
Spectral fit results from the power-law fits to the H.E.S.S.\ data. Both statistical and systematic errors are given for the fit parameters. The systematic uncertainties result from deviations from the nominal parameters of the simulations of the instrument, nonoptimized observation strategy, and the large size of the sources, which lead to substantial susceptibility of the spectral results to potential errors in the background estimation (see Sect.\ \ref{SubSubSect:HessDedicatedAnalysis}), and are estimated to 30\% for $N_0$ and energy flux and to 0.2 for $\Gamma$, respectively. To simplify a comparison between the sources, the normalization at \SI{1}{\tera\electronvolt}, $N_{0,1\,\mathrm{TeV}}$, and the energy flux from \SI{1}{\tera\electronvolt} to \SI{10}{\tera\electronvolt} are given as well.
}
\label{Table:spectralresults}
\centering
\begin{tabular}{lccccc}
        \hline\hline
        \multicolumn{1}{l}{Source} & \multicolumn{1}{c}{$\mathrm{E}_{\mathrm{min}}$}& \multicolumn{1}{c}{$\mathrm{E}_{\mathrm{max}}$} & \multicolumn{1}{c}{$E_0$}& \multicolumn{1}{c}{$N_{0}$} & \multicolumn{1}{c}{$\Gamma$} \\
        HESS & \multicolumn{1}{c}{[\SI{}{TeV}]} &\multicolumn{1}{c}{[\SI{}{TeV}]} & \multicolumn{1}{c}{[\SI{}{TeV}]} & \multicolumn{1}{c}{[\SI{}{\per\square\centi\metre\per\second\per\TeV}]}  \\
        \hline
        J1534$-$571 & 0.42 & 61.90 & 1.40 & $(1.29\pm0.12_\mathrm{{stat}}\pm0.39_\mathrm{{syst}})\times 10^{-12}$  & $2.51 \pm 0.09_\mathrm{{stat}} \pm 0.20_\mathrm{{syst}}$ \\
        J1614$-$518 & 0.32 & 38.31 & 1.15 & $(5.86\pm0.34_\mathrm{{stat}}\pm1.76_\mathrm{{syst}})\times 10^{-12}$ & $2.42 \pm 0.06_\mathrm{{stat}} \pm 0.20_\mathrm{{syst}}$ \\
        J1912$+$101 & 0.68 & 61.90 & 2.25 & $(4.82\pm0.43_\mathrm{{stat}}\pm1.45_\mathrm{{syst}})\times 10^{-13}$ & $2.56 \pm 0.09_\mathrm{{stat}} \pm 0.20_\mathrm{{syst}}$ \\
\hline\hline
\vspace{0.2cm}
\end{tabular}
\begin{tabular}{lcc}
\hline\hline
\multicolumn{1}{l}{Source}  & \multicolumn{1}{c}{$N_{0,1\,\mathrm{TeV}}$} &  \multicolumn{1}{c}{energy flux ($1-10\,\mathrm{TeV}$)}  \\
HESS  & \multicolumn{1}{c}{[\SI{}{\per\square\centi\metre\per\second\per\TeV}]}  &\multicolumn{1}{c}{[\SI{}{\erg\per\square\centi\metre\per\second}]} \\
\hline
J1534$-$571   & $(2.99\pm0.30_\mathrm{{stat}}\pm0.90_\mathrm{{syst}})\times 10^{-12}$  &  $(6.5\pm0.7_\mathrm{{stat}}\pm2.0_\mathrm{{syst}})\times 10^{-12}$\\
J1614$-$518   & $(8.33\pm0.49_\mathrm{{stat}}\pm2.50_\mathrm{{syst}})\times 10^{-12}$ &   $(2.0\pm0.2_\mathrm{{stat}}\pm0.6_\mathrm{{syst}})\times 10^{-11}$\\
J1912$+$101   & $(3.89\pm0.45_\mathrm{{stat}}\pm1.17_\mathrm{{syst}})\times 10^{-12}$ &   $(8.1\pm0.7_\mathrm{{stat}}\pm2.4_\mathrm{{syst}})\times 10^{-12}$\\
\hline\hline
\end{tabular}
\end{table*}
\renewcommand{\arraystretch}{1.0}

\begin{figure}[t!]
\centering
{\includegraphics[width=0.28\paperheight]{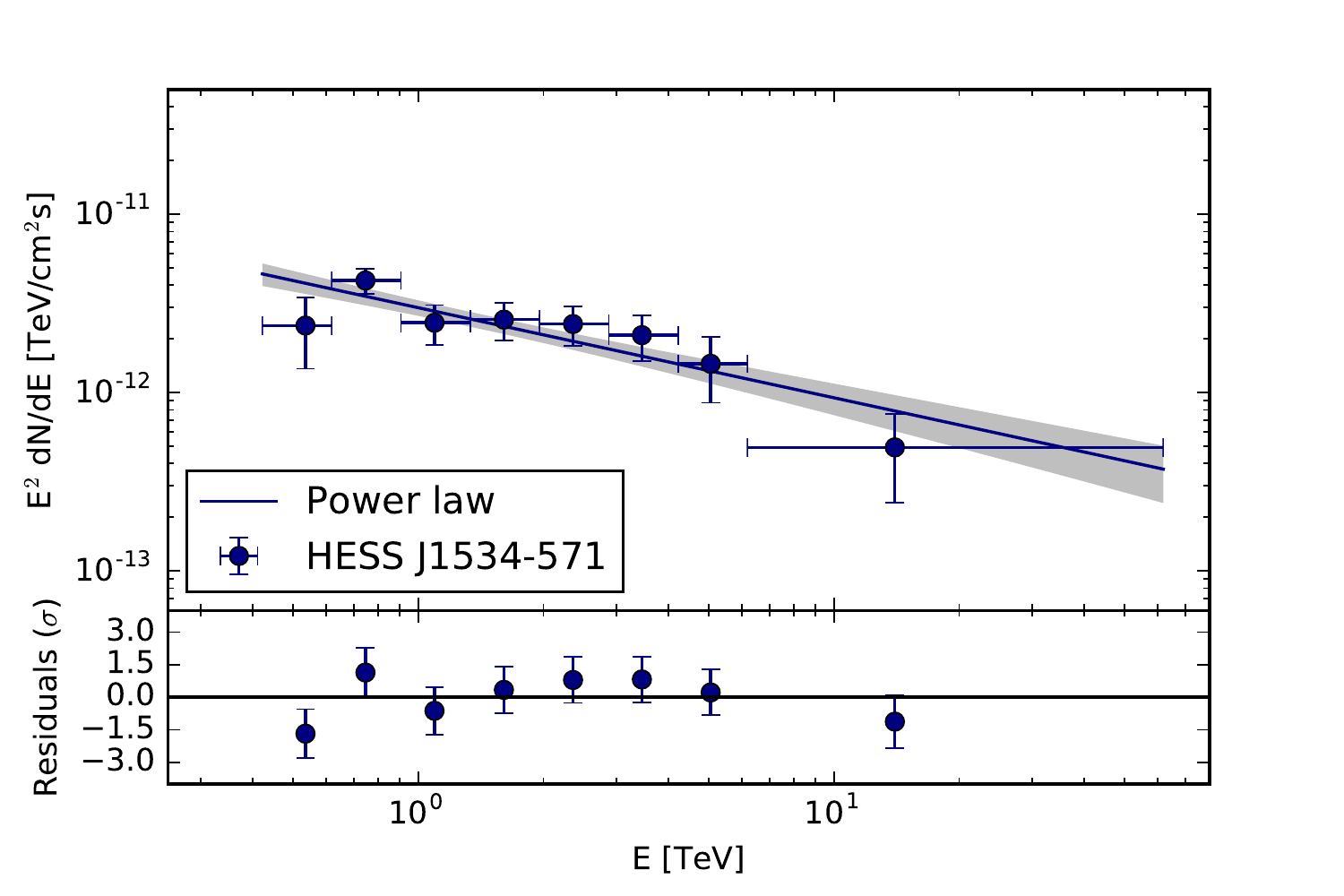}}  \\
{\includegraphics[width=0.28\paperheight]{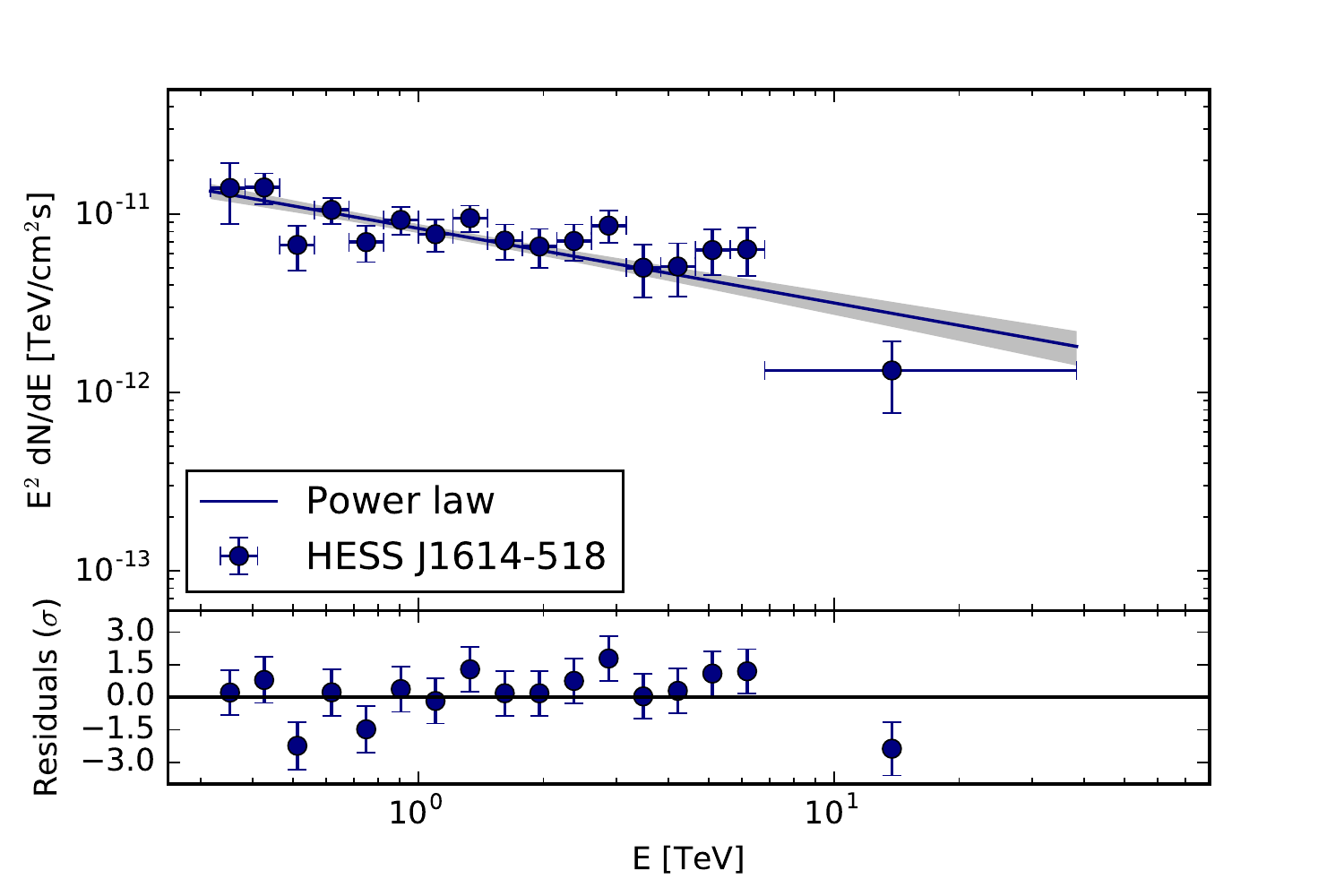}}  \\
{\includegraphics[width=0.28\paperheight]{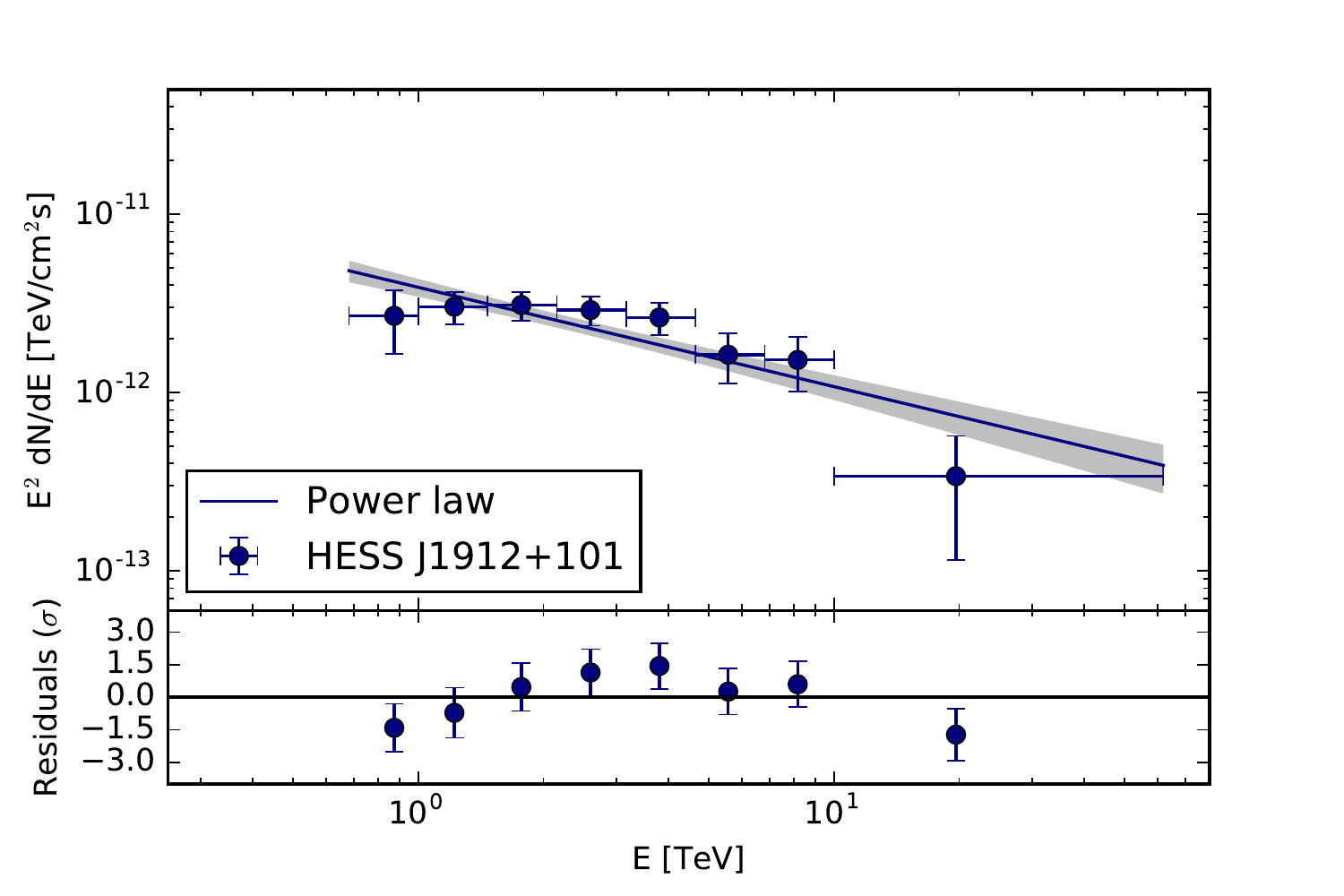}} 
\caption{
Upper boxes show the H.E.S.S.\ energy flux spectra of HESS\,J1534$-$571, HESS\,J1614$-$518, and HESS\,J1912$+$101 (blue data points with \SI{1}{\sigma} statistical uncertainties), respectively. The bin size is determined by the requirement of at least \SI{2}{\sigma} significance per bin. The solid blue lines with the gray butterflies (\SI{1}{\sigma} error of the fit) show the best fit power-law models from Table \ref{Table:spectralresults}. The bottom boxes show the deviation from the respective model in units of sigma, calculated as $(F-F_{\mathrm{model}}) / \sigma_{F}$. Systematic errors do not permit the application of more complex models to describe the data.
}
\label{Fig:tevspectra}
\end{figure}

\SI{}{\tera\eV} energy spectra for each of the new shells were derived using a forward-folding technique. Data were selected and analyzed according to the description in Sect.\,\ref{SubSubSect:HessDedicatedAnalysis}. On-source events are extracted from circular regions centered on the best-fit position and with radius $R_\mathrm{out} + R_{68\%}$ to ensure full enclosure ($R_{68\%}\sim\SI{0.07}{\degree}$ is the 68\% containment radius of the psf). Source, background, and effective area spectra with equidistant binning in log-space are derived runwise and are summed up. Power-law models

\begin{equation}
 \frac{\mathrm{d}N_{\gamma}}{\mathrm{d}E} = N_{0}\left(\frac{E}{E_0}\right)^{-\Gamma}
\end{equation}

\noindent
are convolved with the effective area and fit to the data. The lower fit boundary results from the image amplitude cut and corresponding energy bias cut. The upper fit boundary is dominated by the energy bias cut. The value $E_0$ is the decorrelation energy of the fit, i.e.,\ the energy where the correlation between the errors of $\Gamma$ and $N_0$ is minimal. All spectral parameters are listed in Table\ \ref{Table:spectralresults}.

To display unfolded spectra, bins are background-subtracted and merged to have at least \SI{2}{\sigma} per bin, and spectra are divided by correspondingly binned effective areas. In Fig.\,\ref{Fig:tevspectra}, unfolded spectra and residuals between data and the power-law fits are shown. All three spectra are statistically compatible with power laws. 
Nevertheless, the spectra of the primary analysis of HESS\,J1534$-$571 and HESS\,J1912$+$101 as presented in Fig.\,\ref{Fig:tevspectra} also show indications for curvature. A power-law model with exponential cutoff indeed better fits the data than a simple power law at the 3-4\,$\sigma$ level. However, the curvatures of the spectra (in particular of HESS\,J1912$+$101) are less pronounced in the cross-check analysis and are not significantly preferred there.
The systematic errors that have been discussed in Sect.\,\ref{SubSubSect:HessDedicatedAnalysis} not only lead to an error of the fitted power-law index value, but also to a distortion of the spectral shape. With the current spectral analysis, a more detailed description beyond a power law is not justified.


\section{Counterparts in radio continuum and GeV emission}
\label{Sect:RadioGeV}

\subsection{Radio continuum emission}
\label{SubSect:RadioContinuum}

Several well-known high-energy emitting SNRs have only weak radio counterparts, such as the bright nonthermal X-ray and \SI{}{\tera\eV}-emitting SNRs RX\,J1713.7$-$3946 and RX\,J0852.0$-$4622; cf. e.g.,\ \citet{2015A&ARv..23....3D}. Confusion with thermal emission and Galactic background variations might therefore hamper the detection of radio counterparts of the new SNR candidates as well. We searched publicly available radio catalogs and survey data for counterparts of the new \SI{}{\tera\eV} sources.

\subsubsection{The radio SNR candidate counterpart of HESS\,J1534$-$571}
\label{SubSubSect:radio1534}

\begin{figure}[]
\centering
{\includegraphics[width=0.95\columnwidth]{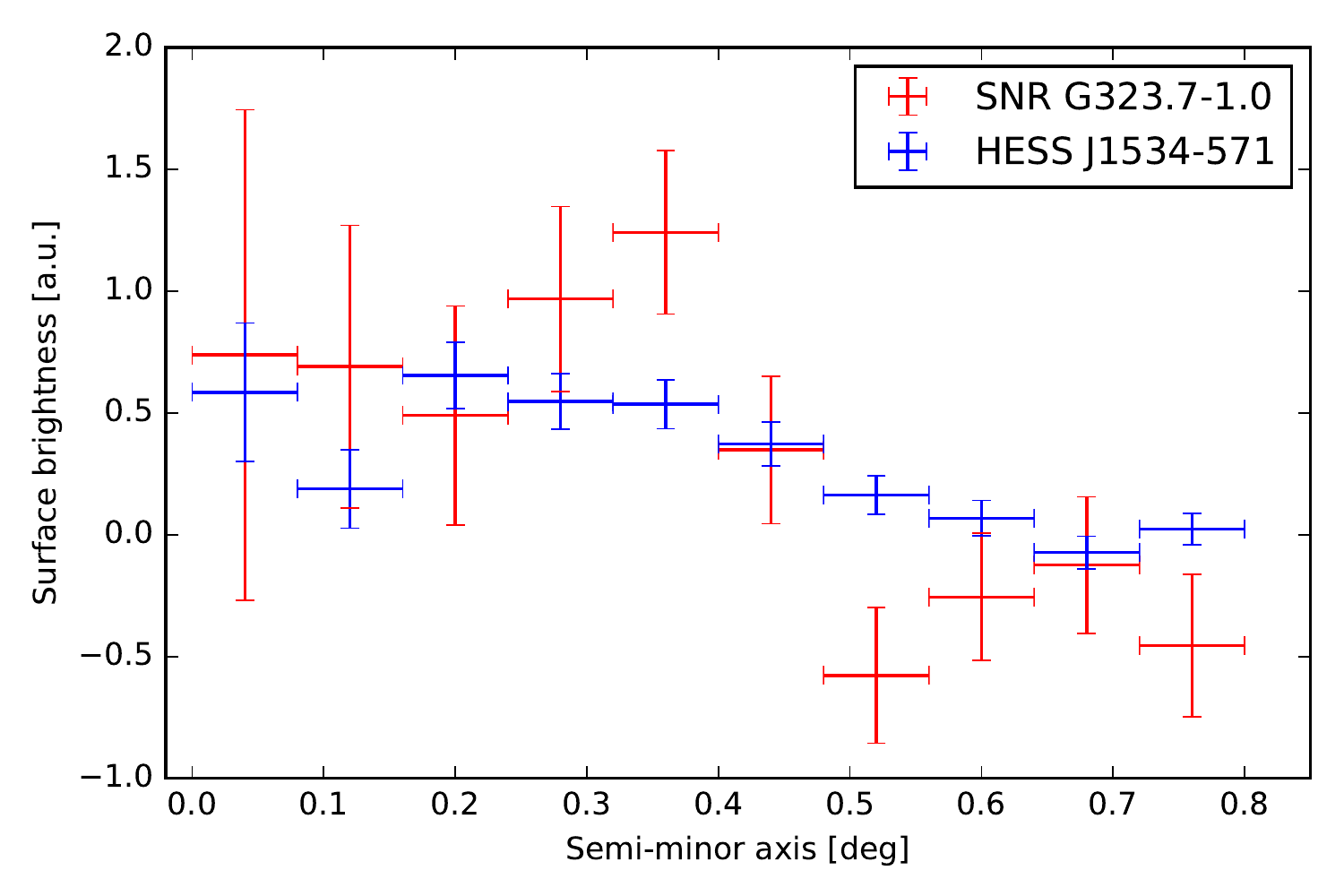}} 
\caption{
Radial profile of HESS\,J1534$-$571 (H.E.S.S. TeV, blue points) and the SNR candidate G323.7$-$1.0 (MGPS-2 radio synchrotron at \SI{843}{MHz}, red points) using elliptical annuli. 
The ratio of minor and major axes of the ellipse and the center position were both taken from \citet{2014PASA...31...42G}. Since \citet{2014PASA...31...42G} do not quote a position angle for the ellipse, an angle of \SI{100}{\degree} of the major axis with respect to north was estimated from the radio map. The MGPS-2 image was convolved with the H.E.S.S.\ point spread function before extraction of the profile. Both profiles were normalized to have the same integral value.
}
\label{Fig:1534radialprofiles}
\end{figure}

In \citeyear{2014PASA...31...42G}, \citeauthor{2014PASA...31...42G} published a catalog of new SNR candidates that have been discovered using the Molonglo Galactic Plane Survey MGPS-2 \citep{2007MNRAS.382..382M}. The data were taken with the Molonglo Observatory Synthesis Telescope (MOST) at a frequency of \SI{843}{MHz} and with a restoring beam of \SI{45}{\arcsec}$\times$\SI{45}{\arcsec}$\mathrm{cosec}\left|\delta\right|$, where $\delta$ is the declination. The newly detected radio SNR candidate G323.7$-$1.0 is in very good positional agreement with the H.E.S.S.\ source. The extension and shell appearance of the two sources are in excellent agreement as well. To compare the two sources on more quantitative basis, radial profiles using elliptical annuli were extracted from the radio and \SI{}{\tera\eV} data.

In order to derive the profile of the radio emission observed in the MGPS-2 image at 843 MHz, a \SI{1.2}{\degree}-large sub-image from the original mosaic\footnote{\url{http://www.astrop.physics.usyd.edu.au/mosaics/Galactic/}}, centered at R.A.(J2000) = \SI{233.572}{\degree} and Dec.(J2000) = \SI{-57.16}{\degree} (i.e.,\ the approximate barycenter of G323.7$-$1.0) was extracted. Each of the 28 sources listed in the MGPS-2 compact source catalog \citep{2007MNRAS.382..382M} was masked out according to its morphological properties (minor/major axes and position angle). Two additional uncataloged source candidates and two extended sources were also removed. The RMS in the resulting source-subtracted image is \SI{1.8}{\milli Jy \per beam}. The flux density of G323.7$-$1.0 was measured after summing all the pixels inside an ellipse and correcting for the beam of \SI{0.90}{\arcmin} $\times$ \SI{0.75}{\arcmin} (FWHM). The ellipse is centered at the above-mentioned position and with the major and minor axes (\SI{51}{\arcmin} and \SI{38}{\arcmin}) as given in \citet{2014PASA...31...42G}, and a position angle of \SI{100}{\degree} of the major axis (with respect to north), estimated from visual inspection of the image. The flux density amounts to \SI[separate-uncertainty = true]{0.49 \pm 0.08}{Jy}, compatible with the flux lower limit of \SI{0.61}{Jy} reported by \citet{2014PASA...31...42G}. The difference is mostly due to one of the two uncataloged sources cited above that lie within the SNR candidate. It should be noted that the MOST telescope does not fully measure structures on angular scales larger than $\sim$ \SI{25}{\arcmin}, so this flux density estimate must be considered as a lower limit, as explained in \citet{2014PASA...31...42G}. The profile of the \SI{843}{\mega\Hz} emission toward G323.7$-$1.0 shown in Figure \ref{Fig:1534radialprofiles} was then derived from elliptical annuli of width \SI{0.08}{\degree} with the same parameters as above, after convolving the MGPS-2 image with the H.E.S.S.\ psf.

The \SI{}{\tera\eV} profile was derived following the procedure described in Appendix \ref{Sect:AppendixTeV}, using elliptical annuli with the same parameters as used for the profile of the radio emission. The resulting \SI{}{\tera\eV} and radio radial profiles are shown in Fig.\,\ref{Fig:1534radialprofiles}. The two profiles are statistically compatible with each other, confirming the association of the two sources as being due to the same object.

The total flux of G323.7$-$1.0 can also be used to provide a rough distance estimate of the SNR candidate, using the empirical surface brightness to source diameter ($\Sigma-D$) relation. Adopting $\Sigma_{1\,\mathrm{GHz}} = 2.07 \times 10^{-17} \times D\left[\mathrm{pc}\right]^{-2.38}\,\mathrm{W\,m^{-2}Hz^{-1}sr^{-1}}$ (omitting the errors on the parameters) from \citet{1998ApJ...504..761C}, $S_{843\,\mathrm{MHz}} = \SI{0.49}{Jy}$ results in a distance estimate of \SI{20}{kpc}. Individual source distance estimates have however typical errors of 40\% \citep{1998ApJ...504..761C}. Moreover, as explained earlier, the radio surface brightness of G323.7$-$1.0 might be underestimated using the MOST data set. A two times higher radio flux would reduce the distance estimate to \SI{15}{kpc}. Nevertheless, even a distance to HESS\,J1534$-$571 at the \SI{10}{\kilo\parsec} scale would imply a TeV luminosity substantially exceeding the values of the known TeV SNRs, as further discussed in Sect.\,\ref{Sect:Discussion}.

\subsubsection{Radio continuum emission at the position of HESS\,J1912$+$101}
\label{SubSubSect:radio1912}

In the sky area covered by HESS\,J1912$+$101, a radio SNR candidate G44.6$+$0.1 was discovered in the Clark Lake Galactic plane survey, at \SI{30.9}{\mega\Hz} \citep{1988ApJ...328L..55K}. The SNR candidate status was confirmed through polarization detected at \SI{2.7}{\giga\Hz} \citep{1990ApJ...364..187G}. However, as already discussed in \citet{2008A&A...484..435A}, the morphology and extension of the radio SNR candidate and the \SI{}{\tera\eV} source do not match. G44.6$+$0.1 is just covering part of the northwestern \SI{}{\tera\eV} shell and has an approximate elliptical shape of $l \times b$ = \SI{41}{\arcmin} $\times$ \SI{32}{\arcmin}.\footnote{The error of the radio position is estimated to be \SI{0.1}{\degree}, and the relative error on the size is estimated to be \SI{30}{\percent} \citep{1988ApJ...328L..55K}.}

The area of HESS\,J1912$+$101 was covered by the NRAO/VLA Sky Survey \citep{1998AJ....115.1693C} at \SI{1.4}{\giga\Hz}. Also, sky maps from the new Multi-Array Galactic Plane Imaging Survey \citep[MAGPIS;][]{2006AJ....131.2525H}  that combine VLA and Effelsberg data at \SI{1.4}{\giga\Hz} were checked. The data suffer to some extent from side lobes from bright emission in the W49A region. No obvious counterpart to the \SI{}{\tera\eV} source was found by inspecting these data.

It is possible that the radio SNR candidate G44.6$+$0.1 is in fact part of a larger radio SNR corresponding to the \SI{}{\tera\eV} SNR candidate, but there is no further evidence from the data. Because of the lack of morphological correspondence between HESS\,J1912$+$101 and G44.6$+$0.1, the radio SNR candidate is not considered a confirmed counterpart to the \SI{}{\tera\eV} source, and G44.6$+$0.1 has therefore not been used to promote HESS\,J1912$+$101 from a SNR candidate to a confirmed SNR.

\subsubsection{Search for radio continuum emission from HESS\,J1614$-$518}
\label{SubSubSect:radio1614}

No cataloged radio SNR candidate exists in the field of HESS\,J1614$-$518. We inspected data from the Southern Galactic Plane Survey \citep[SGPS;][]{2006ApJS..167..230H}, obtained with the Australia Telescope Compact Array (ATCA) at \SI{1.4}{\giga\Hz}, and from the MGPS-2 \citep{2007MNRAS.382..382M} at \SI{843}{\mega\Hz}. No obvious features spatially coincident with HESS\,J1614$-$518 were found, which is consistent with the findings by \citet{2008PASJ...60S.163M} using data from the Sydney University Molonglo Sky Survey \citep[SUMSS;][]{1999AJ....117.1578B}. Much of the radio emission is associated with the HII regions to the western side of HESS\,J1614$-$518 and thus likely to be thermal in nature. 

\subsection{GeV emission with {\it Fermi}-LAT}
\label{SubSect:FermiLat}

\begin{figure}[]
\centering
{\includegraphics[width=0.9\columnwidth]{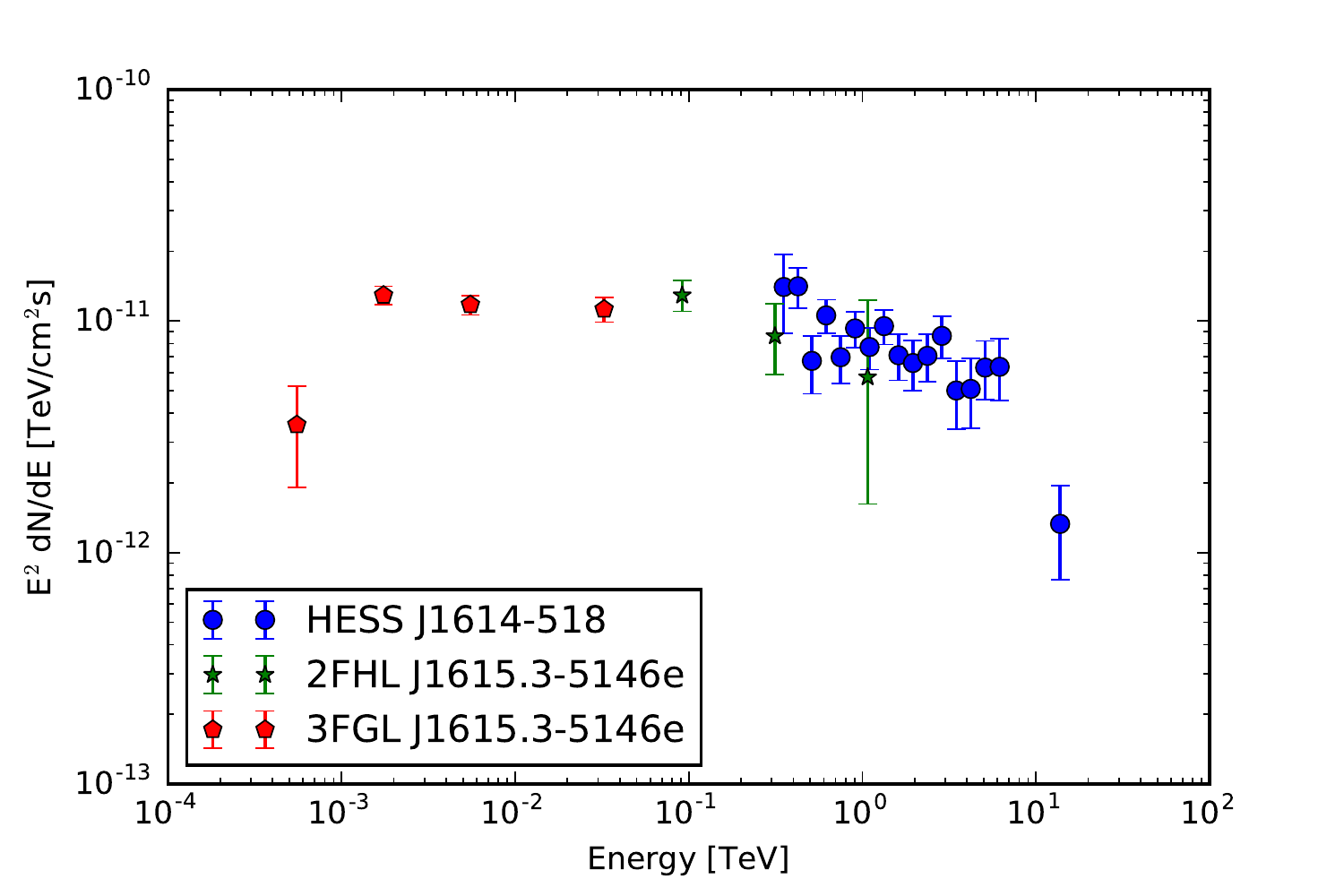}}  \\
\caption{
\SI{}{\tera\eV} spectral energy flux of HESS\,J1614$-$518 plotted together with the \SI{}{\giga\eV} spectral energy flux of \object{2FHL\,J1614.3$-$5146e} \citep{2016ApJS..222....5A} and \object{3FGL\,J1614.3$-$5146e} \citep{2015ApJS..218...23A}. The integral fluxes given in the {\it Fermi}-LAT catalogs were converted using the power-law models given in the catalogs. The energy of the flux points was determined by calculating the geometrical center of the energy boundaries in log-space. 
}
\label{Fig:1614sed}
\end{figure}

Given the flux of the new \SI{}{\tera\eV} SNR candidates, a detection of source photons also in the adjacent \SI{}{\giga\eV} band covered with the {\it Fermi}-LAT instrument \citep{2009ApJ...697.1071A} is plausible. Indeed, HESS\,J1614$-$518 has a known counterpart listed in the LAT source catalogs, namely 3FGL\,J1615.3$-$5146e / 2FHL\,J1615.3$-$5146e. The source is classified in both catalogs as disk-like. Extension and position of the LAT source make it a clear counterpart candidate for HESS\,J1614$-$518; see Fig.\,\ref{Fig:tevsurfacebrightnesses}. Together with the spectral match (see Fig.\,\ref{Fig:1614sed}), we followed \citet{2015ApJS..218...23A} and \citet{2016ApJS..222....5A} and identified 3FGL\,J1615.3$-$5146e / 2FHL\,J1615.3$-$5146e with HESS\,J1614$-$518. However, the identification currently does not add enough information to improve the astrophysical classification of the object.

Triggered by preliminary H.E.S.S. results on HESS\,J1534$-$571, \citet{2017ApJ...843...12A} has recently shown that a disk-like GeV counterpart to HESS\,J1534$-$571 can be extracted from {\it Fermi}-LAT data as well. The GeV source position and size are in good morphological agreement with G323.7$-$1.0 and HESS\,J1534$-$571. The astrophysical classification of the object as a SNR from the TeV and radio data is not affected by the GeV data at this stage.

HESS\,J1912$+$101 does not have a published counterpart in the LAT band. Since the \SI{}{\tera\eV} source is extended and located in the Galactic plane, source confusion and emission from the Galactic plane might so far have prevented discovery of the source using the LAT instrument information alone. An analysis of all three objects using the \SI{}{\tera\eV} sources as prior information for a LAT analysis is ongoing.


\section{Counterpart searches using X-ray, infrared, and radio/sub-mm line emission, and pulsar catalogs}
\label{Sect:XrayIRSubmm}

In this section, all MWL searches are reported that have not resulted in firm positive identifications with the new TeV sources. The following section provides details on these searches; in Sect.\ \ref{Sect:Discussion}  the most relevant information for the discussion of the individual sources (specifically the X-ray results and potential gas densities at the object locations) is summarized.

\subsection{Radio pulsars}
\label{SubSect:RadioPulsars}

The known, well-established TeV SNRs, such as RX\,1713.7$-$3946, RX\,J0852.0$-$4622 (Vela Jr.), RCW\,86, SN\,1006, Cas\,A, and HESS\,J1731$-$347, are not associated with known rotation-powered radio pulsars. RCW\,86 is very likely the SNR of a type Ia supernova \citep[e.g.,][]{2014MNRAS.441.3040B} without compact remainder. Cas\,A \citep{2000ApJ...531L..53P} and HESS\,J1731$-$347 (\citealt{2013A&A...556A..41K}, see also \citealt{2010ApJ...712..790T,2010ApJ...710..941H}) host central compact objects (CCOs), i.e.,\,neutron stars that are located close to the centers of the SNRs and whose X-ray emission is thermally driven; for RX\,1713.7$-$3946 and RX\,J0852.0$-$4622, CCO candidates are known. However, that does not preclude other \SI{}{\tera\eV} SNRs with core collapse SN progenitors to be associated with radio pulsars. While the detection of possibly associated pulsars may not help in confirming the SNR nature of the new \SI{}{\tera\eV} shells, they may be used to elaborate possible SNR scenarios. Also, since energetic pulsars may drive \SI{}{\tera\eV} PWNe, it is important to check for these possible alternative object scenarios that may explain (part of) the \SI{}{\tera\eV} emission. In relic PWN scenarios, significant angular offsets between the powering pulsar and bulk of the TeV emission can be expected \citep[e.g.,][]{special_PWN,2006A&A...460..365A,2008A&A...477..353A}. Therefore, all radio pulsars listed in the ATNF catalog\footnote{http://www.atnf.csiro.au/people/pulsar/psrcat/} \citep{2005AJ....129.1993M}, which are located inside a search radius of $R_{\mathrm{out}}+\SI{0.3}{\degree}$ around the fitted center of the TeV shell, were inspected.

\subsubsection{The radio pulsar \object{PSR\,J1913$+$1011} at the center of HESS\,J1912$+$101}
\label{Subsect:pulsar1912}

The rotation-powered pulsar PSR\,J1913$+$1011 is located close to the geometrical center of the TeV shell HESS\,J1912$+$101, at a distance of \SI{2.7}{\arcmin}. An association of that pulsar with the TeV source has been discussed in the context of a possible PWN scenario to explain the (lower-statistics) \SI{}{\tera\eV} source at the time of its discovery \citep{2008A&A...484..435A}. The pulsar has a spin-down power of $\dot{E} \simeq \SI{2.9e36}{\erg\per\second}$, a (characteristic) spin-down age of $\tau_{\mathrm{c}} \simeq 1.7 \times 10^5$ years, a spin period of \SI{36}{\milli \second}, and a distance of \SI{4.5}{\kilo\parsec} estimated from the dispersion measure (DM). The pulsar is in principle able to power a \SI{}{\tera\eV} PWN with a flux similar to HESS\,J1912$+$101 \citep{2008A&A...484..435A}. However, there is no known radio or X-ray PWN around PSR\,J1913+1011 \citep{2004IAUS..218..225G,2008ApJ...682.1177C}, and the \SI{}{\tera\eV} morphology does not support a PWN scenario at all for the \SI{}{\tera\eV} source. Still, PSR\,J1913$+$1011 may be the remainder of the SN explosion that has created the putative TeV SNR; this is discussed in more detail in Sect.\,\ref{Subsect:discussionj1912}.

None of the other radio pulsars in the field of HESS\,J1912$+$101 are particularly compelling counterpart candidates to the \SI{}{\tera\eV} source. None of the pulsars show known radio or X-ray PWNe. 

\subsubsection{Radio pulsars around HESS\,J1614$-$518 and HESS\,J1534$-$571}
\label{SubSubSect:pulsar16141534}

None of the radio pulsars in the field of HESS\,J1614$-$518 are compelling counterpart candidates in a \SI{}{\tera\eV} PWN scenario. Also in this case, none of the pulsars show known radio or X-ray PWNe. In principle, in relic PWN scenarios high efficiencies of converting current spin-down luminosity into \SI{}{\tera\eV} luminosity \citep[formally even above \SI{100}{\percent},][]{2006A&A...460..365A} can be present. It cannot be excluded that, for example,\ PSR\,J1613$-$5211 ($\tau_{\mathrm{c}} \approx 4\times10^{5}\,\SI{}{yrs}$, $\dot{E} \approx 8\times10^{33}\,\SI{}{\erg\per\second}$) drives the southwestern additional component in HESS\,J1614$-$518. However, there is currently no positive evidence to support such a hypothesis.

Three radio pulsars are located near HESS\,J1534$-$571, but energetics and distances make an association with HESS\,J1534$-$571 (in \SI{}{\tera\eV} PWN scenarios) unlikely.

\subsection{Search for X-ray emission from the new TeV sources}
\label{SubSect:Xrays}

All well-established \SI{}{\tera\eV} SNRs display strong extended nonthermal X-ray synchrotron emission from TeV electrons, typically in filamentary morphologies tracing the forward (and possibly reverse) shocks. Only a few \SI{}{\tera\eV} SNRs (such as Cas\,A and RCW\,86) also exhibit strong thermal thin-plasma X-ray emission. All \SI{}{\tera\eV} SNRs but HESS\,J1731$-$347 are X-ray selected; X-ray emission from these objects was detected before the respective \SI{}{\tera\eV} detection. However, objects such as\ the new \SI{}{\tera\eV} SNR candidates that are located very close to the Galactic plane may not have been detected in soft X-ray surveys, such as that performed with the {\it ROSAT} satellite (energy range \SIrange[range-phrase=\ --\ ]{0.1}{2.4}{\kilo\eV}), because of photoelectric absorption.\footnote{In case of Vela Jr., foreground emission from the Vela SNR inhibits the detection of X-ray emission at low X-ray energies from Vela Jr.} Indeed, none of the three new \SI{}{\tera\eV} SNR candidates have obvious counterparts in {\it ROSAT} survey data. However, current and recent pointed X-ray instruments, such as XMM-{\it Newton} and {\it Suzaku,} may have the sensitivity to detect typical X-ray counterparts of \SI{}{\tera\eV} SNRs that have been detected at current \SI{}{\tera\eV} instrument sensitivities.

In the following, published and unpublished X-ray observations are reviewed that cover the TeV sources, in view of possible shell-like X-ray emission. In addition,  X-ray (mostly point source) catalogs were checked, but (with the exception of \object{XMMU\,J161406.0$-$515225} reviewed below) no compelling counterparts were found there.

\subsubsection{{\it Suzaku} observations of HESS\,J1534$-$571}
\label{SubSect:xrays1534}

After the initial discovery of significant \SI{}{\tera\eV} emission from the position of HESS\,J1534$-$571 with H.E.S.S., the source was observed with {\it Suzaku} XIS \citep{2007PASJ...59S...1M, 2007PASJ...59S..23K} in four pointings with exposure times of \SIlist{36.9;21.2;38.8;24.4}{\kilo\second} (observation IDs 508013010, 508014010, 508015010, and 508016010, respectively; PI A.\ Bamba). Further pointings to complete the coverage of the \SI{}{\tera\eV} source with {\it Suzaku} had already been approved, but could not be performed because of the failure of the satellite and subsequent decommissioning of the observatory.

\begin{figure}[t!]
\centering
{\includegraphics[width=0.8\columnwidth]{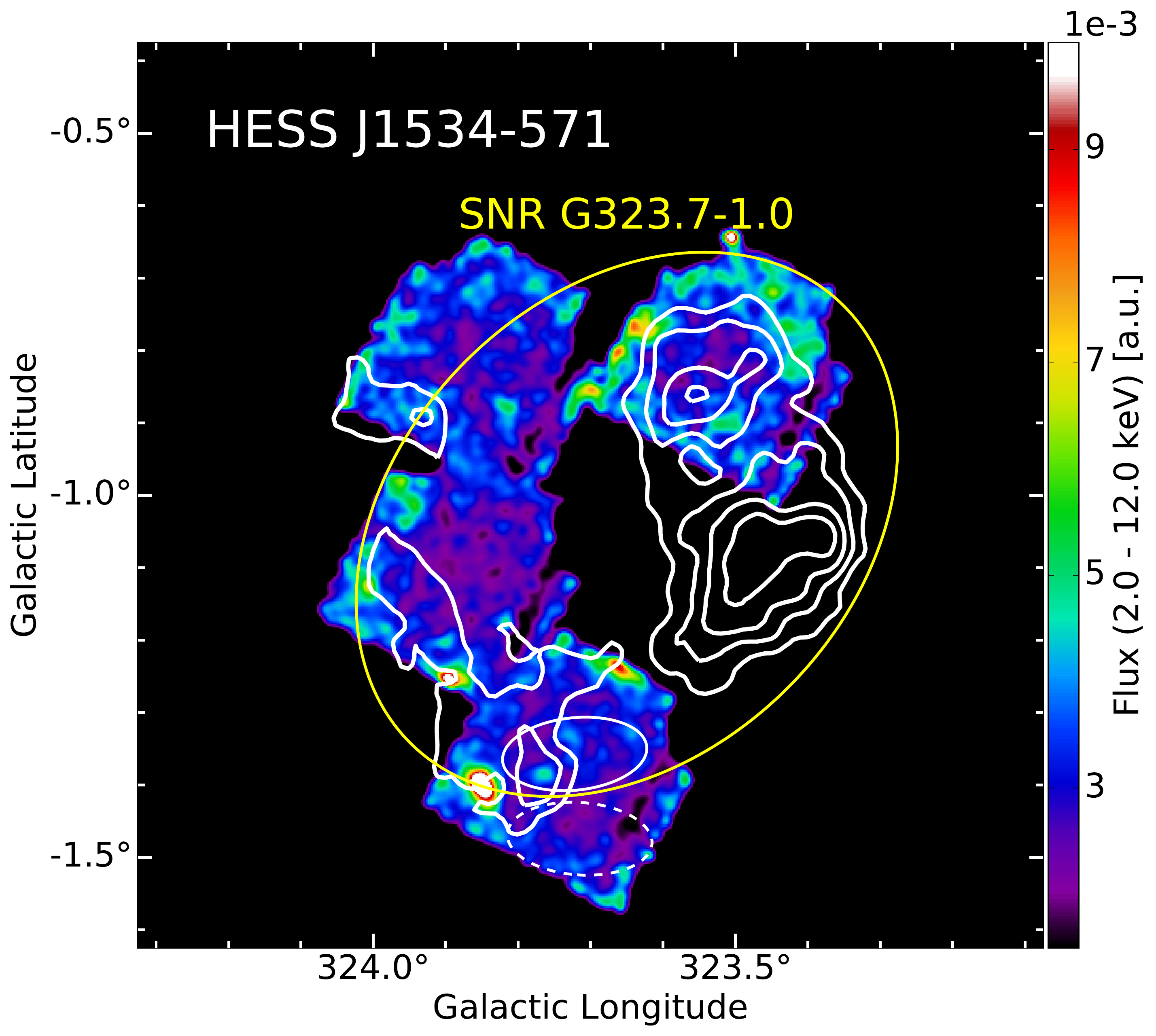}}
\caption{
{\it Suzaku} XIS mosaic of the pointings toward HESS\,J1534$-$571, in a hard band of \SIrange[range-phrase=\ --\ ]{2}{12}{\kilo\eV}, using the XIS0 and XIS3 detectors. Point sources were not removed from the image. Contours denote the TeV surface brightness. The large solid ellipse denotes the outer boundary of the radio SNR. The small solid ellipse is the extraction region to derive an X-ray upper limit estimate from the SNR; the dashed ellipse is the corresponding background extraction region.}
\label{Fig:1534suzaku}
\end{figure}

For the analysis of the data, XSELECT and the {\it Suzaku} FTOOLS ver.\,20 (part of HEASOFT ver.\,6.13) were used. Particle-background subtracted, vignetting- and exposure-corrected mosaic images in the full band and in the harder band of \SIrange[range-phrase=\ --\ ]{2}{12}{\kilo\eV} were created. The harder band is expected to be more sensitive specifically to a nonthermal component of the potential X-ray counterpart because of Galactic absorption. In Fig.\,\ref{Fig:1534suzaku}, this hard-band mosaic image is shown. No significant emission is detected from the source region in both mosaics. To estimate an X-ray upper limit from the area of the \SI{}{\tera\eV} source, spectra from a limited on- and an off-source region are derived; see Fig.\,\ref{Fig:1534suzaku} for the extraction regions that are defined with respect to the radio SNR boundary. In both spectra, there is a soft emission component whose characteristics are consistent with emission from hot thermal interstellar gas, and which is likely due to local X-ray foreground. In order to estimate an X-ray flux upper limit from the SNR, an additional (absorbed) power-law component was included in the on-spectrum model. The (unabsorbed) flux upper limit, derived from this power-law model (assuming two different photon indices $\Gamma$) and scaled to the area of the entire radio SNR, is \SI{2.4e-11}{\erg\per\square\centi\metre\per\second} in the \SIrange[range-phrase=\ --\ ]{2}{12}{\kilo\eV} band for $\Gamma$ = 2 and \SI{1.9e-11}{\erg\per\square\centi\metre\per\second} in the \SIrange[range-phrase=\ --\ ]{2}{12}{\kilo\eV} band for $\Gamma$ = 3.

\begin{figure}[t!]
\centering
{\includegraphics[width=0.8\columnwidth]{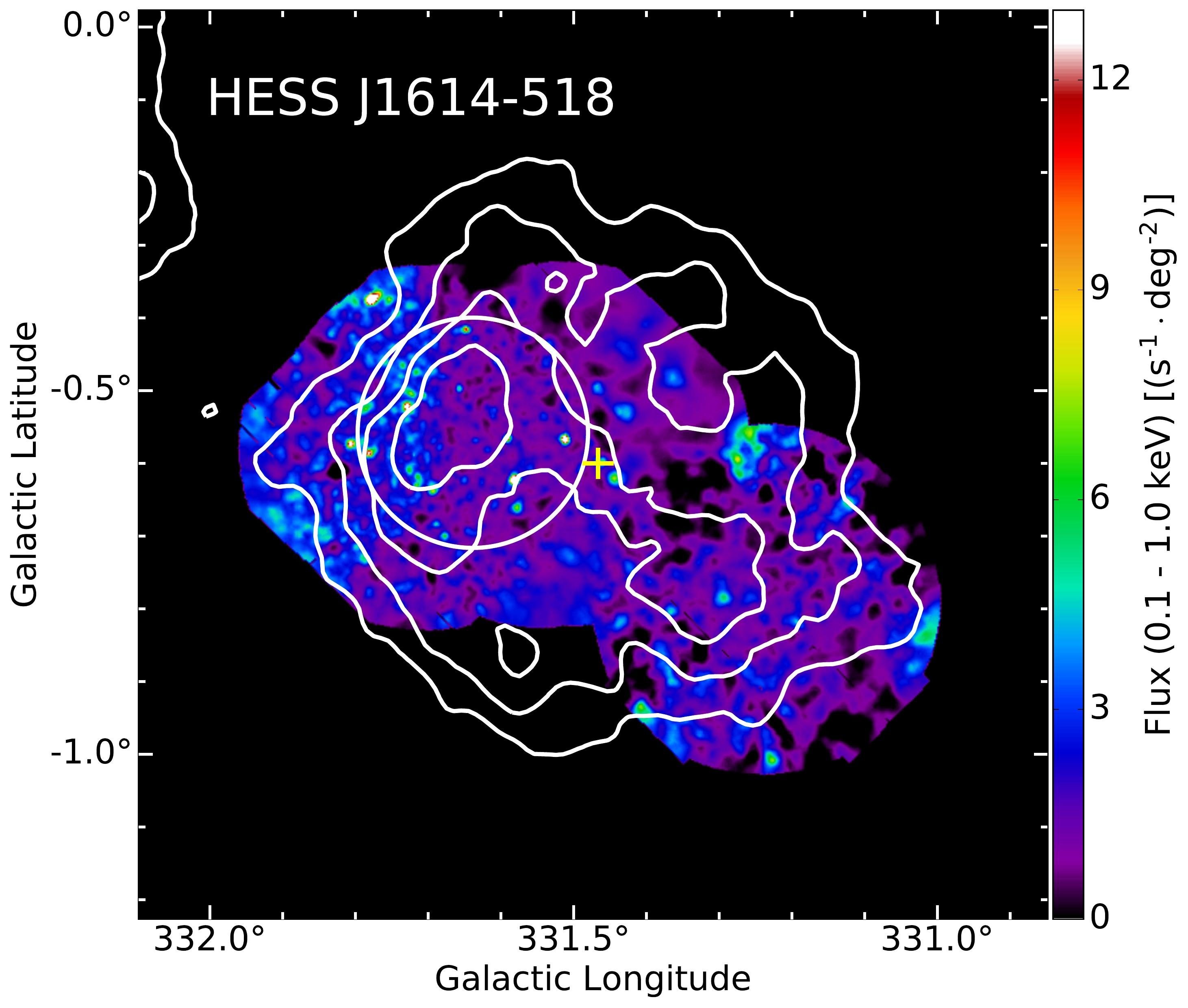}} \\
{\includegraphics[width=0.8\columnwidth]{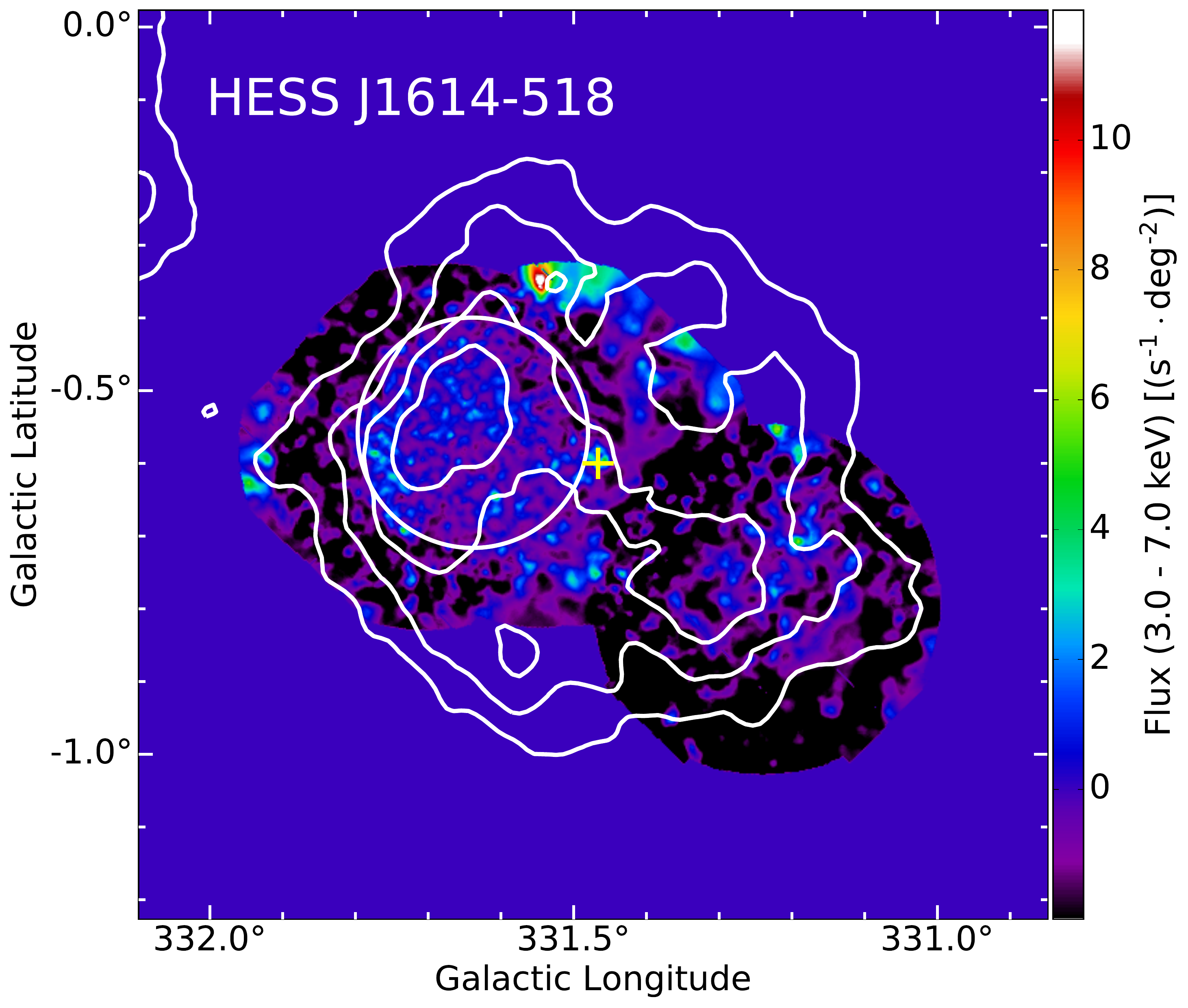}} 
\caption{
\textit{XMM-Newton} mosaics of the pointings toward HESS\,J1614$-$518, in a soft band (top panel, \SIrange[range-phrase=\ --\ ]{0.3}{1.0}{\kilo\eV}) and in a hard band (bottom panel, \SIrange[range-phrase=\ --\ ]{3}{7}{\kilo\eV}). Point sources were removed from the images. Contours denote the TeV surface brightness. The cross indicates the position of XMMU\,J161406.0$-$515225. The soft-band image is dominated by stray light from RCW\,103 (northeastern arc feature). The hard-band image is dominated by a putative diffuse X-ray emission region coincident with the north component of the \SI{}{\tera\eV} source. The solid circle indicates an extraction area used to assess the spectrum for this diffuse component. Different background control regions (not shown) were used to estimate the systematic error induced by the background estimate.
}
\label{Fig:1614xmm}
\end{figure}

\subsubsection{{\it XMM-Newton} and {\it Suzaku} observations of HESS\,J1614$-$518}
\label{SubSect:xrays1614}

HESS\,J1614$-$518 was observed with \textit{Swift}, \textit{Suzaku}, and \textit{XMM-Newton} in several pointings after the announcement of the TeV discovery \citep{2006ApJ...636..777A}. Observations did not cover the entire \SI{}{\tera\eV} source but focused on the northeastern and southwestern components as well as on the central position. \citet{2008PASJ...60S.163M} and \citet{2011PASJ...63S.879S} reported on all {\it Suzaku} observations of the source and on the {\it XMM-Newton} data on the central area. Concerning point sources specifically in the central area, a possibly relevant source for this study is XMMU\,J161406.0$-$515225, at a distance of $\sim$\SI{1}{\arcmin} from the geometrical center of the \SI{}{\tera\eV} shell.\footnote{The central extended {\it Suzaku} source \object{Suzaku\,J1614$-$5152} \citep{2008PASJ...60S.163M} was resolved into several point sources using {\it XMM-Newton}, including the strongest source XMMU\,J161406.0$-$515225 \citep{2011PASJ...63S.879S}.} \citet{2008PASJ...60S.163M} argued that the source might be an anomalous X-ray pulsar related to the \SI{}{\tera\eV} object.  However, XMMU\,J161406.0$-$515225 has an optical point source counterpart, as already noted by \citet{2006ApJ...651..190L} based on the {\it Swift}-XRT detection of the source, and the source was classified as a star candidate in \citet{2012ApJ...756...27L}.

\citet{2008PASJ...60S.163M} also reported on an extended X-ray source (\object{Suzaku\,J1614$-$5141}) with angular scale \SI{5}{\arcmin}, coincident with the northeastern component of HESS\,J1614$-$518. The X-ray absorption column is similar to that of XMMU\,J161406.0$-$515225, thus both objects could be related to the \SI{}{\tera\eV} source, at a distance scale to Earth on the order of $\SI{10}{kpc}$ using the X-ray absorption column as proxy \citep{2011PASJ...63S.879S}. To further probe potential diffuse X-ray emission from the region of HESS\,J1614$-$518, all {\it XMM-Newton} observations on the object have been examined (ObsID 0406650101 (northeast; PI Rowell), 0555660101 (southwest; PI Horns), and 0406550101 (central; PI Bussons), net exposure after event filtering (EPIC-MOS1/MOS2/pn) 21.0/19.9/13.7\,ksec, 17.0/17.6/6.5\,ksec, and 5.6/6.5/3.2\,ksec, respectively). Data were analyzed using XMMSAS ver.\,14.0.0. First, mosaic images of the extended emission were created with ESAS. To this end, images for each observation were created and source detection was performed. Point sources detected by this procedure were masked out of the data. Then the quiet particle background and the soft proton contamination were modeled for each observation. All these images were combined and used to create background-subtracted, exposure-corrected mosaic images. Fig.\,\ref{Fig:1614xmm} shows these mosaics in two different energy bands, \SIrange[range-phrase=\ --\ ]{0.3}{1.0}{\kilo\eV}, and \SIrange[range-phrase=\ --\ ]{3}{7}{\kilo\eV}. The soft-band image demonstrates that the field specifically in the northeast is significantly contaminated by (soft) stray light from the nearby SNR RCW\,103. The hard-band image shows that there is likely a hard diffuse X-ray emission component coincident with the northeastern \SI{}{\tera\eV} component of HESS\,J1614$-$518 with an angular scale of \SI{20}{\arcmin}, possibly indicating that the {\it Suzaku} source Suzaku\,J1614$-$5141 is more extended than seen in the {\it Suzaku} image.

To estimate the flux of the diffuse X-ray emission component, spectra were extracted for the entire hard emission region seen in the northwest above \SI{3}{\kilo\eV} (see circle in Fig.\,\ref{Fig:1614xmm}), and for a region excluding the strongest stray-light impact. Nearby background regions were selected to be representative of the expected background in the respective source extraction regions, and their spectra were fitted simultaneously. In general, a hard power-law component ($\Gamma \lesssim 2.0$) is identified at the source position at high significance, but systematic errors are large specifically due to the stray-light impact. 

A detailed assessment of the parameters of this diffuse X-ray component and its detection significance, including all systematic effects, is beyond the scope of this paper, as is a detailed comparison with the {\it Suzaku} result of Suzaku\,J1614$-$5141.  The emission -- if confirmed -- fills a large portion of the FoVs of the EPIC instruments, and spectral analysis requires a detailed modeling of all background components. It seems very likely that the northwestern component of the TeV source HESS\,J1614$-$518 is accompanied by hard diffuse X-ray emission. However, the results are not sufficient to significantly improve the astrophysical classification of the object at this time.

\subsubsection{X-ray observations of HESS\,J1912$+$101}
\label{SubSect:xrays1912}

HESS\,J1912$+$101 is located at an angular distance of about \SI{47}{\arcmin} to GRS\,1915+105, making observations of the HESS\,J1912$+$101 region with X-ray satellites that are susceptible to stray light very difficult. As already discussed in \citet{2008A&A...484..435A}, archival {\it ASCA} observations coincident with the HESS\,J1912$+$101 area are strongly affected by stray-light artifacts and are therefore of limited use to search for X-ray counterparts to HESS\,J1912$+$101. Archival {\it Chandra} data from observations targeting at PSR\,J1913$+$1011 (and only covering the central region of HESS\,J1912$+$101) were analyzed by \citet{2008ApJ...682.1177C} to explore a potential PWN scenario for HESS\,J1912$+$101 and to search for X-ray counterparts. None of the detected nine point sources seem particularly outstanding. No diffuse emission was detected. In view of the improved \SI{}{\tera\eV} morphology derived in this work, we reanalyzed the {\it Chandra} data (ObsId 3854). Comparison with the \SI{}{\tera\eV} image confirms the lack of compelling counterparts to HESS\,J1912$+$101, but also shows that there is no significant overlap of the {\it Chandra} exposure with the TeV-emitting shell.

At the moment, the shell of HESS\,J1912$+$101 remains largely unexplored at current pointed X-ray satellite sensitivity.

\begin{figure}[t!]
\centering
{\includegraphics[width=0.23\paperheight]{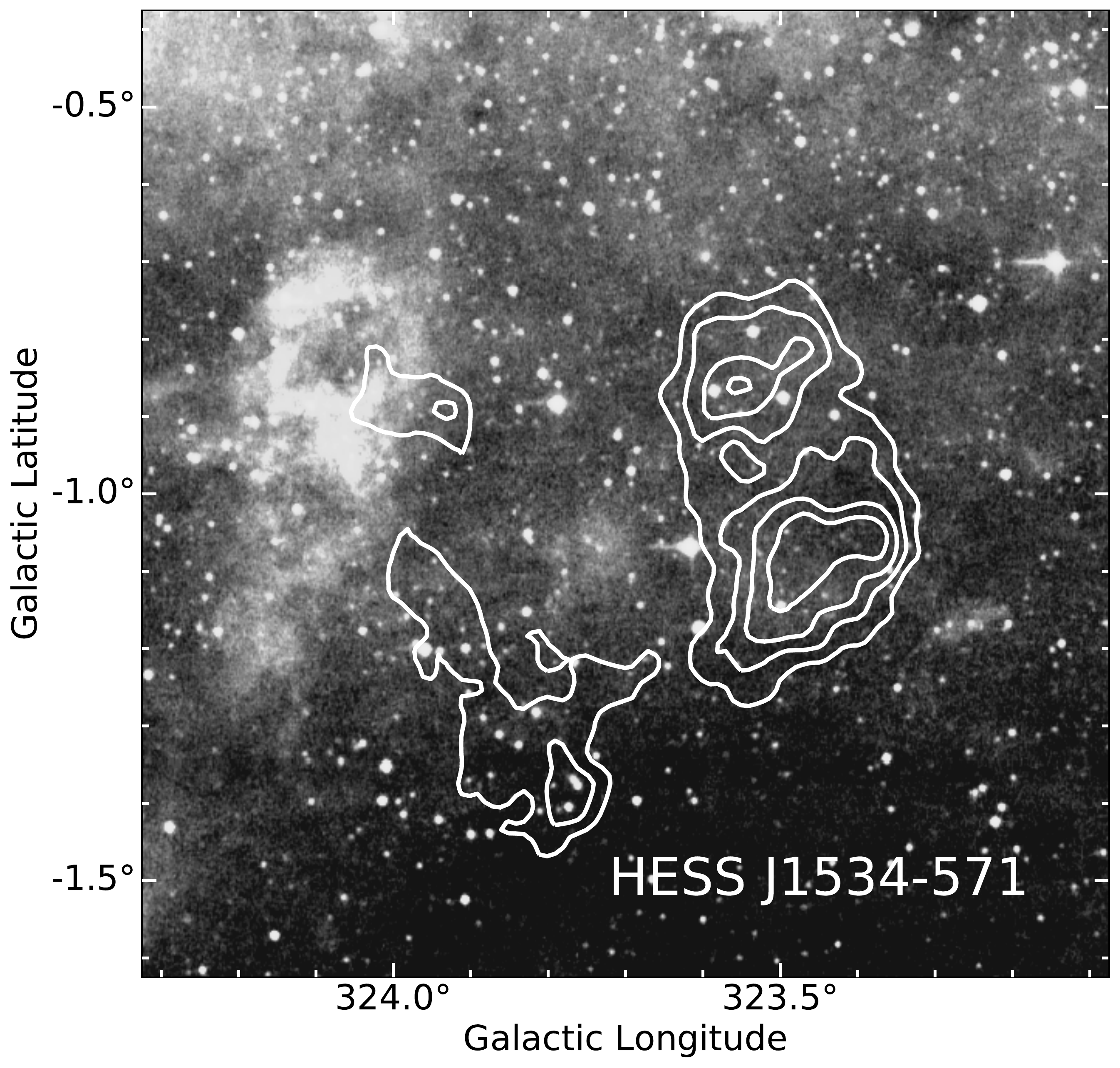}}  \\
{\includegraphics[width=0.23\paperheight]{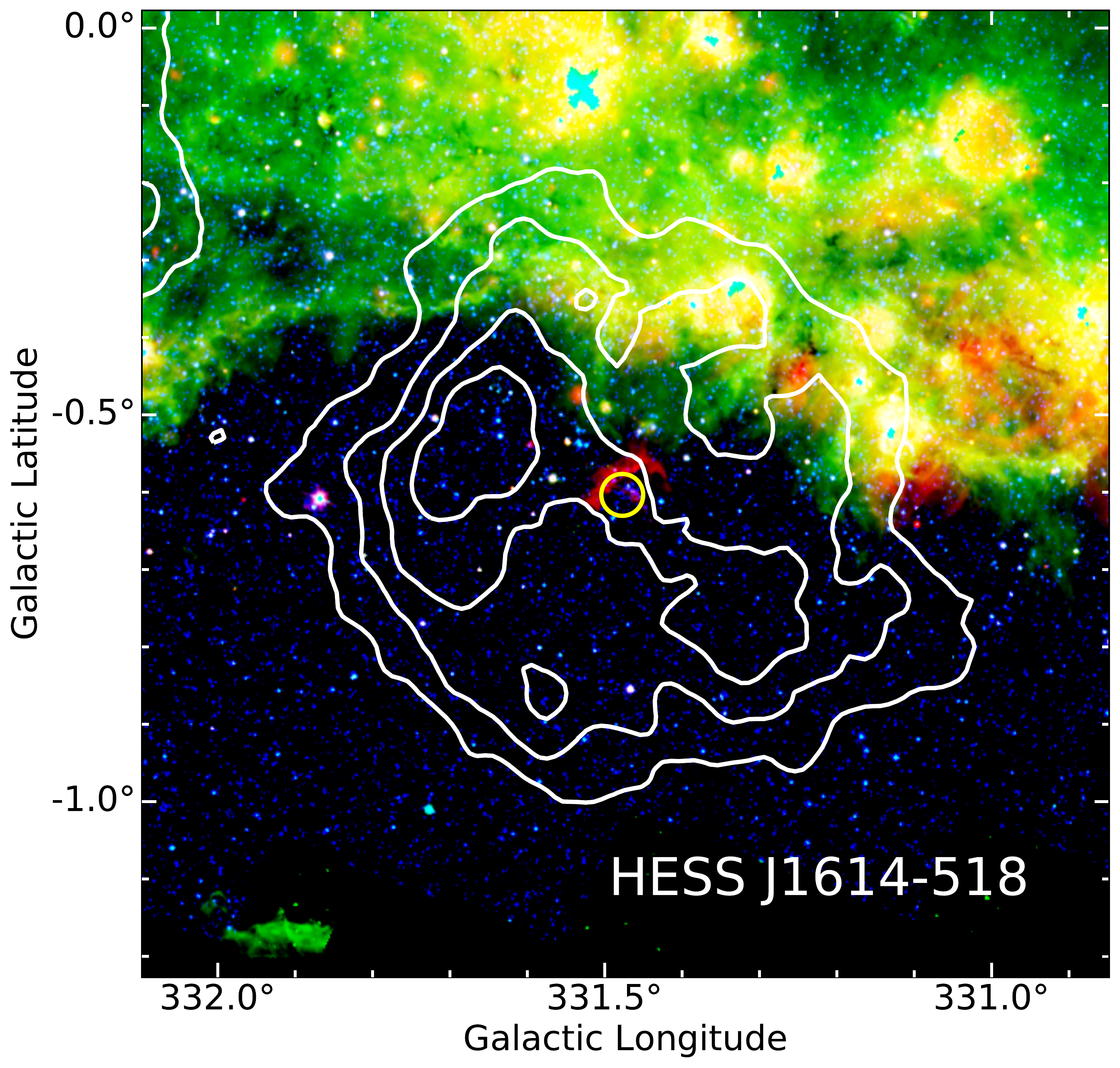}}  \\
{\includegraphics[width=0.23\paperheight]{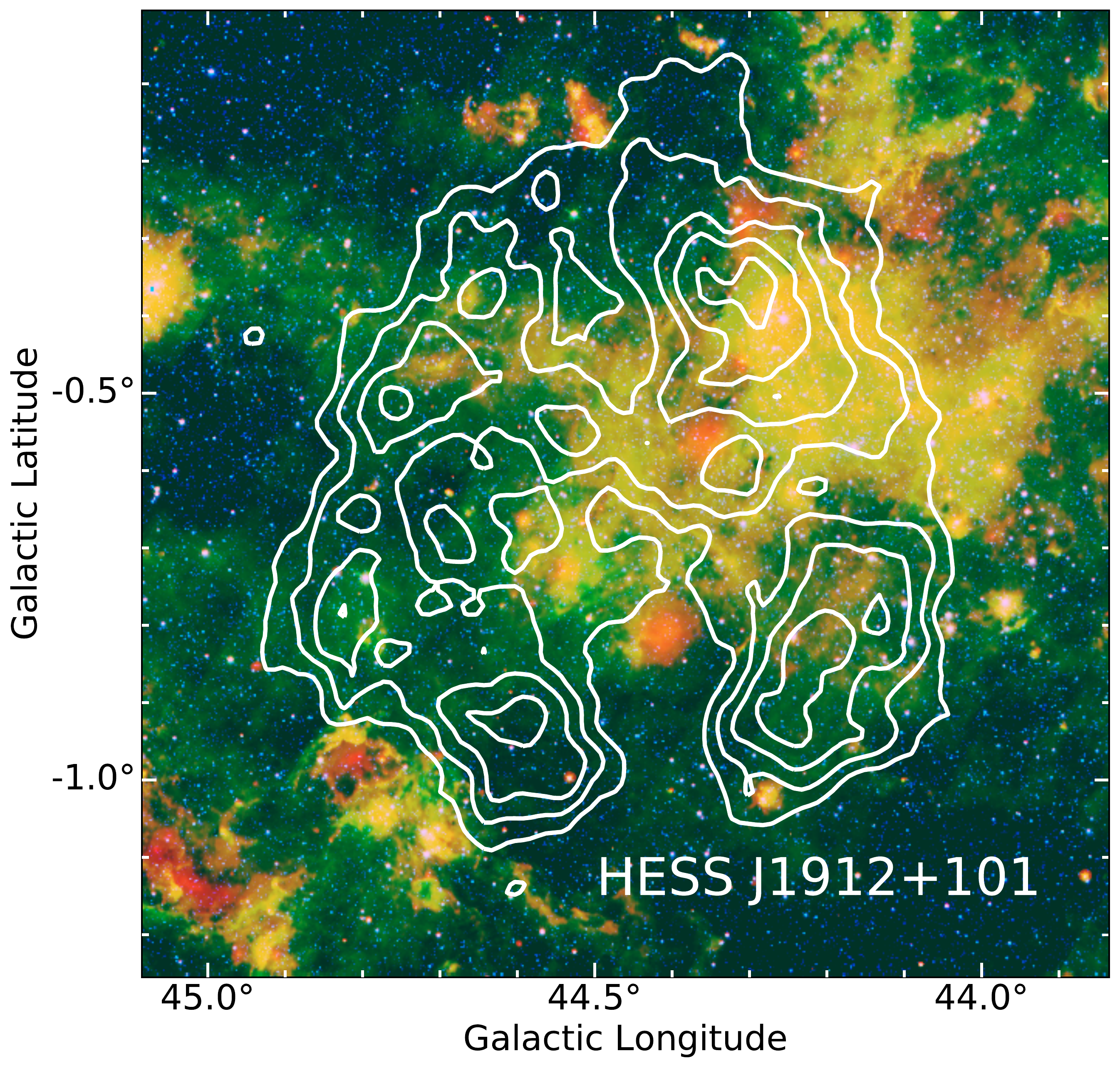}} 
\caption{Archival infrared images toward the fields of the three \SI{}{\tera\eV} sources. The {\it top panel} shows an MSX \citep{2001AJ....121.2819P} image of the region toward HESS\,J1534-571, at \SI{8.28}{\micro\metre}. The {\it middle} and {\it bottom panels} show three-color Spitzer images toward HESS\,J1614$-$518 and HESS\,J1912$+$101, respectively. 
Red, green, and blue colors indicate \SI{24}{\micro\metre} \citep[MIPSGAL,][]{2009PASP..121...76C}, \SI{8}{\micro\metre}, and \SI{3.6}{\micro\metre} \citep[GLIMPSE,][]{2009PASP..121..213C} emission, respectively. Color scales were adjusted individually to emphasize the structures in the images. Contours denote the \SI{}{\tera\eV} surface brightness of the respective source. The circle at the center of HESS\,J1614$-$518 denotes the position and extension of Pismis\,22 \citep{2013A&A...560A..76M}.
}
\label{Fig:irimages}
\end{figure}

\begin{figure*}[ht]
\centering
\includegraphics[width=0.99\textwidth]{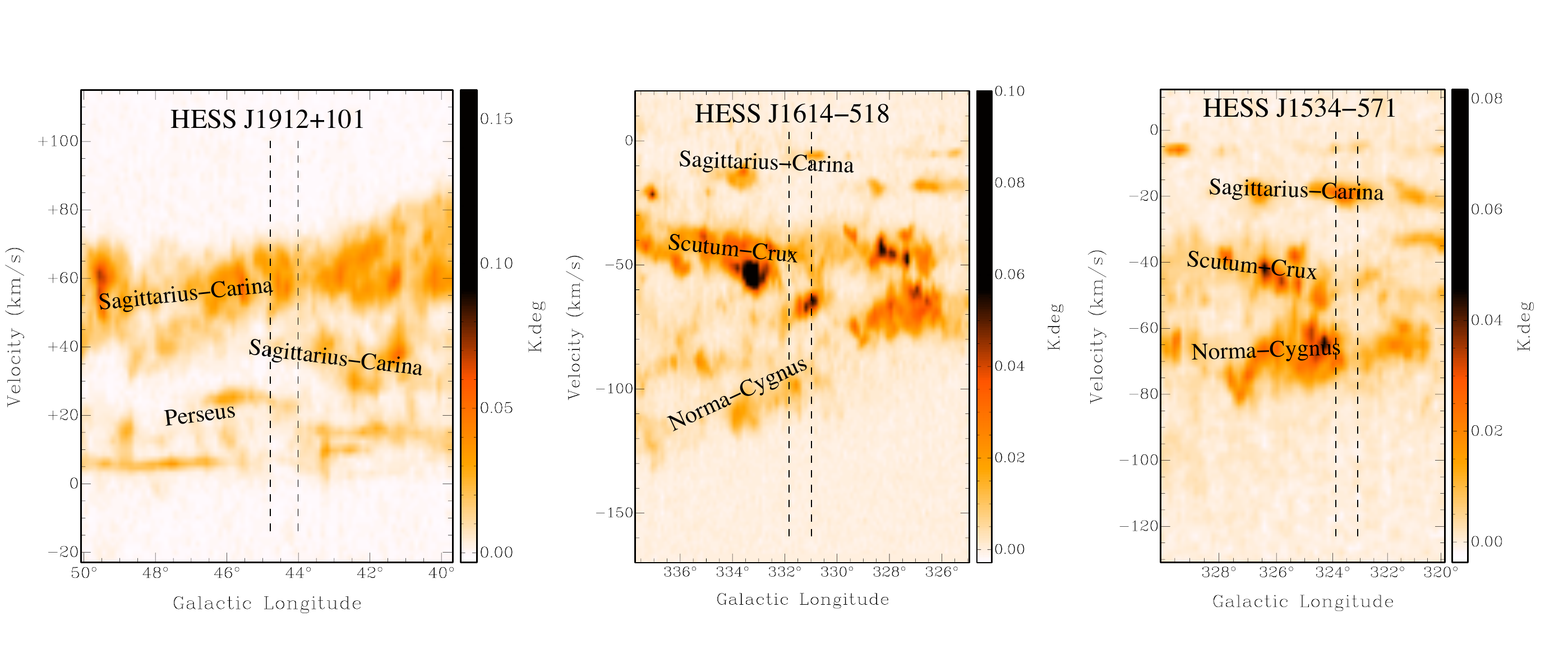}
\caption{Longitude-velocity plots of Columbia CO(1-0) data \citep{2001ApJ...547..792D}, integrated over a latitudinal range consistent with HESS\,J1912$+$101, HESS\,J1614$-$518, and HESS\,J1534$-$571 (left, middle, and right, respectively). The longitudinal extent of each of these SNRs is indicated by dashed lines. The names of Galactic arms are overlaid onto the approximate corresponding map locations, following \citet{2008AJ....135.1301V}.
\label{Fig:pvplot}}
\end{figure*}

\begin{figure*}[ht]
\centering
\includegraphics[width=0.80\textwidth]{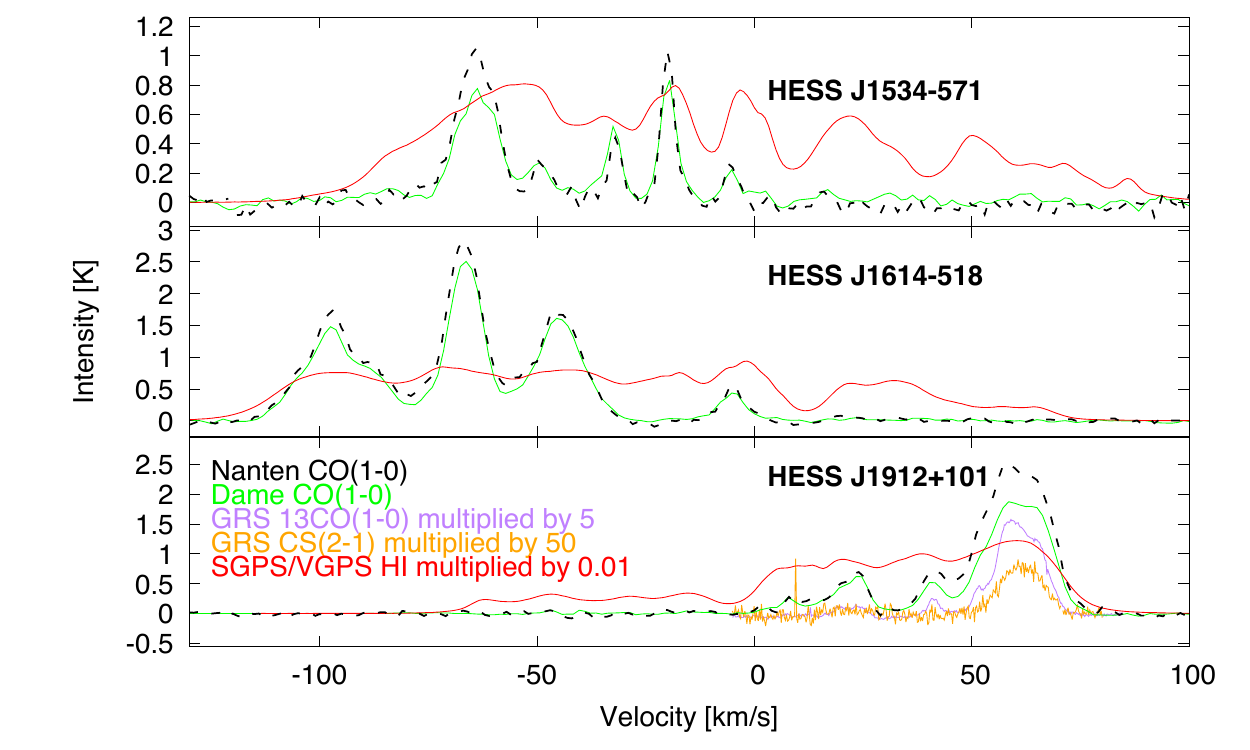}
\caption{Average Nanten/Columbia CO(1-0) \citep{2001PASJ...53.1003M,2001ApJ...547..792D} data within circular regions encompassing HESS\,J1534$-$571 ([\textit{l,b,r}]=[323.70,-1.02,0.4]), HESS\,J1614$-$518 ([\textit{l,b,r}]=[331.473,-0.601,0.42]), and HESS\,J1912$+$101 ([\textit{l,b,r}]=[44.46,-0.13,0.49]). HI data are also shown for all sources \citep[SGPS/VGPS][]{2005ApJS..158..178M,2006AJ....132.1158S}, while GRS $^{13}$CO(1-0) and CS(2-1) \citep{2006ApJS..163..145J} data are shown for HESS\,J1912$+$101. The CS(2-1) feature at \SI{+9.5}{\kilo\metre\per\second} in the bottom image is likely an artifact caused by an offset in one frequency channel.  
\label{Fig:linespectra}}
\end{figure*}

\subsection{Infrared emission, possible associations with HII regions and stellar clusters}
\label{SubSect:Infrared}

Infrared maps obtained with Spitzer at \SIlist{24;8;3.6}{\micro\metre} are used to illustrate the projected distribution of warm gas, HII regions, and stellar clusters in the field of the new \SI{}{\tera\eV} sources (Fig.\,\ref{Fig:irimages}). HESS\,J1534$-$571 is only partially covered by Spitzer data, therefore a Midcourse Space Experiment (MSX) map at \SI{8.2}{\micro\metre} is used for that source. 

The maps illustrate HII emission regions in apparent spatial coincidence with all three \SI{}{\tera\eV} sources. Such emission regions could consist of stellar wind material \citep[e.g.,][]{2007A&A...468..993K} from star-forming regions that could have also hosted the SNR progenitor stars. The formation of HII-emitting regions might also be triggered by the interaction with one or several SNRs. A morphological correlation between the IR and the \SI{}{\tera\eV} maps is not seen in the images and would also not necessarily be expected even if the \SI{}{\tera\eV} sources were associated with the HII emission regions. 

The open stellar cluster \object{Pismis\,22} is located close to the geometrical center of HESS\,J1614$-$518. Its age is estimated to be \SI[separate-uncertainty]{40(15)}{\mega yrs}, at a distance of \SI[separate-uncertainty]{1.0+-0.4}{\kilo\parsec} \citep{2000A&A...360..529P}. Pismis\,22 could be the host of the progenitor star of a SNR, if the SNR interpretation of HESS\,J1614$-$518 is confirmed.

It is also interesting to evaluate the energy from the cluster as a whole. The total cluster mass is unconstrained, since most of the stars in the vicinity of Pismis\,22, which are likely to be member stars, lack a spectral type determination. The expected kinetic energy in the system was estimated using the Starburst 99 cluster evolution model \citep[][and references therein]{2000ESASP.445...37L}. The model results scale with the initial cluster mass $M_{\mathrm{SC}}$. The total kinetic energy including stellar winds and SNe over a time span of \SI{40}{\mega yrs} is $E_{\mathrm{kin}} \simeq 1.3 \times 10^{52} \left( M_{\mathrm{SC}}/10^{3}M_{\sun}\right) \mathrm{erg}$. Such a system with its output in kinetic energy would certainly leave its imprint in the cluster surroundings. Models such as that by \citet{2005ApJ...635.1116S} predict a large cluster-wind driven void. Interestingly, the candidate HII region G331.628$-$00.926 with a radius of $\sim$\SI{0.8}{\degree} is encompassing in projection both HESS\,J1614$-$518 and Pismis\,22. The current cluster luminosity at the adopted age of \SI{40}{\mega yrs} is $L_{\mathrm{kin}}(t = \SI{40}{\mega yrs}) \simeq 5.2 \times 10^{36} M_{\mathrm{SC}}/10\,M_{\sun}\,\mathrm{erg\,s^{-1}}$. Therefore, a fraction of this luminosity would be sufficient to explain the TeV luminosity of HESS\,J1614$-$518 in terms of its energy requirement. Lacking any observational evidence, this possibility remains hypothetical for the moment.

In conclusion, the presented IR data themselves do not contribute to the astrophysical identification of the new \SI{}{\tera\eV} sources, but the HII and stellar data add information to a possible SNR scenario for HESS\,J1614$-$518.

\subsection{Atomic and molecular gas density around the sources}
\label{SubSect:Gas}

Archival radio and sub-mm atomic and molecular line data were investigated to search for signatures of gas associations corresponding to the new shell-type $\gamma$-ray sources. Voids in HI data would be suggestive of stellar wind bubbles blown by massive progenitor stars, while arcs of HI emission or asymmetric spectral line profiles may be attributable to gas shocked by a SNR. Such features would deliver a SNR kinematic distance solution. A positive correlation between $\gamma$-ray emission and gas density would be suggestive of a spatial connection that would yield a kinematic distance estimate, while at the same time lending support to the hypothesis that a $\gamma$-ray source is of hadronic origin \citep[typically better seen in molecular gas, e.g., toward SNR W28;][]{2008A&A...477..353A}.

Nanten CO(1-0) data \citep{2001PASJ...53.1003M} are available toward the new TeV shells and have an angular resolution of $\sim$\SI{3}{\arcmin}. Additionally, Galactic Ring Survey \citep[GRS;][]{2006ApJS..163..145J} $^{13}$CO(1-0) and CS(2-1) data with an angular resolution of \SI{46}{\arcsecond} are available toward HESS\,J1912$+$101 for molecular gas components with positive velocities. To trace atomic gas (as opposed to the aforementioned molecular gas tracers) toward the new H.E.S.S.\ shells, $\sim$\SI{2}{\arcmin}-resolution Southern Galactic Plane Survey \citep[SGPS;][]{2005ApJS..158..178M} HI data are available for the southern sources HESS\,J1534$-$571 and HESS\,J1614$-$518, while HESS\,J1912$+$101 is covered by the \SI{1}{\arcmin} resolution HI data from the VLA Galactic Plane Survey \citep[VGPS;][]{2006AJ....132.1158S}.

Longitude-velocity plots of Columbia CO(1-0) data \citep{2001ApJ...547..792D} are used to illustrate the large-scale Galactic structure toward the new TeV shells in Fig.\,\ref{Fig:pvplot}. For translating radial velocities relative to the local standard of rest (LSR) into distance estimates, the prescriptions of \citet{2008AJ....135.1301V,2013IJAA....3...20V} are used. In general, there are often ambiguities in associating specific gas components with specific Galactic arms, but associations indicated in Fig.\,\ref{Fig:pvplot} are considered the most likely according to current literature. HESS\,J1534$-$571 and HESS\,J1614$-$518 are within \SI{8}{\degree} of each other and are both in projection coincident with the Sagittarius-Carina, Scutum-Crux, and Norma-Cygnus arms \citep{2008AJ....135.1301V,2013IJAA....3...20V}. Regarding HESS\,J1912$+$101, the CO data show two velocity components meeting near the tangent point of the Sagittarius arm \citep{2008AJ....135.1301V,2013IJAA....3...20V}, so significant uncertainty exists in distance estimations to specific atomic and molecular gas components there. 

CO, $^{13}$CO, CS, and HI spectral profiles toward the three new \SI{}{\tera\eV} shell sources are shown in Fig.\,\ref{Fig:linespectra}. Multiple line-of-sight gas components may potentially be associated with the \SI{}{\tera\eV} sources. No unambiguous match has been found in the comparisons of the ISM sky maps with the \SI{}{\tera\eV} sky maps. For these comparisons, CO and HI sky maps were produced in a series of velocity bands covering the entire velocity range shown in Fig.\,\ref{Fig:linespectra}, and were then compared to the \SI{}{\tera\eV} maps. The only noteworthy indication found in this investigation is a void or dip in the SGPS HI data set centered on HESS\,J1614$-$518 in the distance range between \SI{1.2}{\kilo\parsec} and \SI{1.5}{\kilo\parsec} ($v_{\mathrm{lsr}}=$\ \SIrange{-15}{-22}{\kilo\metre\per\second}) (see Fig.\,\ref{Fig:1614hicoslices} top left panel), which may suggest an association with HESS\,J1614$-$518. The HI emission appears as a partial shell toward the rim of the TeV source, hinting at the possibility of a blown-out bubble related to a shell-type SNR (from the progenitor star) or to an  alternative central energy source. This distance range encompasses that of the Pismis\,22 open cluster (cf.\ Sect.\,\ref{SubSect:Infrared}), and thus may signal the mechanical influence of this cluster on the atomic ISM, in a similar fashion to the HI feature identified by \citet{2007A&A...468..993K} toward the WR cluster Westerlund\,1, which is possibly associated with a TeV $\gamma$-ray source \citep{2012A&A...537A.114A}.

Since no unambiguous association was found for any of the three sources, two representative distances with a possible gas match have been chosen per source. The goal is to demonstrate that reasonable hadronic emission scenarios can be constructed for each of the sources, using possible target gas densities in which the sources could be embedded (deviating from the canonically assumed particle density of $1\,\mathrm{cm^{-3}}$). No attempt was made to quantify whether the matches themselves are statistically significant, from the gas data alone. The associations are detailed in Appendix \ref{SubSect:ShellsGasAssociations}. The way gas densities are derived is explained in Appendix \ref{SubSect:ShellsGasMasses}. Results are shown in Table\ \ref{Table:NewSNRShells_2} and are discussed further in the following discussion section.


\section{Discussion}
\label{Sect:Discussion}

\renewcommand{\arraystretch}{1.5}

\begin{table*}[t]
\caption{Overview of confirmed (HESS\,J1534$-$571) and possible (HESS\,J1614$-$518, HESS\,J1912$+$101) association scenarios.}
\label{Table:Associations}
\centering
\begin{tabular}{lllc}
\hline\hline
H.E.S.S. source & Association, distance estimate method, and paper section & Galactic spiral arm & Distance scale [\SI{}{\kilo\parsec}]  \\
\hline
HESS\,J1534$-$571 & G323.7$-$1.0 ($\Sigma - D$ underluminous, \ref{SubSubSect:radio1534} and \ref{Subsect:discussionj1534})   & Scutum-Crux & 3.5 \\
                  & G323.7$-$1.0 ($\Sigma - D$, \ref{SubSubSect:radio1534})                    & Norma-Cygnus        & 8   \\ 
\hline
HESS\,J1614$-$518 & Pismis 22 ({\citet{2000A&A...360..529P}}, \ref{SubSect:Infrared}), HI void (line velocity, \ref{SubSect:Gas})               & Sagittarius-Carina  & 1.5 \\
                  & XMMU\,J161406.0 and Suzaku\,J1614 ($N_{\mathrm{H}}$, \ref{SubSect:xrays1614})   & Norma-Cygnus        & 5.5 \\ 
\hline
HESS\,J1912$+$101 & PSR\,J1913$+$1011 (DM, \ref{Subsect:pulsar1912})   & Sagittarius-Carina  & 4.5 \\
                  &                                                          & Perseus      & 10 \\ 
\hline\hline
\end{tabular}
\end{table*}

\renewcommand{\arraystretch}{1.0}

From the presented morphological studies using H.E.S.S. data, the three TeV sources HESS\,J1534$-$571, HESS\,J1614$-$518, and HESS\,J1912$+$101 have been classified as SNR candidates. The identification of HESS\,J1534$-$571 with the radio SNR candidate G323.7$-$1.0 has led to the classification of HESS\,J1534$-$571 as SNR. The nondetection of radio synchrotron emission from the other two sources is not in conflict with the SNR hypothesis for these objects. The GeV counterpart situation (i.e.,\ the {\it Fermi-}LAT counterparts for HESS\,J1534$-$571 and HESS\,J1614$-$518 and no known counterpart for HESS\,J1912$+$101) is compatible with the respective TeV fluxes but does not contribute to the classification of the sources. Table\ \ref{Table:Associations} gives an overview of the confirmed (HESS\,J1534$-$571) and possible (HESS\,J1614$-$518, HESS\,J1912$+$101) association scenarios and corresponding distance scale estimates, as discussed in Sections \ref{Sect:RadioGeV} and \ref{Sect:XrayIRSubmm}.

For the new SNR candidates, Table\ \ref{Table:NewSNRShells_2} lists derived parameters (diameter, luminosity) for three different assumed distances, namely a generic \SI{1}{kpc} distance and the distances listed in Table\ \ref{Table:Associations}. The values can be compared to the corresponding parameters of known TeV SNR shells as shown in Table\ \ref{Table:KnownSNRShells}. On average, the photon indices of the new shells seem slightly softer than those of the known SNR shells. However, for individual sources the respective errors are too large to draw any conclusion. Luminosities and diameters of the new SNR candidates are compatible with the known SNR shells for assumed nearby distances of \SI{\sim 1}{kpc}. The new sources would be larger and more luminous than the known SNRs already for moderate distances of \SI{\sim 3}{kpc}. Distances at a \SIrange{8}{10}{\kilo\parsec} scale are either disfavored or would indicate a different, substantially more luminous new TeV SNR source population. At these distances, the physical diameters of the SNRs would also be substantially larger than those of the known TeV SNRs, and well beyond a cutoff in the general SNR diameter distribution at \SI{\sim 60}{pc} as derived from SNRs in the Magellanic clouds and M\,33 \citep[see][and references therein]{Badenes_2010}. If this cutoff is indeed attributed to a lower limit in the ambient density distribution, as argued by \citet[][]{Badenes_2010}, then outliers beyond \SI{\sim 60}{pc} (e.g., the new TeV SNRs if indeed at \SIrange{8}{10}{\kilo\parsec} distance) might be connected to remnants of core collapse SNe with modified ambient medium \citep[cf. the discussion in][]{Badenes_2010}. This would be consistent with arguments that have been made to explain the sizes\ of, for example, the TeV SNRs RX\,J1713.6$-$3946 and HESS\,J1731$-$347. To investigate whether this is a realistic scenario and whether specific wind progenitor bubbles could lead to SNR sizes at the \SI{\sim 100}{pc} diameter scale is beyond the scope of this paper.

\subsection{TeV emission from leptons or protons}
\label{Subsect:discussionlepton}

One of the fundamental questions of SNR research is what fraction of the ejecta energy goes into thermal (typically X-ray-emitting) plasma, and what fraction goes into a nonthermal component consisting of relativistic particles (best visible in hard X-rays and high to very high $\gamma$-ray energies). The energy share between relativistic hadrons and leptons and the maximum particle energy determine whether SNRs contribute significantly to the generation of Galactic CRs up to the knee in the CR particle spectrum. Unlike other identified TeV SNRs, the new TeV sources do not have confirmed X-ray counterparts. The lack of thermal X-rays may be interpreted as a signature for low density environment and therefore leptonically dominated TeV emission processes. However, the absorption column to the new sources is not constrained from the TeV data. Depending on the actual distance to the source, and on the temperature and density of the emitting plasma, foreground absorption might have prevented detection of soft thermal X-rays, for example,\ in the {\it ROSAT} survey. There is also sufficient uncertainty about which level of thermal heating could be expected, for example,\ in a clumpy environment \citep[e.g.,][]{2014MNRAS.445L..70G}. 

Nonthermal X-rays above $\sim 2\,\mathrm{keV}$ are however expected in shock-compressed (and possibly further amplified by CR streaming) magnetic fields, if the TeV emission stems from relativistic electrons. As discussed in Sect.\,\ref{SubSect:xrays1614}, within the presented study it has not been possible to constrain conclusively the hard X-ray emission likely associated with HESS\,J1614$-$518. HESS\,J1912$+$101 lacks sensitive pointed X-ray coverage. The nondetection of X-ray emission from HESS\,J1534$-$571 from a (partial) {\it Suzaku} coverage seems interesting at first
glance, but may not yet be constraining (see next section).

\renewcommand{\arraystretch}{1.5}

\begin{table*}[t]
\caption{Parameters of known \SI{}{\tera\eV} SNR shells. For each of the sources, diameter and $L_{\gamma, \SIrange[range-phrase = -]{1}{10}{\tera\eV}}$ are calculated based on the parameters quoted in the respective papers. Where power laws with cutoffs better fit the spectra and are used to compute luminosities, the corresponding fit values are also reported. The distance to HESS\,J1731$-$347 is debated in the literature \citep[see, e.g.,][]{2014ApJ...788...94F,2015A&A...573A..53K}; the values reported in the table correspond to the two most probable distance solutions. Flux errors are dominated by their systematic errors of typically 20\%, but luminosity errors are dominated by the distance uncertainties and are therefore on the order of 30\% or more. Spectral indices have statistical errors of $\Delta \Gamma \approx 0.2$ or better.
}
        \label{Table:KnownSNRShells}
        \centering
        \begin{tabular}{llccccccc}
                \hline\hline
                HESS source name & Identification & Dist. & Diameter & Age & $L_{\gamma, \SIrange[range-phrase = -]{1}{10}{\tera\eV}}$& $\Gamma_{\gamma,\mathrm{PL fit}}$ & $\Gamma_{\gamma}$/$E_{\mathrm{cutoff}}$  & Ref.\\
                &   & [\SI{}{\kilo\parsec}] & [\SI{}{\parsec}] & [\SI{}{\kilo yr}]  & [$10^{33}$\SI{}{\erg\per\second}]   &  -         & -/[\SI{}{\TeV}]         \\
                \hline
                HESS\,J0852$-$463 & RX\,J0852.0$-$4622          & $0.75$      & $26.2$      & $1.7-4.3$    & $5.7$        & $2.3$ & $1.8$/$6.7$  & (1)\\
                HESS\,J1713$-$397 & RX\,J1713.7$-$3946          & $1$         & $20.2$      & $\approx1$   & $7.2$        & $2.3$ & $2.1$/$12.9$ & (2)\\
                HESS\,J1731$-$347 & G353.6$-$0.7                & $3.2$/$5.2$ & $30.2$/$49$ & $\approx2.5$ & $8.5$/$22.4$ & $2.3$ &              & (3)\\
                HESS\,J1442$-$624 & G315.4$-$2.3 (RCW\,86)      & $2.5$       & $\approx30$ & $\approx1.8$ & $6.3$        & $2.3$ & $1.6$/$3.5$  & (4)\\
                HESS\,J1502$-$418 & SN\,1006\,(NE) & \multirow{2}{*}{$2.2$} & \multirow{2}{*}{22.3}  & \multirow{2}{*}{$\approx1$} & $0.46$ &  $2.4$ & & \multirow{2}{*}{(5)}\\
                HESS\,J1502$-$421 & SN\,1006\,(SW) & & & & 0.31 &  2.3  & &\\ 
                \hline\hline
        \end{tabular}
        \tablebib{
                (1) \citet{2007ApJ...661..236A,special_velajr}; (2) \citet{2006A&A...449..223A,special_1713}; (3) \citet{2011A&A...531A..81H,2015A&A...573A..53K}; (4) \citet{2016arXiv160104461H}; (5) \citet{2010A&A...516A..62A}.
        }
\end{table*}

\begin{sidewaystable*}
        \caption{Parameters of the new \SI{}{\tera\eV} SNR candidates, assuming a generic \SI{1}{\kilo\parsec} distance and two other distances coming from possible association scenarios. The gas parameters are derived from HI and CO(1-0) emission, from circular regions toward HESS\,J1534$-$571, HESS\,J1614$-$518, and HESS\,J1912$+$101 in the respective velocity ranges; the centers and extensions of these regions are the best fit centers and $R_{\mathrm{out}}$ presented in Table \ref{Table:morphologyresults} as derived from the morphological study of the TeV sources. The resulting total encircled gas mass is given in $M_{\mathrm{HI+H2}}$. Density ranges are calculated by assuming that the line-of-sight thickness is between a value equal to the SNR diameter (upper limit) and the approximate thickness of a Galactic arm (0.5\,kpc, lower limit; see appendix \ref{SubSect:ShellsGasMasses} for a more detailed description). The values $n_{\mathrm{p,HI}}$, $n_{\mathrm{p,H2}}$ indicate the estimated gas density coming from atomic (HI) and molecular (H2) gas contributions, respectively. The value $W_{\mathrm{p}}^{*}$ gives the energy content of relativistic protons in the range \SIrange[range-phrase=\ --\ ]{10}{100}{\tera\eV}, which emit $\gamma$-rays in the range \SIrange[range-phrase=\ --\ ]{1}{10}{\tera\eV}; the proton spectral index is the same as the photon index $\Gamma$ in the \SI{}{\tera\eV} range (see Table \ref{Table:spectralresults}). The value $W_{\mathrm{p}}^{**}$ assumes a proton spectral index equal to the \SI{}{\tera\eV} photon index from \SIrange[range-phrase=\ --\ ]{10}{100}{\tera\eV} and an index of 2 from \SI{10}{\tera\eV} down to \SI{1}{\giga\eV} in proton energy.}
        \label{Table:NewSNRShells_2}
        \centering
        \resizebox{\columnwidth}{!}{
                \begin{tabular}{lcccccccccccc}
                        \hline\hline
                        HESS source name & distance & velocity & diameter & $L_{\gamma, \SIrange[range-phrase = -]{1}{10}{\tera\eV}}$  &  $M_{\mathrm{HI+H2}}$ & $n_{\mathrm{p,HI}}$ & $n_{\mathrm{p,H2}}$ & $n_{\mathrm{p,HI+H2}}$ & $W_{\mathrm{p},\SI{10}{\tera\eV} - \SI{100}{\tera\eV}}^{*}$  & $W_{\mathrm{p},\SI{1}{\giga\eV} - \SI{100}{\tera\eV}}^{**}$  \\
                        & [\SI{}{\kilo\parsec}] & [\SI{}{\kilo\metre\per\second}] & [\SI{}{\parsec}] &  [\SI{1e33}{\erg\per\second}]  & [$\mathrm{M}_{\sun}$] & [\SI{}{\per\cubic\centi\metre}] & [\SI{}{\per\cubic\centi\metre}] & [\SI{}{\per\cubic\centi\metre}] & [\SI{1e51}{\erg}] & [\SI{1e51}{\erg}] \\
                        \hline
                        HESS\,J1534$-$571 & 1.0 &  & 14.0 & 0.78 &  &  & & 1 & 0.004 & 0.03   \\  
                        & 3.5 & \SIrange{-55}{-45} &48.9 & 9.6  & \SI{3.0e4}{} & 3..29 & 1..10 & 4..39 & 0.01..0.001 & 0.08..0.009  \\ 
                        & 8 & \SIrange{-88}{-78} & 111.7 & 50 & \SI{6.4e4}{} & 1..6 & 0.3..1 & 1.3..7 & 0.17..0.03 & 1.34..0.25  \\
                        \hline
                        HESS\,J1614$-$518 & 1.0 & & 14.7 & 2.4 & & & & 1 & 0.01 & 0.08   \\
                        & 1.5 & \SIrange{-30}{-12} & 22 &5.3 & \SI{4.9e3} & 3..71 & 0 & 3..71 & 0.008..0.0003 & 0.06..0.002   \\
                        & 5.5 & \SIrange{-110}{-80} & 80.6 & 71 & \SI{3.7e5} & 5..33 & 12..77 & 17..110 & 0.02..0.003 & 0.14..0.02   \\
                        \hline
                        HESS\,J1912$+$101  & 1.0 &  & 17.1 & 0.97 &  & &  & 1 & 0.004 & 0.04  \\
                        & 4.5 & \SIrange{45}{73} & 77.0 & 19.6 & \SI{4.7e5} & 7..44 & 21..134 & 28..178 & 0.003..0.0005 & 0.03..0.004   \\
                        & 10 & \SIrange{+5}{+15} & 171.0 & 97  &\SI{2.2e5} & 3..6 & 1..2 & 4..8 & 0.11..0.05 & 0.88..0.44  \\
                        \hline\hline
                \end{tabular}
        }
\end{sidewaystable*}

\renewcommand{\arraystretch}{1.0}

\subsection{HESS\,J1534$-$571}
\label{Subsect:discussionj1534}

A confirmed upper limit on the X-ray emission from HESS\,J1534$-$571 in the \SIrange[range-phrase=\ --\ ]{2}{10}{keV} range would present the first case of a \SI{}{\tera\eV} SNR without an X-ray counterpart at current satellite sensitivity. The absence of nonthermal X-ray emission is generally used to challenge the interpretation that a relativistic electron population in the source is responsible for the $\gamma$-ray emission because a certain minimum level of X-ray synchrotron emission is expected in a minimum Galactic magnetic field of \SI{3}{\micro\gauss}. More quantitatively, a ratio $R = F_{\gamma}(1-10\,\mathrm{TeV})/F_{\mathrm{X}}(2-10\,\mathrm{keV})$
can be defined following, for example,\ \citet{2006MNRAS.371.1975Y}. Young known SNRs have a ratio $R$ significantly less than 2, while $R$ larger than 2 could be indicative of a largely evolved (so-called X-ray dark) SNR, whose \SI{}{\tera\eV} emission should then be dominated by proton-induced $\pi^{0}$-decay \citep{2006MNRAS.371.1975Y}. Assuming the current {\it Suzaku} limit holds for the entire HESS\,J1534$-$571 shell, $R > 0.25$. A detection of nonthermal X-ray emission is therefore within reach if HESS\,J1534$-$571 is similar to other \SI{}{\tera\eV} SNRs. To establish a sensitive upper limit (indicating that the \SI{}{\tera\eV} emission stems from protons) is however challenging even for current X-ray instruments because of the large extension of the source.

The TeV data do not permit a distance estimate to HESS\,J1534$-$571 (as for any of the new sources). Using the SNR radio surface brightness to diameter ($\Sigma - D$) relation, the distance to G323.7$-$1.0 is estimated to be $20\,\mathrm{kpc}$ (cf.\ Sect.\,\ref{SubSubSect:radio1534}). However, individual distances derived from $\Sigma - D$ have large errors, which are typically 40\% for the normal SNR population. Assuming that the radio counterpart of HESS\,J1534$-$571 is similarly underluminous as that of RX\,J1713.7$-$3946 with respect to its $\Sigma - D$ expectation value at $1\,\mathrm{kpc}$ would reduce the distance estimate of G323.7$-$1.0 to $\sim$$5\,\mathrm{kpc}$. This distance would imply a TeV luminosity of HESS\,J1534$-$571 in reasonable agreement with TeV luminosities from the other known TeV SNRs (cf.\ Table\ \ref{Table:KnownSNRShells}).

\subsection{HESS\,J1614$-$518}
\label{Subsect:discussionj11614}

Also for HESS\,J1614$-$518, distance estimates only come from possible associations. One such is the open stellar cluster Pismis\,22, located close to the center of HESS\,J1614$-$518 (cf.\ Sect.\,\ref{SubSect:Infrared}), which may have hosted the SNR progenitor star. The distance estimate to this stellar cluster, \SI[separate-uncertainty]{1.0+-0.4}{\kilo pc} \citep{2000A&A...360..529P}, is in accordance with a possible HI void seen in projection at a distance of \SIrange{1.2}{1.5}{\kilo\parsec} (cf.\ Sect.\,\ref{SubSect:Gas}).

Another association was suggested by \citet{2011PASJ...63S.879S}. The X-ray source XMMU\,J161406.0$-$515225, a point source close to the center of HESS\,J1614$-$518, and the diffuse source {\it Suzaku} Src\,A defined in \citet{2008PASJ...60S.163M} may be associated with the TeV source (cf.\ Sect.\,\ref{SubSect:xrays1614}). Using the X-ray absorption column as proxy, a rough distance scale of \SI{10}{\kilo pc} was estimated, which is likely compatible with a possible spiral arm association at \SI[]{\sim 5.5}{\kilo pc} distance (cf.\ Appendix \ref{SubSect:ShellsGasAssociations}). 

\subsection{HESS\,J1912$+$101}
\label{Subsect:discussionj1912}

As discussed in Sect.\,\ref{Subsect:pulsar1912}, the pulsar PSR\,J1913$+$1011 located at the center of HESS\,J1912$+$101 may be the remainder of the SN explosion that has created the putative \SI{}{\tera\eV} SNR. At a distance of \SI{4.5}{\kilo\parsec} estimated from the dispersion measure of the pulsar, the \SI{}{\tera\eV} shell size would then correspond to a SNR radius of $\sim$\SI{40}{\parsec}. As for all sources presented in the paper, the \SI{}{\tera\eV} shell morphology implies that particle acceleration to supra-TeV energies is likely still ongoing or has been ongoing until the recent past; otherwise, diffusion of the no longer confined particles would have washed out the current morphology. To provide the necessary high shock speed, the age of the putative SNR HESS\,J1912$+$101 should then be at least an order of magnitude lower than the characteristic age of the pulsar ($\tau_{\mathrm{c}} \simeq 1.7 \times 10^5$ years), if the association with PSR\,J1913$+$1011 was confirmed. To maintain the association, the birth spin period of the pulsar therefore needs to have been close to the current spin period, which is in principle possible.

\subsection{Proton scenarios}
\label{Subsect:discussionproton}

Assuming that the $\gamma$-ray emission seen in the TeV regime is purely due to hadronic processes, an estimate of the fraction of the SNR explosion energy going into accelerated CR protons can be given.

In the delta-function approximation \citep{2006PhRvD..74c4018K}, $\gamma$-ray photons of energy $E_{\gamma}$ are produced by protons with energy $E_{\mathrm{p}} = 10 \times E_{\gamma}$. Following the arguments in \citet{2006A&A...449..223A}, the total energy in accelerated protons in the \SIrange[range-phrase=\ --\ ]{10}{100}{\tera\eV} range can be estimated from the $\gamma$-ray luminosity in the range \SIrange[range-phrase=\ --\ ]{1}{10}{\tera\eV}, using

\begin{equation}
W_{\mathrm{p}}^{\mathrm{tot}}\left(\mathrm{10-100\,TeV}\right) \approx \tau_{\mathrm{pp}\rightarrow \pi^{0}} L_{\gamma}\left(\mathrm{1-10\,TeV}\right);
\end{equation}
\begin{equation}
\tau_{\mathrm{pp}\rightarrow \pi^{0}} \approx 4.5 \times 10^{15} \left(\frac{n}{\mathrm{cm}^{-3}}\right)^{-1}\mathrm{s}
\end{equation}

\noindent
is the characteristic cooling time of protons by $\pi^{0}$ production and

\begin{equation}
L_{\gamma}(\mathrm{1-10\,TeV}) = 4\pi d^{2}\int_{\mathrm{1\,TeV}}^{\mathrm{10\,TeV}}E_{\gamma}\frac{\mathrm{d}N_{\gamma}}{\mathrm{d}E}\mathrm{d}E,
\end{equation}

\noindent
where $d$ is the distance to the source and $\mathrm{d}N_{\gamma}/\mathrm{d}E = N_{0,\SI{1}{\tera\eV}}\left(E/\SI{1}{\tera\eV}\right)^{-\Gamma}$, as described in Sect.\,\ref{SubSect:HessAnalysis}. The values $N_{0,\SI{1}{\tera\eV}}$ and $\Gamma$ are derived from the fit to the TeV data as presented in Table\ \ref{Table:spectralresults}.

Knowing $W_{\mathrm{p}}^{\mathrm{tot}}(\mathrm{10-100\,TeV})$ and assuming a power-law spectrum for the accelerated protons ($\mathrm{d}N_{\mathrm{p}}/\mathrm{d}E = N_{\mathrm{p}}E^{-\alpha}$, with $\alpha = \Gamma$ and $N_{\mathrm{p}}$ a normalization factor), $W_{\mathrm{p}}^{\mathrm{tot}}$ can be calculated in an arbitrary proton energy range. It is assumed that the proton energy spectrum can be described by a broken power law with a break energy at $10\,\mathrm{TeV}$,

\begin{equation}
\frac{\mathrm{d}N_{\mathrm{p}}}{\mathrm{d}E} = \begin{cases} N_{\mathrm{p},1}E^{-2} & 1\ \mathrm{GeV} \leq E_{\mathrm{p}} \leq\ 10\  \mathrm{TeV} \\ N_{\mathrm{p},2}E^{-\alpha} & 10\ \mathrm{TeV} < E_{\mathrm{p}} \leq\ 100\  \mathrm{TeV} \end{cases}
,\end{equation}

\noindent
which is roughly compatible with the TeV spectra of all three new shell sources and the GeV spectrum of the {\it Fermi}-LAT source associated with HESS\,J1614$-$518.

To illustrate the possible energy contents in accelerated protons, three scenarios are given in Table\ \ref{Table:NewSNRShells_2} for each source. The first is a generic case with a distance of \SI{1}{\kilo\parsec} and a target gas density of $1\,\mathrm{cm^{-3}}$. The other two are derived from the possible gas association scenarios as introduced in Sect.\,\ref{SubSect:Gas} and Appendix \ref{SubSect:ShellsGasAssociations} (cf.\ Table\ \ref{Table:Associations}), where error ranges are propagated from the estimated ranges of gas densities. The table lists both the energy contents of the protons in the TeV-emitting energy range and for an extrapolated spectrum down to \SI{1}{\GeV}. The latter values show that the available data are compatible with the expected energy content of 10\% of $10^{51}\mathrm{erg}$ for assumed nearby distances of \SI{\sim 1}{kpc} and for moderate distances of \SI{\sim 3}{kpc}. Distances at a \SIrange{8}{10}{\kilo\parsec} scale and beyond are disfavored in hadronic emission scenarios.


\section{Conclusions}
\label{Sect:Conclusion}

A dedicated search for new SNR shells in the H.E.S.S.\ Galactic plane survey data has revealed three new SNR candidates. HESS\,J1534$-$571 was confirmed as a SNR from an identification with a radio SNR candidate, while HESS\,J1614$-$518 and HESS\,J1912$+$101 remain SNR candidates for the time being. From the current knowledge of multiwavelength data, both leptonic or hadronic (or a blend of both) TeV emission scenarios are possible. Distances to the objects on the \SIrange{8}{10}{\kilo\parsec} scale or beyond seem unlikely from a comparison of the TeV fluxes to the luminosities of known TeV SNR shells. Such large distances are also difficult to accommodate in hadronic TeV emission scenarios. Distances to the new objects at the \SIrange{1}{3}{\kilo\parsec} scale seem most likely, while distances below the \SI{1}{\kilo\parsec} scale might indicate unusual properties regarding their X-ray emissivity.

The analysis has demonstrated that current Imaging Atmospheric Cherenkov Telescope Arrays have the power to discover new SNRs. The future Cherenkov Telescope Array will substantially increase the available sensitivity to detect TeV SNRs in the entire Galaxy. Source confusion may however become a more severe problem than in the study presented here.


\begin{acknowledgements}
The support of the Namibian authorities and of the University of Namibia in facilitating the construction and operation of H.E.S.S. is gratefully acknowledged, as is the support by the German Ministry for Education and Research (BMBF), the Max Planck Society, the German Research Foundation (DFG), the Alexander von Humboldt Foundation, the Deutsche Forschungsgemeinschaft, the French Ministry for Research, the CNRS-IN2P3 and the Astroparticle Interdisciplinary Programme of the CNRS, the U.K. Science and Technology Facilities Council (STFC), the IPNP of the Charles University, the Czech Science Foundation, the Polish National Science Centre, the South African Department of Science and Technology and National Research Foundation, the University of Namibia, the National Commission on Research, Science \& Technology of Namibia (NCRST), the Innsbruck University, the Austrian Science Fund (FWF), and the Austrian Federal Ministry for Science, Research and Economy, the University of Adelaide and the Australian Research Council, the Japan Society for the Promotion of Science, and by the University of Amsterdam.\newline
We appreciate the excellent work of the technical support staff in Berlin, Durham, Hamburg, Heidelberg, Palaiseau, Paris, Saclay, and in Namibia in the construction and operation of the equipment. This work benefited from services provided by the H.E.S.S. Virtual Organisation, supported by the national resource providers of the EGI Federation.\newline
The shell search performed on the entire HGPS data set as presented in Section \ref{SubSect:TeVShellSearch} employed Gammapy routines as presented in \citet{2015arXiv150907408D}.\newline
This research has made use of the VizieR catalog access tool, CDS, Strasbourg, France. The original description of the VizieR service was published in \citet{2000A&AS..143...23O}.\newline
This research has made use of the SIMBAD database,
operated at CDS, Strasbourg, France \citep{2000A&AS..143....9W}.
\end{acknowledgements}


\bibliographystyle{aa} 
\bibliography{puehlhofer_hess_newshells_references} 

\newpage
\clearpage


\appendix

\section{TeV surface brightness profiles}
\label{Sect:AppendixTeV}

\begin{figure*}
  \centering
        \includegraphics[width=0.44\textwidth]{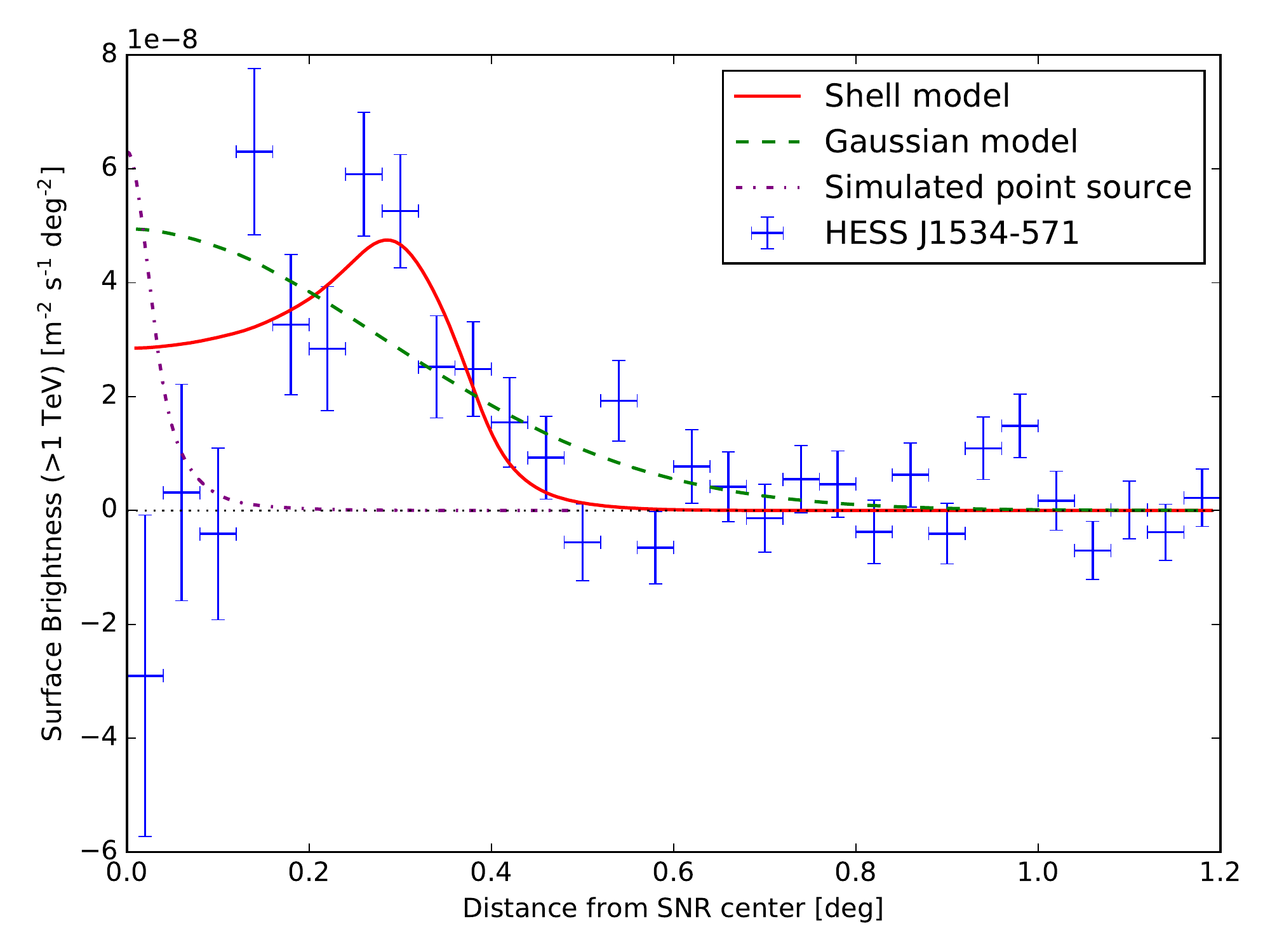} \includegraphics[width=0.47\textwidth]{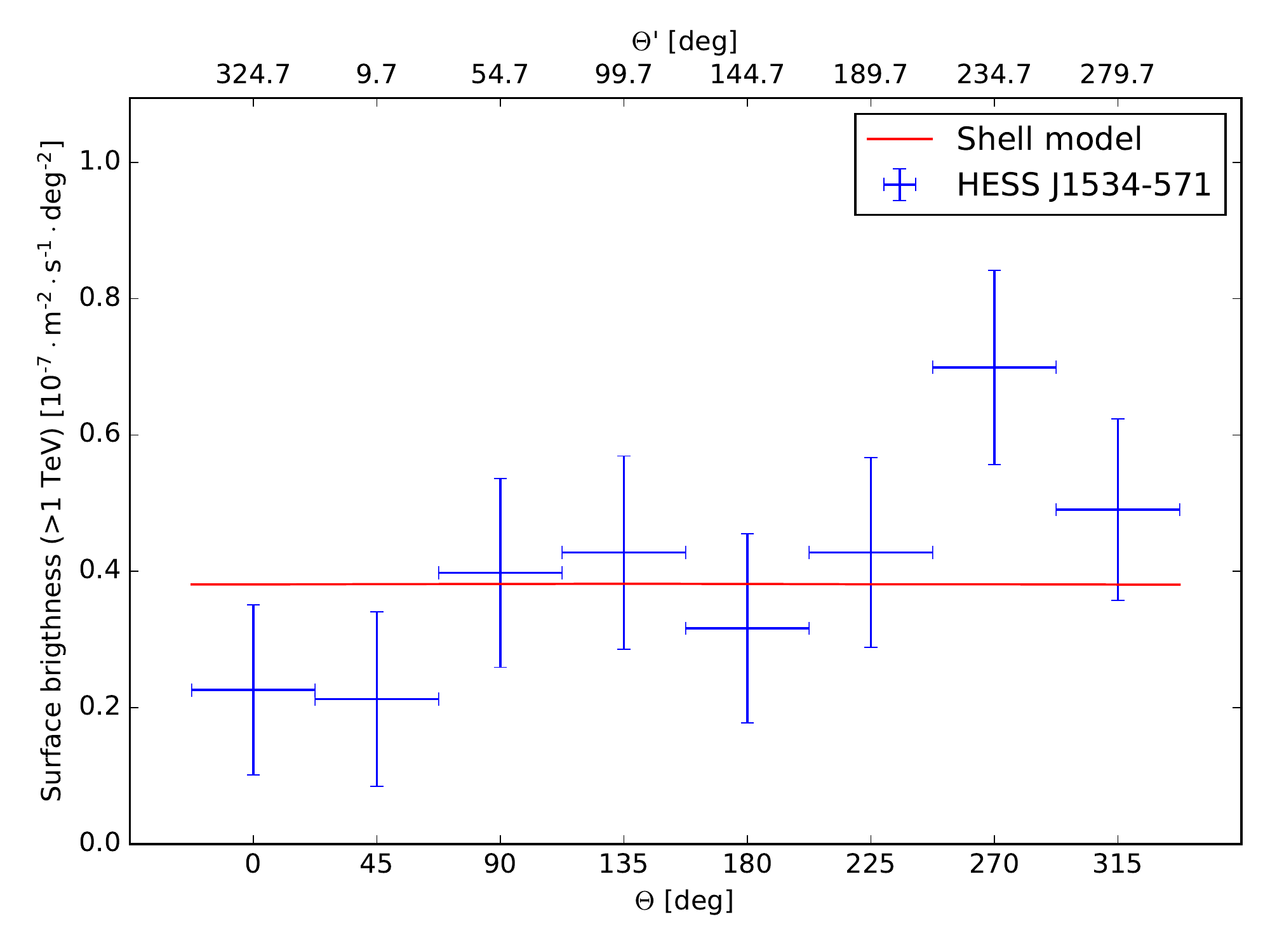}\\
        \includegraphics[width=0.44\textwidth]{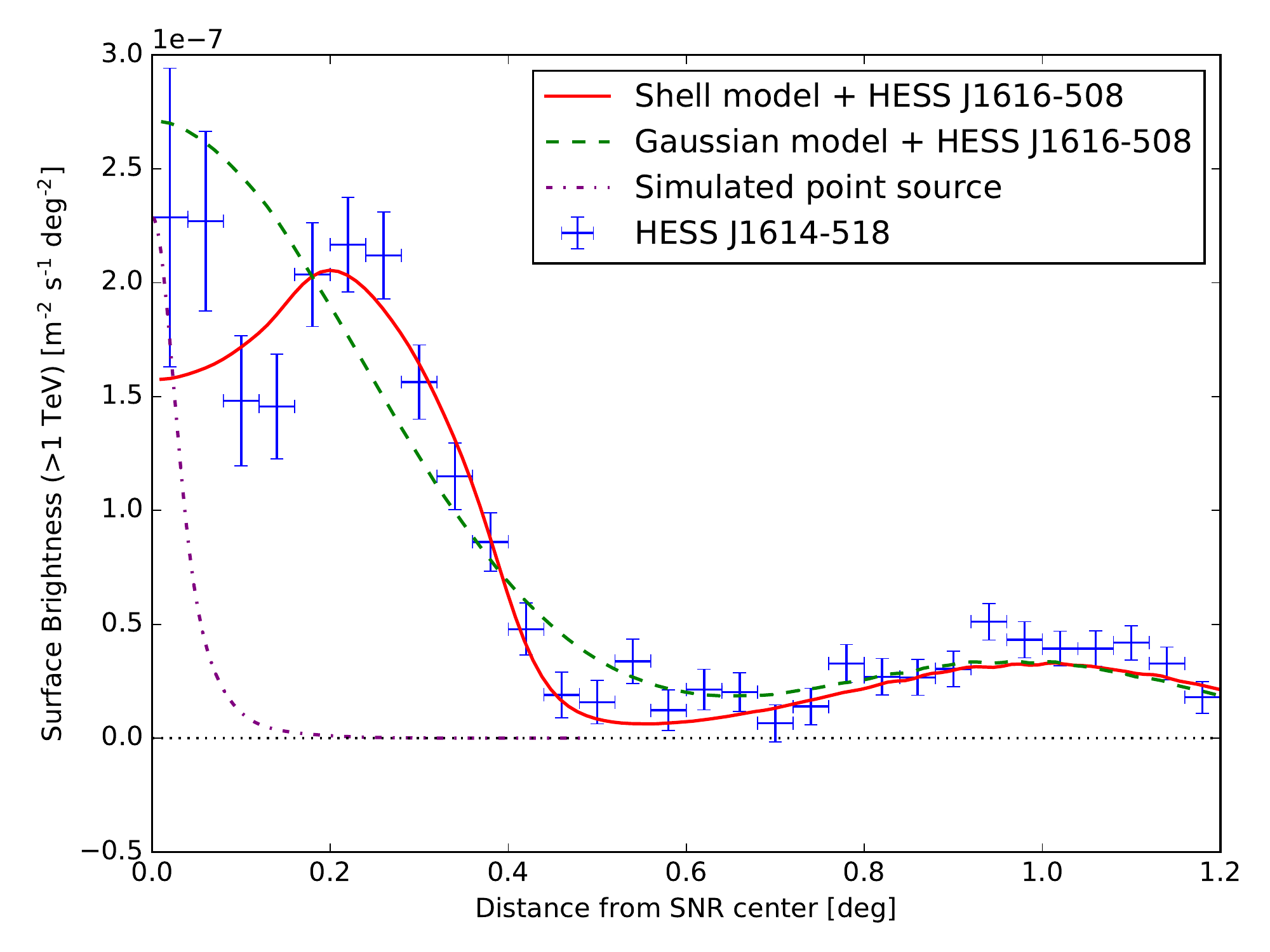} \includegraphics[width=0.47\textwidth]{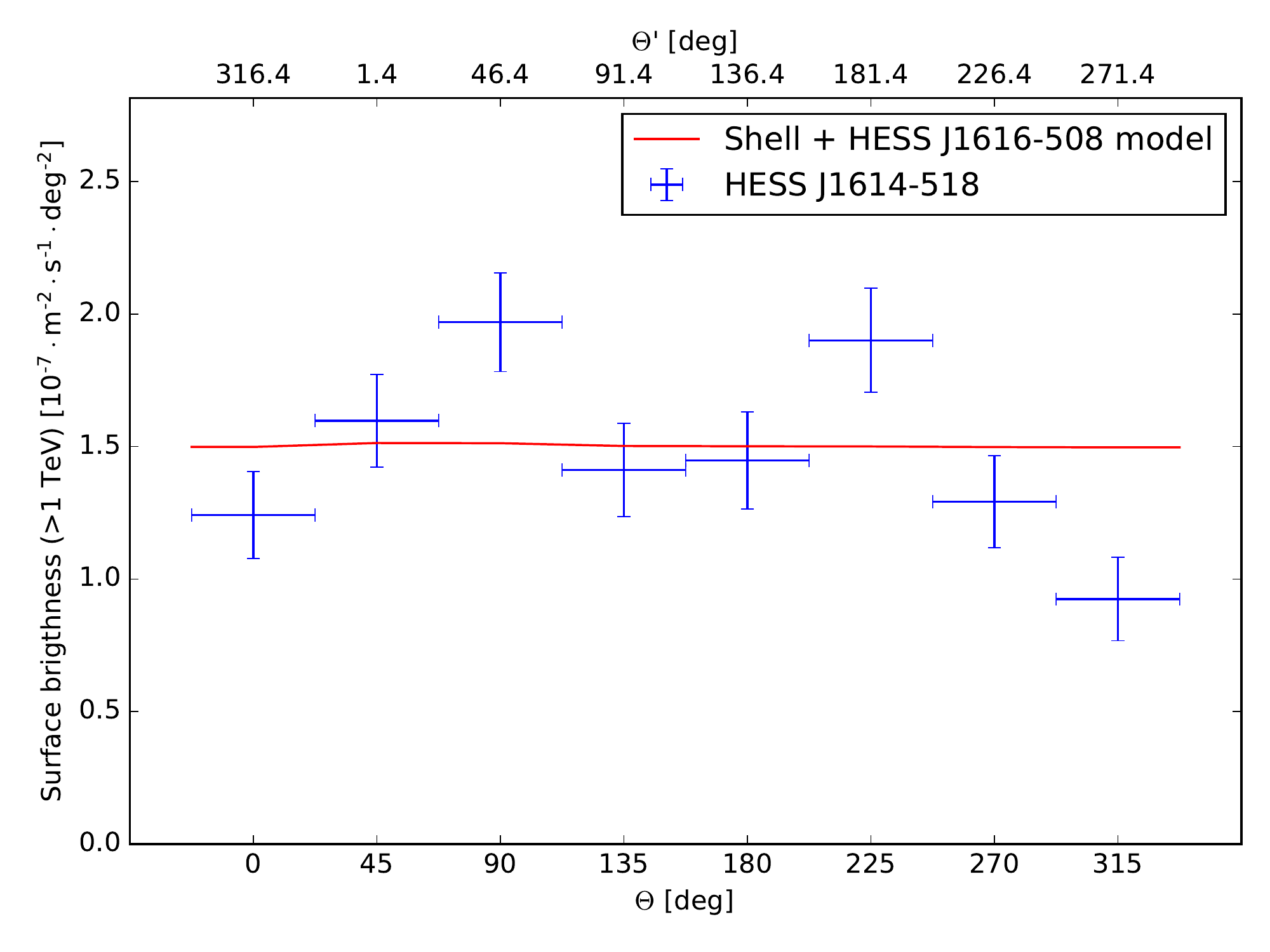}\\
        \includegraphics[width=0.44\textwidth]{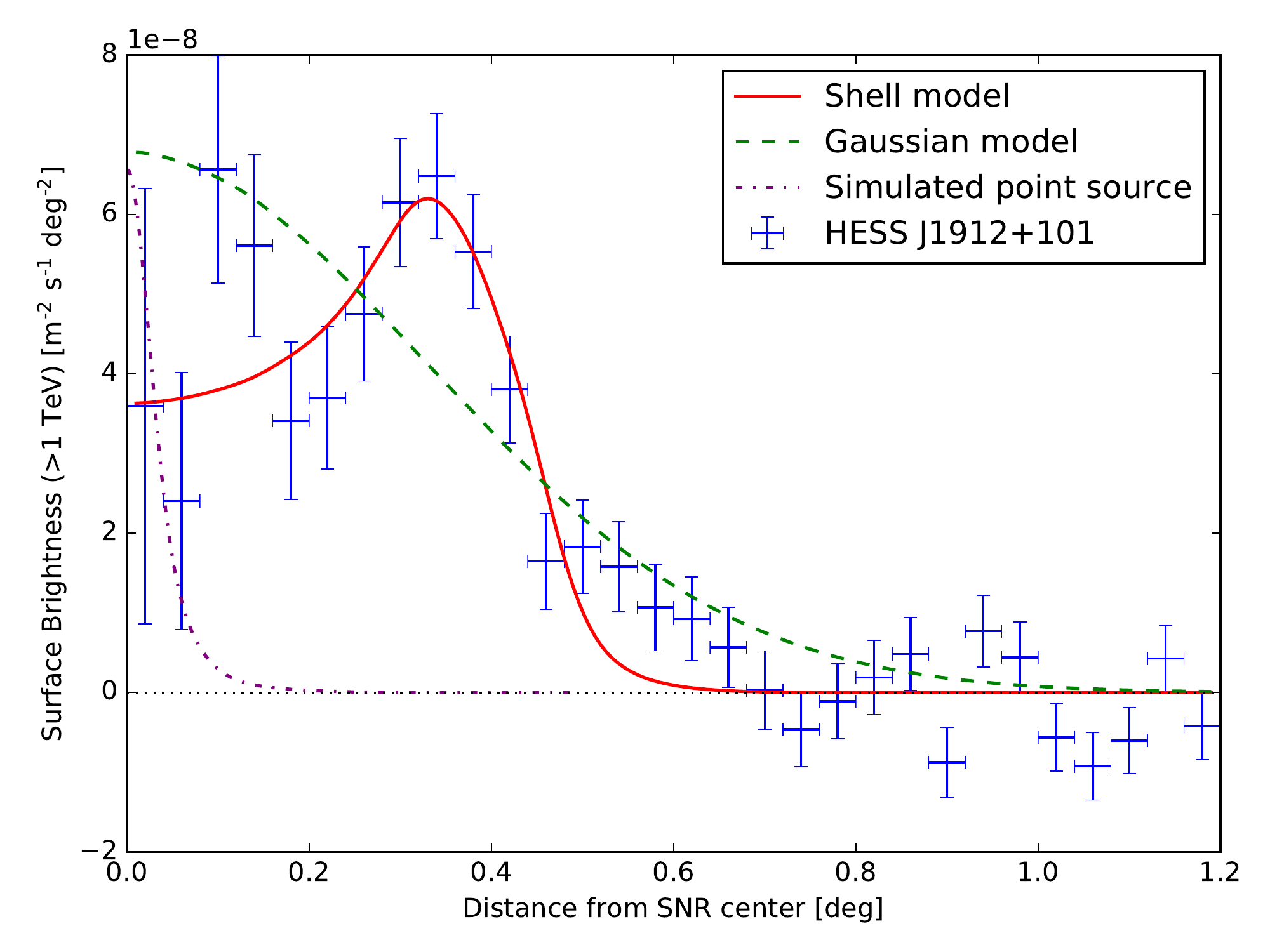} \includegraphics[width=0.47\textwidth]{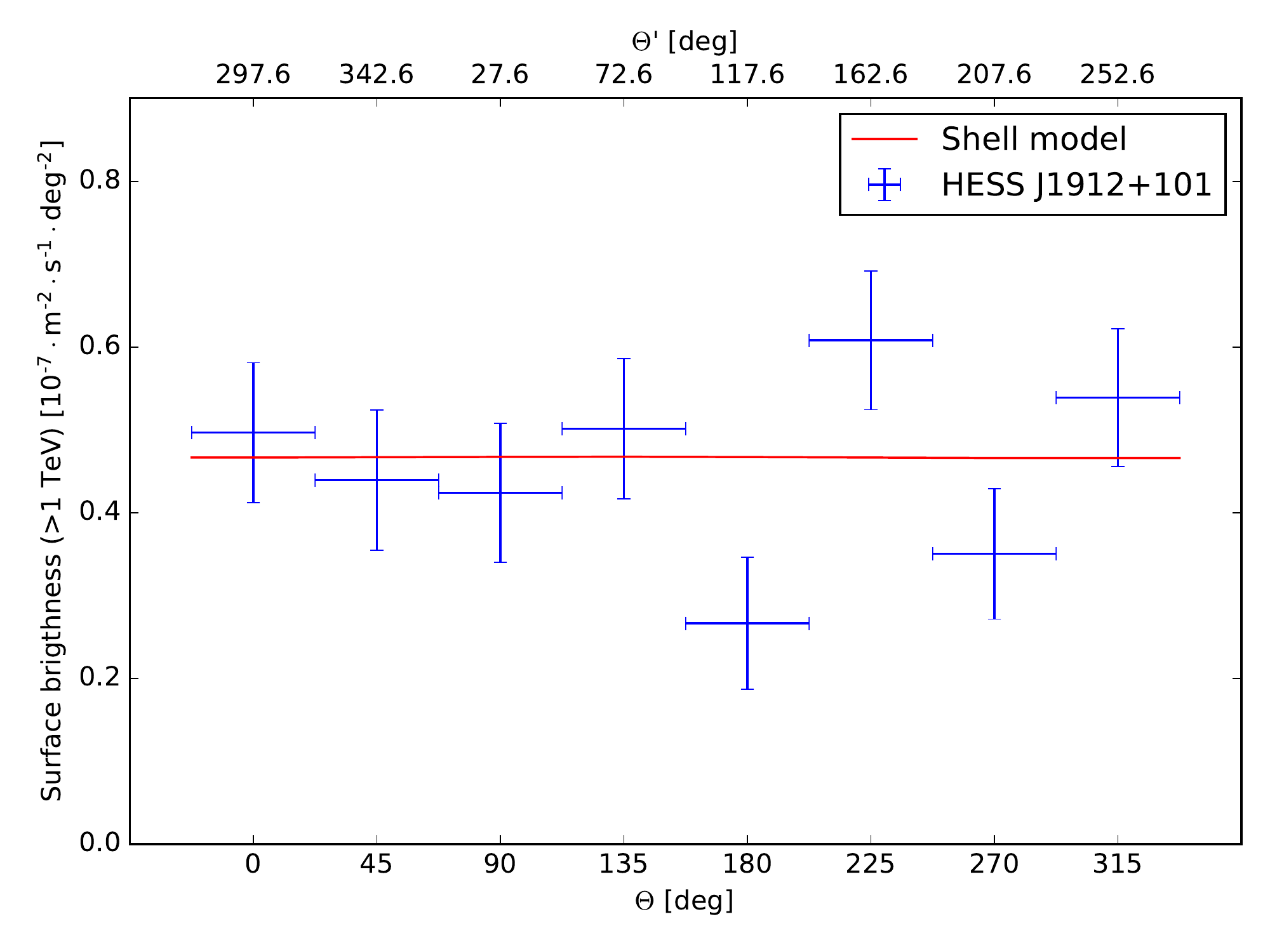}\\
\caption{TeV surface brightness profiles of HESS\,J1534$-$571, HESS\,J1614$-$518, and HESS\,J1912$+$101. The left column shows radial profiles and the right column shows azimuthal profiles, respectively. The shell models and Gaussian models (shown in the left panels) are from fits to the data. The point source simulation shown in the left panels is derived from the sum of three Gaussians fitted to a simulated point source as it would appear in the actual data set (cf.\ Sect.\,\ref{SubSubSect:TeVShellSearch_Motivation}), arbitrarily normalized for good visual representation. The value of $\Theta$ in the azimuthal profiles represents the angle with respect to the Galactic latitude. The first wedge is at $\Theta = \SI{0}{\degree}$, following wedges are added counter-clockwise. The angle with respect to north in equatorial coordinates is represented by $\Theta^\prime$.
\label{Fig:TeVprofiles}}
\end{figure*}

After fitting the two-dimensional models to the TeV count sky maps as described in Sect.\,\ref{SubSect:TeVShellSearch}, profiles were extracted in units of surface brightness from the observations and the models to verify the fit results. The azimuthal profiles were also used to quantify possible deviations of the source maps from azimuthal symmetry, as discussed in Sect.\,\ref{SubSubSect:TeVShellSearch_FittingProcedure}.
The measured profiles were derived from the uncorrelated excess maps of the dedicated source analysis (cf.\ Sect.\,\ref{SubSubSect:HessDedicatedAnalysis}). Surface brightness profiles are derived by dividing the excess within the respective profile bin area by the expected counts derived from the simulated instrument response \citep{special_HGPS} and by the area of the profile bin. The profiles are shown in Fig.\,\ref{Fig:TeVprofiles}. Shell and Gaussian model profiles were scaled to corresponding surface brightness units.

For the radial profiles of the observed data, circular annuli around the centroids of the shell models, and with a width of \SI{0.04}{\degree} were used. The profiles of the shell and Gaussian models were derived analogously but with a two times finer binning. As expected from the fitting results, the shell model describes the data best in all three cases. A positive excess is visible in the profile of HESS\,J1614$-$518 from \SI{\sim 0.7}{\degree} outward. This emission stems from the nearby source HESS\,J1616$-$508 and is taken into account in the fit procedure with an additional Gaussian component in the model of HESS\,J1614$-$518, both when fitting the Gaussian model and the shell model. The separation of both sources is large enough so that the emission of HESS\,J1616$-$508 does not significantly affect the spectral results of HESS\,J1614$-$518.

To derive the azimuthal profiles, the sources were divided into eight wedges of equal angular size with outer radii of $R_\mathrm{out}$, respectively, and with inner radii slightly smaller than $R_\mathrm{in}$  to increase the photon statistics of the profiles (because of the projection of the shells, the peak of the radial profiles is close to $R_\mathrm{in}$). In the azimuthal model profile of HESS\,J1614$-$518, the modeled emission of HESS\,J1616$-$508 is barely visible above the flat shell profile.

\section{Details on gas association}
\label{Sect:AppendixCo}

\subsection{SNRs associated with molecular clouds}
\label{SubSect:SNRMCAssociations}

Besides the well-identified TeV SNR shells, several resolved \SI{}{\tera\eV} sources are likely driven by SNR particle acceleration processes as well, but the \SI{}{\tera\eV} morphology is determined by molecular clouds close to or partially coincident with the SNRs. The underlying assumption is that \SI{}{\tera\eV} emission is created in these clouds through collisions of SNR-accelerated hadronic particles with dense gas. Examples comprise IC\,443, W51C, CTB\,37A, or Tycho's SNR \citep{2007ApJ...664L..87A,Fiasson:2009,2008A&A...490..685A,2011ApJ...730L..20A}. The firm confirmation of such a scenario is often challenging. The \SI{}{\tera\eV} images, which are statistically limited, need to be correlated with sub-mm line data (tracing molecular gas), which provide moderate distance resolution and thus a large number of possible projections over different distance ranges, resulting in a large number of trials. In some cases, possible alternative PWN scenarios also exist for the interpretation of the TeV emission. Because of the expected ambiguities as experienced from the analysis of these known sources, molecular gas information is not used as criterion for the morphological identification of the \SI{}{\tera\eV} sources as SNR candidates for the work presented in this paper.

\subsection{Choices for gas/spiral arm associations for the new TeV shells}
\label{SubSect:ShellsGasAssociations}

\begin{figure*}
\centering
\includegraphics[width=0.75\textwidth]{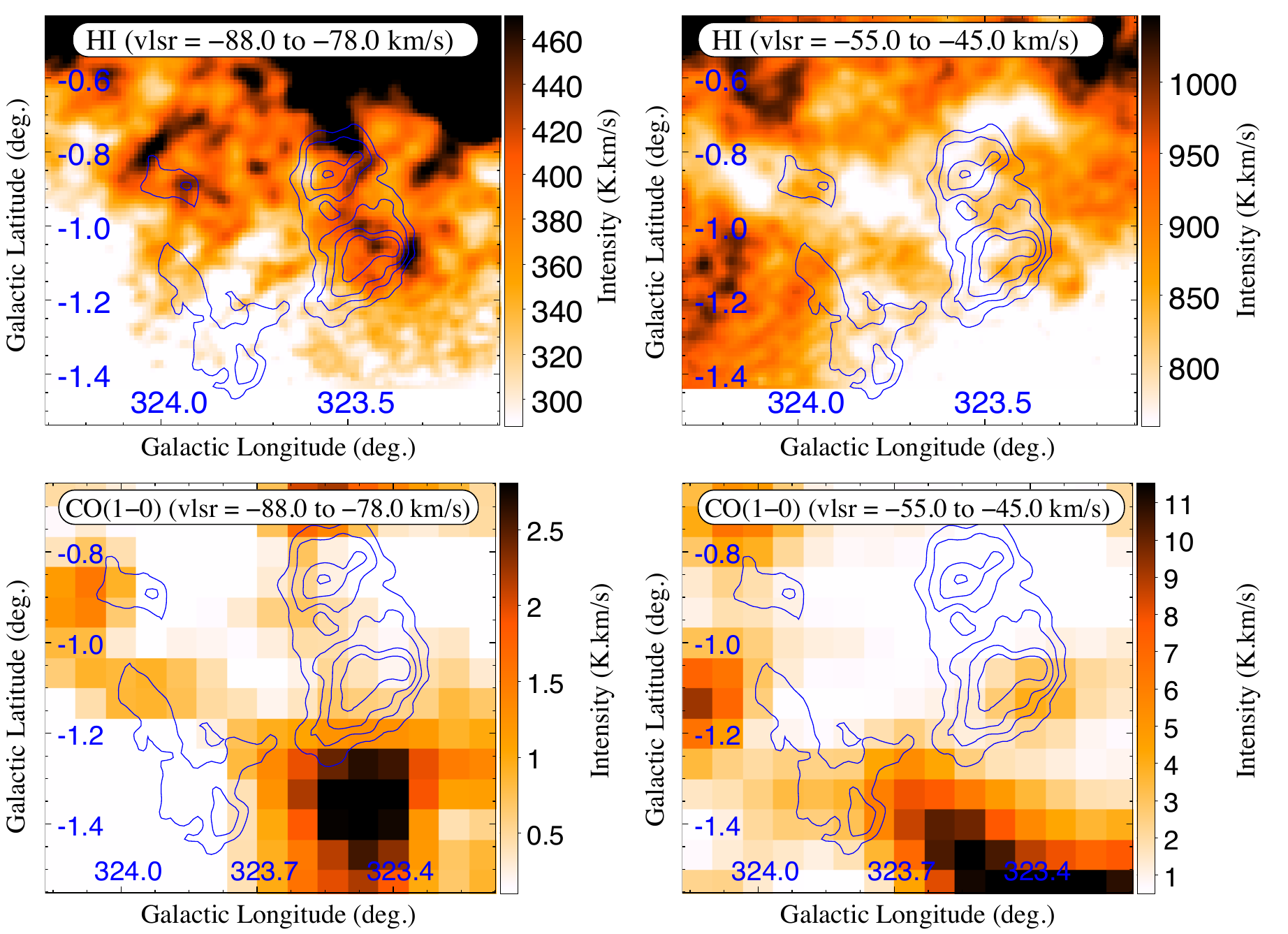}\\
\caption{HESS\,J1534$-$571. HI (top panels) and CO (bottom panels) intensity images toward the direction of HESS\,J1534$-$571 in velocity ranges as given on the images. HI data are from the Southern Galactic Plane Survey \citep{2005ApJS..158..178M}, CO data are from Nanten \citep{2001PASJ...53.1003M}. The slices in the left two panels correspond to a location in the Norma-Cygnus arm at a distance of \SI{8}{\kilo\parsec}, while the right two panels are from a location in the Scutum-Crux arm at a distance of \SI{3.5}{\kilo\parsec}. 
\label{Fig:1534hicoslices}}
\end{figure*}
        
\begin{figure*}
\centering
\includegraphics[width=0.75\textwidth]{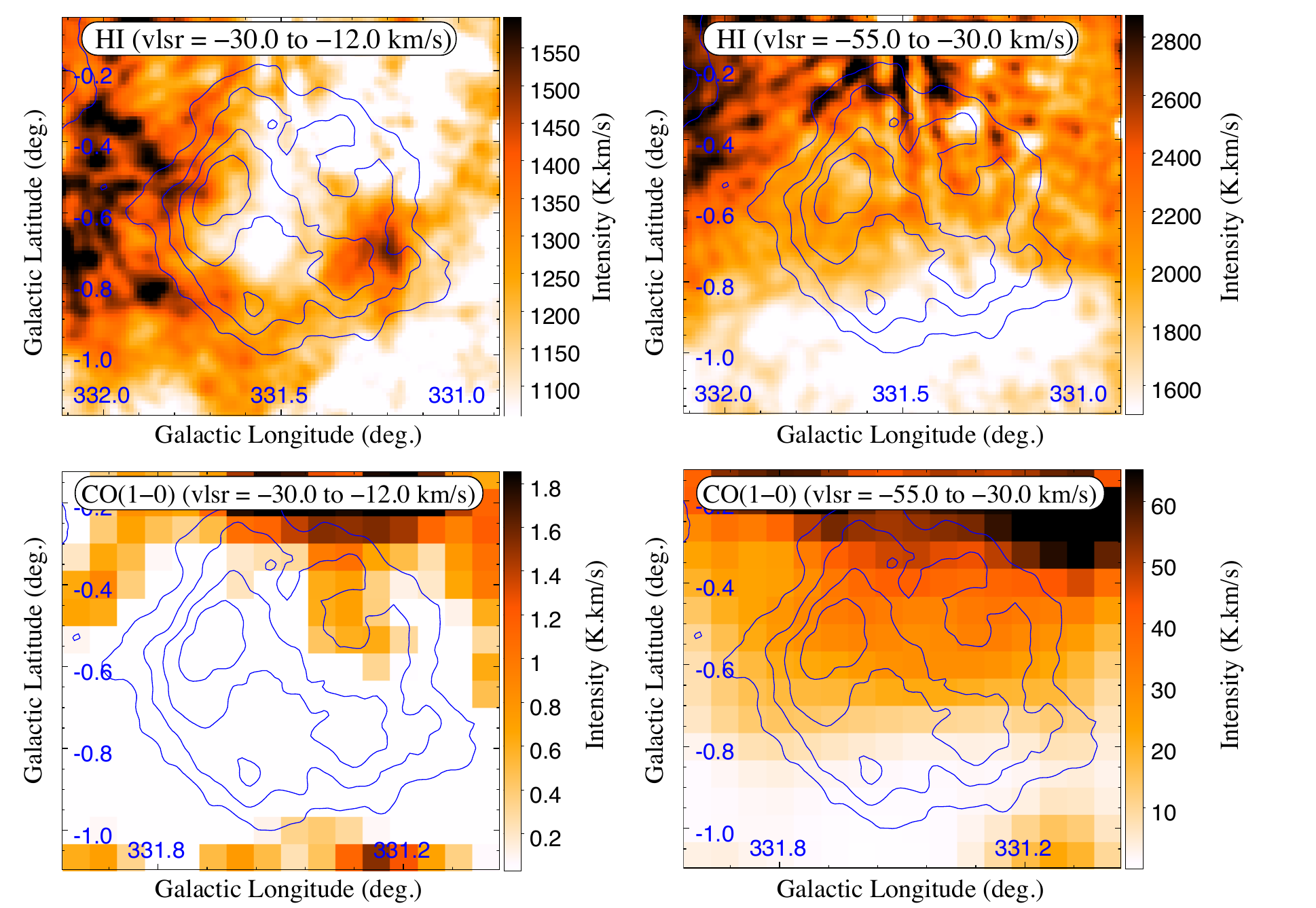}
\caption{HESS\,J1614$-$518. HI (top panels) and CO (bottom panels) intensity images toward the direction of HESS\,J1614$-$518, in velocity ranges as given on the images. HI data are from the Southern Galactic Plane Survey \citet{2005ApJS..158..178M}; CO data are from Nanten \citep{2001PASJ...53.1003M}. The slices in the left two panels correspond to a location in the Sagittarius-Carina arm at a distance of \SI{1.5}{\kilo\parsec}, while the right two panels are from a location in the Norma-Cygnus arm at a distance of \SI{5.5}{\kilo\parsec}. The HI images at the velocity range of \SIrange{-30}{-12}{\kilo\metre\per\second} indicate a possible void in the gas in directional coincidence with HESS J1614$-$518. At the center of this position and in compatible distance to Earth, the stellar cluster Pismis\,22 is located.
\label{Fig:1614hicoslices}}
\end{figure*}
 
\begin{figure*}
\centering
\includegraphics[width=0.75\textwidth]{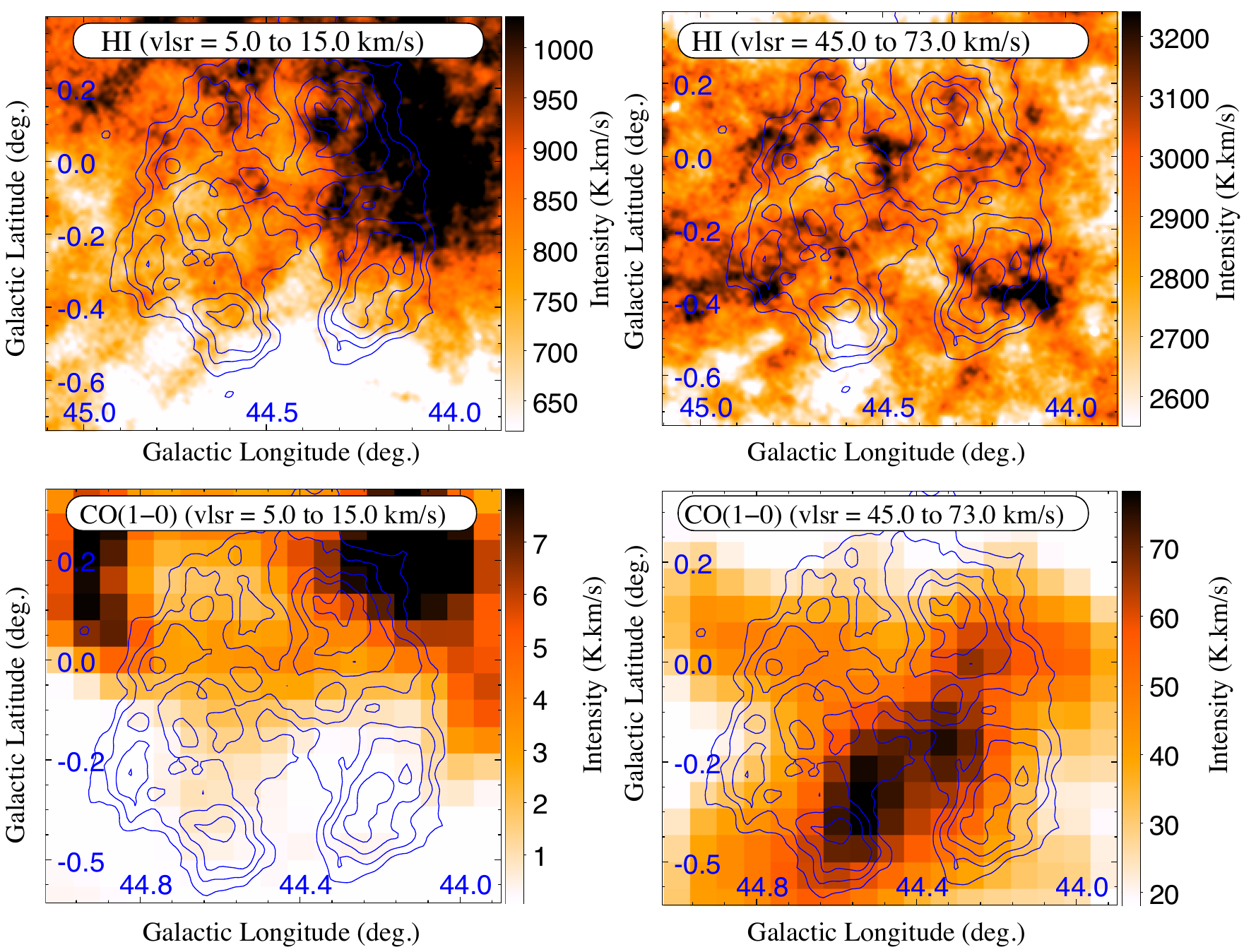}
\caption{HESS\,J1912$+$101. HI (top panels) and CO (bottom panels) intensity images toward the direction of HESS\,J1912$+$101, in velocity ranges as given on the images. HI data are from the VLA Galactic Plane Survey \citep{2006AJ....132.1158S}; CO data are from Nanten \citep{2001PASJ...53.1003M}. The slices in the left two panels correspond to a location in the Perseus arm at a distance of \SI{10}{\kilo\parsec}, while the right two panels are from a location in the Sagittarius-Carina arm at a distance of \SI{4.5}{\kilo\parsec}.
\label{Fig:1912hicoslices}}
\end{figure*}

For HESS\,J1534$-$571, one gas match around \SI{3.5}{\kilo\parsec} (right panels of Fig.\,\ref{Fig:1534hicoslices}) and one around \SI{8}{\kilo\parsec} (left panels of Fig.\,\ref{Fig:1534hicoslices}) were chosen. The latter distance might be compatible with the distance estimate of \SI{20}{\kilo\parsec} derived for the radio counterpart from the $\Sigma - D$ relation, given the uncertainties of the method. No attempt was made to construct a case with a distance far beyond the Galactic center distance.

For HESS\,J1614$-$518, the HI void distance range between \SIlist{1.2;1.5}{\kilo\parsec} as discussed in Sect.\,\ref{SubSect:Gas} has been chosen as a first possibility (left panels of Fig.\,\ref{Fig:1614hicoslices}). The distance is compatible with the estimated distance to Pismis\,22 by \citet{2000A&A...360..529P}. As a second possibility, a \SI{5.5}{\kilo\parsec} distance was selected (right panels of Fig.\,\ref{Fig:1614hicoslices}). The integrated proton column density to this distance from atomic and molecular hydrogen ($3-4 \times 10^{22}\,\mathrm{cm^{-2}}$) is in fact of similar order to the absorption column $N_{\mathrm{H}} \sim 1-2 \times 10^{22}\,\mathrm{cm^{-2}}$ that has been derived for XMMU\,J161406.0$-$515225 and Suzaku\,J1614$-$5141 \citep{2011PASJ...63S.879S}, which were suggested by these authors to possibly be associated with HESS\,J1614$-$518.\footnote{Part of the gas integrated along the line of sight might actually be attributed to a far distance solution beyond \SI{5.5}{\kilo\parsec} and might artificially increase the estimated column density.}  

Regarding HESS\,J1912$+$101, the dispersion distance of the possibly associated pulsar PSR\,J1913$+$1011 (cf.\ Sect.\,\ref{Subsect:pulsar1912}) of $\sim$\SI{4.5}{\kilo\parsec} is consistent with the range of distances expected to be found within the Sagittarius arm in this direction. One of the two distance ranges was therefore chosen as the approximate mid-range of the Sagittarius arm line-of-sight velocities, matching \SI{4.5}{\kilo\parsec} (right panels of Fig.\,\ref{Fig:1912hicoslices}). As a second possibility, another possible match within the Perseus arm at \SI{10}{\kilo\parsec} distance was chosen (left panels of Fig.\,\ref{Fig:1912hicoslices}). 

Gas tracers toward HESS\,J1912$+$101 were also recently investigated by \cite{2017ApJ...845...48S}. The authors have argued that $^{12}$CO and HI line velocity features found in their data are indicative of shock-gas interaction, possibly originating from an old ($\sim$\,$10^5$\,years) SNR that could be associated with HESS\,J1912$+$101. Here, we note that the line velocity range of the bulk of the corresponding molecular gas (around $v_\mathrm{lsr} \simeq 60\,\mathrm{km\,s^{-1}}$) is consistent with one of the two line velocity ranges chosen for HESS\,J1912$+$101 in this work.

All distances used were again estimated using figures from \citet{2008AJ....135.1301V,2013IJAA....3...20V}. 

\subsection{Gas density estimates for the new TeV shells}
\label{SubSect:ShellsGasMasses}

In order to derive the total (atomic and molecular) ISM density possibly associated with each respective \SI{}{\tera\eV} source, the gas emission is assumed to come from a homogeneous cylindrical region defined by the SNR outer boundary for a given velocity range. Molecular densities are derived using the CO-to-H$_2$ mass conversion factor $1.5 \times 10^{20}\mathrm{cm}^{-2}\,[\mathrm{K\,km\,s^{-1}}]^{-1}$ from \citet{2004A&A...422L..47S}, while atomic H densities are derived using the HI brightness temperature to column density conversion factor of $1.8 \times 10^{18}\,\mathrm{cm}^{-2}\,[\mathrm{K\,km\,s^{-1}}]^{-1}$ from \citet{1990ARA&A..28..215D}. For each case, a density range was calculated by assuming that the line-of-sight thickness is between a value either equal to the SNR candidate diameter or equal to the approximate thickness of a Galactic arm (\SI{0.5}{\kilo\parsec}). The first (upper) density value is therefore in all likelihood an overestimation. Moreover, it cannot be excluded that the density at the position of the source is well below the second (lower) value, for example,\ if the SNR is fully contained in a wind-blown bubble of the progenitor star. In such a case, however, a hadronic TeV emission scenario is anyway unlikely. Results are shown in Table\,\ref{Table:NewSNRShells_2}.

\end{document}